\documentclass[%
reprint,
amsmath,amssymb,
aps,
]{revtex4-2}

\usepackage{graphicx}
\usepackage{dcolumn}
\usepackage{bm}

\usepackage{amsmath}
\usepackage[export]{adjustbox}
\usepackage{enumitem}
\usepackage[usenames, dvipsnames]{color}
\usepackage{relsize}
\usepackage{enumitem}
\usepackage{lipsum}
\usepackage{varwidth}

\usepackage{bbm}
\usepackage{physics}
\usepackage{mathtools}
\usepackage{marvosym}

\newcommand\myMAPTOe{\stackrel{\mathclap{\normalfont\mbox{$O_{E_{}}(n_z) $}}}{\longmapsto}}

\newcommand\myMAPTOh{\stackrel{\mathclap{\normalfont\mbox{$O_{H_{}}(n_z+1) $}}}{\longmapsto}}

\usepackage{xcolor}
\usepackage{soul}

\definecolor{LightGreen}{rgb}{0.85,1.00,0.96}

\definecolor{MyBlue}{HTML}{e0fbf6}
\definecolor{MyGreen}{HTML}{f3f9e2}
\definecolor{MyGrayGreen}{HTML}{f0f4e3}
\definecolor{MyLightGrayBlue}{HTML}{f1f6fb}

\usepackage{physics}

\usepackage{fourier-orns}

\usepackage{siunitx}
\usepackage{esvect}

\begin{document}

\preprint{APS/123-QED}

\title{Closed-form analytical solution for the transfer matrix based on Pendry-MacKinnon discrete Maxwell's equations}

\author{Ovidiu-Zeno Lipan}%
 \email{olipan@richmond.edu}
\affiliation{
Department of Physics, Gottwald Center for the Sciences 138 UR Drive, University of Richmond, Richmond, VA 23173, USA
}%
\author{Aldo De Sabata}
\affiliation{
 Department of Measurements and Optical Electronics, Politehnica University of Timisoara, 300006 Timisoara, Romania}

\date{\today}

\begin{abstract}
Pendry and MacKinnon  meaningful discretization of  Maxwell's equations was put forward specifically as part of a finite-element numerical algorithm. By contrast with a numerical approach, in the same spirit evoked by the relationships between the difference and the differential equations, we provide an analytical solution for the transfer matrix elements generated by this discretization of  Maxwell's equations. The method, valid for all modes and any non-magnetic pattern with frequency-dependent permittivities, is exemplified for a bilaminar structure. The results from hundreds of  path-operators simplify to a small number of propagation channels, which transfer the electromagnetic field between input and output. A specific bilaminar device is design-optimized and used to generate  a topological map from permittivities to the number of modes and very high Q-factor resonances.
\begin{description}
\item[PACS Numbers]
78.67.Pt, 42.25.Fx
\end{description}
\end{abstract}

\maketitle



\section{\label{sec:RDIntro}Introduction}

What follows came about as an overlap of a present technological need with a perennial scientific quest. The current need is represented by photonic devices based on thin laminae,  be they in the form of waveguides for augmented reality \cite{xiong2021augmented}, \cite{kress2021waveguide} or of frequency-selective surfaces for the 5 and 6G networks \cite{shlezinger2021dynamic}, of wearable applications \cite{lorenzo2016modulated} or of surface wave-mediated sensing applications \cite{descrovi2010guided}, \cite{lova2018advances}. The perennial quest is to understand the  interplay between discrete and continuous models \cite{feynman2018simulating}, \cite{hagar2014discrete}. The focus on this association has varied since the 19th century, \cite{norlund1924vorlesungen},  piquing the interest of many in, for example, the theory of solitons \cite{nijhoff1995discrete},  universality in chaos \cite{feigenbaum}, \cite{cvitanovic2017universality}, fractals \cite{mandelbrot1982fractal}.  The manifestation that interests us  is  the relationship between the difference and the differential equations. For example, the discrete version of the equation of Korteweg and de Vries for soliton wave motion,\cite{korteweg1895xli}, shows a distinct asymmetric  richness  since it contains a series of universal  properties that are not readily visible in the continuum. Particularly, the discrete form of soliton equations proposed by  Hirota \cite{hirota1977nonlinear},\cite{hirota1981discrete} is connected with quantum integrable systems  \cite{krichever1997quantum}, \cite{lipan1997fusion}.
It is essential to emphasize that it is not at all trivial to discretize an equation in a way that preserves the prime properties of the original continuous equation. 
 
Pendry and MacKinnon obtained a meaningful discretization of  Maxwell's equations \cite{pendry1992calculation} on a carefully chosen simple cubic lattice \cite{pendry1994photonic}, by retaining the essential properties of the longitudinal modes present in the continuum. This discretization was put forward specifically  as part of a finite-element numerical algorithm \cite{pendry1996transfer}.
  
By contrast with a numerical approach, in the same spirit evoked by the relationships between the difference and the differential equations, we asked the question of whether a complete analytical solution can be found to the transfer matrix of the Pendry and MacKinnon discrete Maxwell's equations. 

The transfer matrix, \cite{li2003photonic}, is widely used well beyond the study of the propagation of electromagnetic fields \cite{sanchez2012transfer}, \cite{rhazi2010simple}. 
Its common appearance is in the form of a $2\times 2$ matrix, reduced to one propagation mode \cite{yeh1977electromagnetic}, \cite{fan2002sharp},\cite{koufidis2022mobius}.
However, we are looking here for an  analytical solution for the transfer matrix that is valid for any non-magnetic pattern and that has to encompass all Bloch-Floquet modes, either propagative or evanescent. As a consequence, our procedures differ from those employed in theories where the number of parameters are few and symmetries are many, \cite{faddeev1996algebraic},\cite{grammaticos2004discrete}.

In the ensuing sections we derive this  analytical solution and present the techniques we used, noteworthy among them being the definition of path-operators, Fig.\ref{fig:Flowchart}, and the appearance of propagation channels for the transfer of the electromagnetic field between input and output, Fig.\ref{fig:AllChannels}. This analytical solution is valid for filling the lattice's nodes with frequency-dependent non-magnetic isotropic material of any kind.  We present the general solution for two laminae, although the method can be extended to additional laminae. 

 To be more specific, in Sections \ref{sec:DiscreteMaxwell} and \ref{sec:Basis} we rewrite the Pendry and MacKinnon discrete version of  Maxwell's equations to take shape in a tensor product of two 2-dimensional linear spaces, (\ref{eq:14}) to (\ref{eq:16}) and (\ref{eq:20}) to (\ref{eq:22}). As a consequence, the electromagnetic states and the propagation operators are decomposed as a tensor product, (\ref{eq:pS+}). This course of action proves to be essential in simplifying a large volume of analytic computations. 
 Sections \ref{sec:Dispersion} and \ref{sec:Varphi} are dedicated to the discrete dispersion relation in vacuum and to the on-shell scattering method, \cite{pendry1974low}, from which a function along the z-coordinate is extracted, (\ref{eq:phizWitharg}). The Bloch-Floquet periodic boundary conditions along the $x$ and $y$ axes, but not along $z$, are introduced in Section \ref{sec:BF}. The building up of the transfer matrix from its elements is covered in Section \ref{sec:TM}.
 
  A key step towards the solution, the path-operators, is defined and discussed in Section \ref{sec:Paths}. The summation over the entire set of path-operators greatly simplifies the transfer matrix to reveal that the propagation run along a small number of channels, Section \ref{sec:Channels}. 
  
  To concretize the proposed method, which hinges on path-operators and channels, we completely solve for the transfer matrix elements of a bilaminar structure. The presentation of the solution starts in Section \ref{sec:Bilaminae} from a bird's eye view, and dives into the details in Sections \ref{sec:Channel3SS} and \ref{sec:ZandXYfactors}.  The classification of the channels by the  input-output polarization and the direction of propagation is summarized in Table \ref{tab:Channels}, Figs.\ref{fig:AllChannels} and \ref{fig:InSOutSPartitura.pdf} and Appendix B. 
  
  In Section \ref{sec:Continuum} we segue into the continuum limit, and then apply it to the specific bilaminar pattern from Fig.\ref{fig:PiPiRig.pdf}.
   The limit is in the $x$ and $y$ directions and it reveals that channels can be classified by a set of universal functions. It also calls for computing 2-dimensional Fourier integrals of products of these functions, each function's argument being defined by the permittivities of a lamina, Section \ref{sec:Continuum} and \ref{sec:Cscaling}. For polygonal tessellated laminae, Section \ref{sec:Tessellation}, these integrals can be computed in a closed form, Section \ref{sec:FourierChi} and \ref{sec:Fourier11}. The formulas contain a gauge freedom and are arranged as a sum of terms over the vertices of the tessellated polygons. These terms hold all the information about the non-magnetic materials and the geometric orientations of the structure, as can be seen in (\ref{eq:ResultIntegralC3SS}).
  
Lastly, from the formulas for the transfer matrix, an optimization procedure is devised in Section \ref{sec:Optim} to create a dielectric bilaminar structure, Fig.\ref{fig:PiPiRig.pdf}, on which Bloch-Floquet modes propagate along the laminae with large amplitudes, being evanescent in the z-direction \cite{png2017nanophotonics}. These resonances display a series of relevant properties.  One is the robustness with respect to the change of the dielectric constants, Section \ref{sec:Map}. The resonant frequencies are located close to the zeroes of the diagonal elements of the transfer matrix. As the dielectric permittivities pass through different values, the zeroes are created or disappear  allowing us to define a topological map that counts the number of resonant modes, Fig.\ref{fig:NumberOfModes}.
  
On a separate note, in Section \ref{sec:ResonantFreq} and \ref{sec:HighQ} we find that the same bilaminar structure, but with specifically selected  dielectric constants, resonates with a very large  Q-factor, Fig.\ref{fig:ResonanceEpsilonHighQ}. Finally, in Section \ref{sec:ComplexOmega}, the complete analytical formula offers the possibility to obtain the Fano-Lorentz spectral line shape for a resonant mode, \cite{fano1935sullo},\cite{avrutsky2013linear}, through a zero-pole approximation (\ref{eq:ZeroPole}), in the complex frequency plane Fig.\ref{fig:ComplexPlaneFrequency}.

\section{Discrete Maxwell's equations on a lattice}\label{sec:DiscreteMaxwell}

\begin{figure}[h]
\includegraphics[scale=1.0]{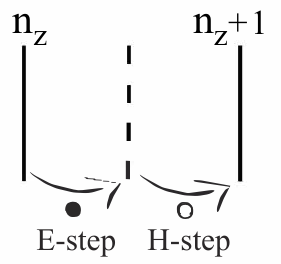}
\caption{\label{fig:FigC1} Propagation between two planes. The dotted plane is fictitious and is used to place the field configuration transformed by $O_E(n_z)$. The argument $n_z$ emphasizes that $O_E$ depends exclusively on the material properties $\Gamma_{E/H}(n_\perp,n_z)$  on the $(x,y)$-plane located at $n_z$.}
\end{figure}

Space is an orthogonal lattice with lattice constants $a,b$ and $c$ on the $x,y$ and $z$ directions, respectively. The z-axis is singled-out as the propagation axis for the transfer matrix. The discrete space is thus viewed as a sequence of $x,y$-parallel planes labeled by the $z$-axis coordinate. The discrete position vector is $\vv{n}=(a n_x,b n_y,c n_z)=(n_{\perp},c n_z)$, where the notation $n_{\perp}=(a n_x,b n_y)$ emphasizes the decomposition of the space as a collection of planes perpendicular to the $z$-axis, as in \cite{schwinger1998classical}. If there is a preference to view the collection of planes as central, then $\perp$ should be changed to $\parallel$ everywhere in the paper. At each point of the lattice, the electromagnetic field is described by six components, $\vv{E}$ and $\vv{H}$. The material attached to each lattice point is characterized by two discrete scalar fields which connect $\vv{D}$ to $\vv{E}$ and $\vv{B}$ to $\vv{H}$, respectively:
\begin{eqnarray}
  \vv{D}&=&\epsilon_0\epsilon(n_{\perp},c n_z)\vv{E}, \\
  \vv{B}&=&\mu_0\mu(n_{\perp},c n_z)\vv{H}.
\end{eqnarray}
In the discrete electrodynamics that we are taking forward, the $\nabla$ operator becomes a pair, $(\vv{T}^E ,\vv{T}^H )$, of lattice operators,  where $\vv{T}^E=\left(T_x^E(a),T_y^E(b),T_z^E(c)\right)$ and $\vv{T}^H=\left(T_x^H(a),T_y^H(b),T_z^H(c)\right)$. These discrete operators are defined by their action on a lattice scalar function $f(\vv{n})$. For the x-axis
\begin{align}
	T_x^E(a)f(\vv{n})&=(i a)^{-1}(f(\vv{n}+a \hat{x})-f(\vv{n})),\\
	T_x^H(a)f(\vv{n})&=-(i a)^{-1}(f(\vv{n}-a \hat{x})-f(\vv{n})).
\end{align}
with $(y,b)$ and $(z,c)$ taking place of $(x,a)$, for the other axes.
The Maxwell's equations on the lattice are
\begin{eqnarray}
  \vv{T}^E \times \vv{E}&=& {\Gamma_H}^{-1} \vv{H}^\prime , \\
  \vv{T}^H\times \vv{H}^\prime &=& {\Gamma_E}^{-1} \vv{E},
\end{eqnarray}
where
\begin{eqnarray}
 {\Gamma_H}^{-1} &=& -i\omega^2 \epsilon_0\mu_0 \mu(n_{\perp},c n_z) c,\\
 {\Gamma_E}^{-1} &=&-i \epsilon(n_{\perp},c n_z)c^{-1},\\
  \vv{H}^\prime &=&i(\omega \epsilon_0 c)^{-1}\vv{H}.
\end{eqnarray}
Eliminating $E_z$ and ${H^\prime}_z$ from  Maxwell's equations, the propagation of the  4-dimensional vector field 
\begin{equation}\label{eq:ketF}
   \ket{F(n_\perp,n_z)}=
          \begin{bmatrix}
           E_x(n_{\perp},c n_z) \\
           E_y(n_{\perp},c n_z) \\
           H^{\prime}_x(n_{\perp},c n_z)\\
           H^{\prime}_y(n_{\perp},c n_z)
          \end{bmatrix}
 \end{equation}
is carried on, as usual, by two operators. The first operator $O_E$, (\ref{eq:TransferedByOE}), advances only the $E$-components Fig.(\ref{fig:FigC1})

\begin{align}\label{eq:TransferedByOE}
        \begin{matrix}
          \begin{bmatrix}
           E_x(n_{\perp},c n_z) \\
           E_y(n_{\perp},c n_z) \\
           H^{\prime}_x(n_{\perp},c n_z)\\
           H^{\prime}_y(n_{\perp},c n_z)
          \end{bmatrix}&\;& \myMAPTOe &\;&
          \begin{bmatrix}
           E_x(n_{\perp},c n_z+c) \\
           E_y(n_{\perp},c n_z+c) \\
           H^{\prime}_x(n_{\perp},c n_z)\\
           H^{\prime}_y(n_{\perp},c n_z)
          \end{bmatrix}
    \end{matrix}.
  \end{align}
The material properties $\Gamma_H$ and $\Gamma_E$ embedded in the $O_E$ matrix elements depend only on the spatial position $\vv{n}$ located on the plane $n_z$.  The operator $O_E$ acts on the internal index $\alpha$ of the vector field $F_{\alpha}$, (\ref{eq:ketF}), as a 4 by 4 matrix multiplication, whereas the matrix elements of $O_E$ are operators on the external space variable $n_{\perp}$ of the field, but not on the $n_z$-position.
The $E$-step carried by $O_E$ is followed by an $H$-step carried by $O_H$
\begin{align}
        \begin{matrix}
          \begin{bmatrix}
           E_x(n_{\perp},c n_z+c) \\
           E_y(n_{\perp},c n_z+c) \\
           H^{\prime}_x(n_{\perp},c n_z)\\
           H^{\prime}_y(n_{\perp},c n_z)
          \end{bmatrix}&\;& \qquad & \myMAPTOh &\qquad & \; &
          \begin{bmatrix}
           E_x(n_{\perp},c n_z+c) \\
           E_y(n_{\perp},c n_z+c) \\
           H^{\prime}_x(n_{\perp},c n_z+c)\\
           H^{\prime}_y(n_{\perp},c n_z+c)
          \end{bmatrix}
    \end{matrix}.
  \end{align}
In contrast to $O_E$, the material properties $\Gamma_H$ and $\Gamma_E$ for the $O_H$ matrix elements depend on the spatial position on plane $n_z+1$.
The complete propagation from the plane located at $n_z$ to the one located at $n_{z}+1$ is accomplished by $O_H(n_\perp,c n_z+c) O_E(n_\perp,c n_z)$.

It is apparent from the structure of the operators $O_E$ and $O_H$ that the internal $4$-dimensional space can be viewed as a tensor product of two $2$-dimensional spaces. This decomposition proves to be efficient in classifying different paths taken by the electromagnetic propagation through materials, which will be one of our main instruments to obtain closed analytic formulas for the transfer matrix elements. So, departing from \cite{pendry1996transfer}, the path viewpoint is expressed by writing the operators as
\begin{eqnarray}\label{eq:SumEOperators}
  O_E (n_z)&=& O_E^{(1)}+O_E^{(2)}+O_E^{(3)}, \\\label{eq:14}
   O_E^{(3)} &=& \mathbbm{1} \otimes  \mathbbm{1},
    \\\label{eq:15}
   O_E^{(2)} &=& I_E\otimes \sigma_2 \Gamma_H^{-1}(-c),\\\label{eq:16}
   O_E^{(1)} &=&I_E\otimes \ket{T_E} \Gamma_E \bra{T_E}\sigma_2(-c),
\end{eqnarray}
where $\sigma_2$ is the Pauli matrix and
\begin{equation}
\begin{matrix}
\mathbbm{1}=\begin{bmatrix}
                     1 & 0 \\
                     0 & 1 \\
                   \end{bmatrix}
&,
I_E=\begin{bmatrix}
                     0 & 1 \\
                     0 & 0 \\
                   \end{bmatrix}
&,
I_H=\begin{bmatrix}
                     0 & 0 \\
                     1 & 0 \\
                   \end{bmatrix}
\end{matrix}.
\end{equation}

The ket and the bra operators form (\ref{eq:16}) act  on the internal space as 2-dimensional vectors and on the external space variable  $n_{\perp}$ as operators on functions
\begin{equation}
\begin{matrix}\label{eq:TEketbra}
\ket{T_E}=\begin{bmatrix}
T_x^E\\ \\T_y^E
\end{bmatrix}
&,
\bra{T_E}=\begin{bmatrix}
T_x^H&T_y^H
\end{bmatrix}
\end{matrix}.
\end{equation}

The ket to bra conjugation is defined as the change of $E$ into $H$. There is no  additional conjugation over the complex unit $i$. The labels, as is usually the case, do not change under conjugation.

Similarly, but different to the E-step, the operator for the H-step is
\begin{eqnarray}\label{eq:SumHOperators}
  O_H(n_z+1) &=& O_H^{(1)}+O_H^{(2)}+O_H^{(3)}, \\\label{eq:20}
   O_H^{(3)} &=& \mathbbm{1} \otimes  \mathbbm{1},
    \\\label{eq:21}
   O_H^{(2)} &=& I_H\otimes \sigma_2 \Gamma_E^{-1}(-c),\\\label{eq:22}
   O_H^{(1)} &=&I_H\otimes \ket{T_H} \Gamma_H \bra{T_H}\sigma_2(-c),
\end{eqnarray}
\begin{equation}\label{eq:THketbra}
\begin{matrix}
\ket{T_H}=\begin{bmatrix}
T_x^H\\ \\T_y^H
\end{bmatrix},
&
\bra{T_H}=\begin{bmatrix}
T_x^E&T_y^E
\end{bmatrix}
\end{matrix}.
\end{equation}

Concordant to the tensor product decomposition of the operators, the field will be represented also as a tensor product,
\begin{gather}\label{eq:ketFtensor}
	\begin{aligned}
   \ket{F(n_\perp,n_z)}=&\\
   \ket{e}\otimes
          \begin{bmatrix}
           E_x(n_{\perp},c n_z) \\
           E_y(n_{\perp},c n_z)
          \end{bmatrix}+\ket{h}\otimes&\begin{bmatrix}
           H^{\prime}_x(n_{\perp},c n_z)\\
           H^{\prime}_y(n_{\perp},c n_z)
          \end{bmatrix},
      \end{aligned}
 \end{gather}
where the 2-dimensional vectors $\ket{e}$ and  $\ket{h}$, 
\begin{equation}
\begin{matrix}
\ket{e}=\begin{bmatrix}
1\\ \\0
\end{bmatrix}
&,
\ket{h}=\begin{bmatrix}
0\\ \\1
\end{bmatrix}
\end{matrix},
\end{equation}
select the electric part $(E_x,E_y)$ and the magnetic part  $(H^{\prime}_x,H^{\prime}_y)$  of the 4-dimensional vector $(E_x,E_y,H^{\prime}_x,H^{\prime}_y)$, respectively.

\section{Basis vectors for fields on a lattice}\label{sec:Basis}

The solutions to  Maxwell's equations on a lattice filled with a homogeneous medium, i.e. constant $\Gamma_H(n_\perp,c n_z)=\Gamma_H$ and $\Gamma_E(n_\perp,c n_z)=\Gamma_E$, are plane waves.
 These solutions, $\ket{\Psi_{S/P,p_\perp}^{\pm}} e^{\pm i p_z n_z c} e^{- i \omega t}$, are distinguished by the type of polarization $S$ or $P$ and by the direction of propagation $\pm$ along the positive or negative $z$-direction. Here
\begin{equation}\label{eq:PropagationS}
  \ket{\Psi_{S/P,p_\perp}^{\pm}}=\ket{p_\perp,S/P,\pm}e^{i p_\perp n_\perp},
\end{equation}
with the propagation in the $(x,y)$-plane  described by $p_\perp=(p_x,p_y)$.

The internal $4$-dimensional vectors of the plane waves are 
\begin{gather}\label{eq:pS+}
\begin{aligned}	
\ket{p_\perp,S,+}&=\ket{e}\otimes i \sigma_2 \ket{p_\perp^H}+\ket{h}\otimes \ket{p_\perp^H} \Gamma_H k_z^E,\\
\ket{p_\perp,P,+}&=\ket{e}\otimes \ket{p_\perp^E} \Gamma_E k_z^H+\ket{h}\otimes i \sigma_2 \ket{p_\perp^E},
\end{aligned}
\end{gather}
where, consistent with (\ref{eq:TEketbra}) and (\ref{eq:THketbra}), the column vectors  $\ket{p_{\perp}^{E}}$ and $\ket{p_{\perp}^{H}}$ are the transpose and $E\rightleftarrows H$ conjugate of the row vectors $\bra{p_{\perp}^{E}}=[k_x^{H},k_y^{H}]$ and $\bra{p_{\perp}^{H}}=[k_x^{E},k_y^{E}]$, respectively.
 The space being discrete, the  plane wave depends on the lattice constants $(a,b,c)$ through the vectors $\vv{k}^E=(k_x^E,k_y^E,k_z^E)$ and $\vv{k}^H=(k_x^H,k_y^H,k_z^H)$,
\begin{gather}\label{eq:KxE}
	k_x^E = (i a)^{-1}(e^{i a p_x}-1),\\
	k_x^H = -(i a)^{-1}(e^{-i a p_x}-1), 
\end{gather}
with the pairs $(b,p_y)$ and $(c,p_z)$ instead of $(a,p_x)$, for the other axes. To obtain the other two basis vectors, $\ket{p_\perp,S/P,-}$, change $ k_z^E$ from  $\ket{p_\perp,S/P,+}$ into $- k_z^H$.

To derive the transfer matrix elements, the field (\ref{eq:ketFtensor}), at a given $n_z$, is decomposed in the basis $\ket{\Psi_{S/P,p_\perp}^{\pm}}$. The decomposition is carried out by projecting on an orthogonal set of bra vectors (\ref{eq:braPHI})

\begin{gather}\label{eq:braPHI}
	\bra{\Phi_{S/P,p_\perp}^{\pm}}=\bra{p_\perp,S/P,\pm}e^{-i p_\perp n_\perp},
\end{gather}	
where
\begin{gather}\label{eq:braPHIvectors}
\begin{aligned}	
	\bra{p_\perp,S,+}&=-\bra{e}\otimes  \bra{p_\perp^H}i \sigma_2 k_z^H+\bra{h}\otimes \bra{p_\perp^H} \Gamma_H^{-1},
\\
	\bra{p_\perp,P,+}&=\bra{e}\otimes \bra{p_\perp^E}\Gamma_E^{-1}-\bra{h}\otimes \bra{p_\perp^E}i \sigma_2 k_z^E.
\end{aligned}	
\end{gather}
Multiply  $\bra{p_\perp,S/P,+}$ by $-1$ and change $k_z^H$ into $-k_z^E$ to attain  $\bra{p_\perp,S/P,-}$. The orthogonal basis is not normalized. The norm,  $\bra{p_{\perp}^H}\ket{p_{\perp}^H}( k_z^H+k_z^E)=\bra{p_{\perp}^E}\ket{p_{\perp}^E}( k_z^E+k_z^H)$, is invariant to $E\leftrightarrows H$ conjugation. 

\section{Dispersion relation for a homogeneous medium}\label{sec:Dispersion}

The requirement that the plane wave basis satisfy  Maxwell's equations on the lattice filled with a homogeneous medium imposes that
\begin{equation}\label{eq:DispersionRelation}
  -(\Gamma_H \Gamma_E)^{-1}=\bra{p_{\perp}^H}\ket{p_{\perp}^H}+k_z^E k_z^H.
\end{equation}
This is the dispersion relation for the homogeneous medium
\begin{eqnarray}\label{eq:DispersionRelation}
 \frac{\omega^2}{v_0^2}=k_x^E k_x^H+k_y^E k_y^H+k_z^E k_z^H,
\end{eqnarray}
where the velocity of light in that medium is $ v_0$. 
Through the following c-scaling
\begin{equation}\label{eq:DefinitionOmega}
	\frac{\omega}{v_0}=\frac{\Omega}{c}=\frac{2 \pi}{\lambda},
\end{equation}
the dispersion relation is written in terms of a unitless frequency $\Omega$
\begin{equation}\label{eq:DispersionInOmega}
	\frac{\Omega^2}{c^2}=k_x^Ek_x^H+k_y^Ek_y^H+k_z^Ek_z^H,
\end{equation}
which will become useful for expressing the transfer matrix elements in terms of unitless parameters.
Note that the term $k_x^Ek_x^H+k_y^Ek_y^H$, which is equal to $\bra{p_{\perp}^{\text{E}}}\ket{p_{\perp}^{\text{E}}}=\bra{p_{\perp}^{\text{H}}}\ket{p_{\perp}^{\text{H}}}$, tends in the continuum limit $a \to 0,b\to 0$ to $p_x^2+p_y^2$. In what follows, we use vacuum as the homogeneous medium for the basis. Interchangeably, this medium will be called also as the reference medium for the basis.

\section{Solving for $\varphi _{z}=e^{i c p_z}$ for given $\Omega$, $p_x$ and $p_y$}\label{sec:Varphi}

The approach in \cite{pendry1992calculation} is to work on-shell which means solve the dispersion relation for $p_z$, given the frequency $\Omega$ and  $p_{\perp}$.
To this end, we introduce the function $\varphi _{z}$
\begin{equation}\label{eq:PhiVersusPz}
	\varphi _{z}=e^{i c p_z},
\end{equation}
which is the solution to the equation
\begin{equation}\label{eq:PhizEquation}
	\varphi _{z}+\varphi _{z}^{-1}=2\zeta,
\end{equation}
where
\begin{equation}\label{eq:PhizEquation2}
	\zeta=1-\frac{\Omega^2}{2}+\frac{c^2}{2}\bra{p_{\perp}^{\text{H}}}\ket{p_{\perp}^{\text{H}}}.
\end{equation}

There are two distinct and non-overlapping constraints that we impose on $p_z$ from (\ref{eq:PhiVersusPz}). One condition is $\text{Im}(p_z)\geqslant 0$ so that evanescent waves generated away from sources placed at $z=-\infty$ or at $z=\infty$  are represented by a decaying $\varphi_z$ or $\varphi_z^{-1}$, respectively. The other condition is $p_z\geqslant0$ enforcing the velocity of   propagating waves coming from $z=-\infty$ or $z=\infty$ to be oriented along  $\hat{z}$ or 
 $-\hat{z}$, respectively. The formal solution to the equation (\ref{eq:PhizEquation}), $	\varphi_{z}=\zeta \pm \sqrt{\zeta^2-1}$
divides the real axis  into three intervals, $(-\infty,-1)$, $(-1,1)$ and $(1,\infty)$. The requirement $\text{Im}(p_z)\geqslant0$  for the interval $(-\infty,-1)$ selects the  $\varphi_{z}=\zeta - \sqrt{\zeta^2-1}$
but does not impose specific values for $\varphi_z$ on $(-\infty,-1)$. Both walks in the $\zeta$-complex plane  above or below $(-\infty,-1)$ satisfy  $\text{Im}(p_z)\geqslant 0$. 
The second constrain, $p_z\geqslant 0$,  requires the values of  $\varphi_z$ on $(-1,1)$ to be those obtained from walking below the segment $(-1,1)$. For the last interval the constrain  $\text{Im}(p_z)\geqslant 0$ is fulfilled and, like for  $(-\infty,-1)$, we are free to define $\varphi_z$ on  $(1,\infty)$ from walking above or below the interval. 
 We opt to construct a function $\varphi_z$ continuous from below, $\text{Im}(\zeta)\leqslant 0$, and so select the values of $\varphi_z$ on the real axis by walking below it in the $\zeta$-complex plane.
 
\begin{figure}[h]
	\includegraphics[scale=1.0]{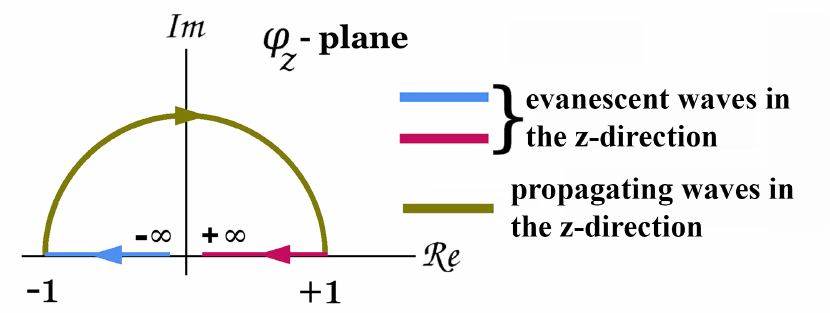}
	\caption{\label{fig:FigPlanePhiZ}The complex plane of $\varphi_z$. As $\zeta$ variable runs from $-\infty$ to $-1$,  $\varphi_z$ goes from $0$ to $-1$. Then, as  $\zeta$ runs from $-1$ to $1$ below the real $\zeta$-axis,  $\varphi_z$ follows the upper semicircle from $-1$ to $1$. Finally, the variable $\varphi_z$ moves back toward $0$ from $1$ as $\zeta$ covers the range from $1$ to $\infty$.}
\end{figure}

 This analysis leads to the following solution
\begin{equation}\label{eq:phizWitharg}
 	\varphi_z(\zeta)=\zeta-\sqrt{\abs{\zeta^2-1}}e^{\frac{i}{2}\big(\text{arg}(\zeta-1)+\text{arg}(\zeta+1)\big)}.
 \end{equation}
The $\text{arg}(\zeta)$ takes values on $[-\pi,\pi)$ in such a way  that $\text{arg}(\zeta)=-\pi$ if $\text{Im}(\zeta)=0$ and $\zeta<0$.
In Fig.\ref{fig:FigPlanePhiZ} the point $\varphi_z=1$ defines the Rayleigh frequency, at which, for a given $p_{\perp}$, the waves turn from evanescent to propagating.

\section{Periodic boundary conditions}\label{sec:BF}

Consider that the electromagnetic field is subjected to the Bloch-Floquet boundary conditions
\begin{align}
\ket{F(n_{\perp}+(a,0)L_x,c n_z)}=	e^{2\pi i \xi_x }\ket{F(n_{\perp},c n_z)},\\
\ket{F(n_{\perp}+(0,b)L_y,c n_z)}=	e^{2\pi i \xi_y }\ket{F(n_{\perp},c n_z)},
\end{align}
where $L_x\geqslant 2$ and $L_y\geqslant 2$ are two integer numbers selected to describe the dimensions of a unit cell in the $(x,y)$-plane. The material properties are taken to be periodic from one unit cell to another
\begin{eqnarray}
	\Gamma_H(n_\perp+(a,0)L_x,n_z) &=& \Gamma_H(n_\perp,n_z) \\
	\Gamma_H(n_\perp+(0,b)L_y,n_z) &=& \Gamma_H(n_\perp,n_z) 
\end{eqnarray}
with a similar periodicity for $\Gamma_E.$ The Bloch-Floquet boundary condition reduces the possible values for $(p_x,p_y)$ to
\begin{gather}\label{eq:MXMY}
	p_x = 2 \pi \frac{M_x}{a L_x}+ 2 \pi \frac{\xi_x}{a L_x},\\\label{eq:MXMY2}
	p_y = 2 \pi \frac{M_y}{b L_y}+ 2 \pi \frac{\xi_y}{b L_y},
\end{gather}
where  $M_x=0,1,\dots,L_x-1$ and $M_y=0,1,\dots,L_y-1$, with $0\leqslant\xi_x<1$, $0\leqslant\xi_y<1$.

The indices of the basis, written as $\ket{\psi_\text{POL,BF}^{\text{DIR}}}$, are thus of four categories. The first, POL, is the polarization index, which takes two values S and P. The second, DIR, is the direction of propagation of the plane wave along the z-axis, which also take two values, $+$ and $-$. The third, BF, is the Bloch-Floquet index consisting of pairs of integer numbers $(M_x,M_y)$ and  it takes a finite number of values. The last index, $(\xi_x,\xi_y)$  takes only one value and can be viewed as a label for the entire Bloch-Floquet linear basis.

The parameters $(\xi_x,\xi_y)$ specify the direction of the wave in the $(x,y)$-plane. For example, if the direction of the incoming mode $(M_x=0,M_y=0)$ is stated in spherical coordinates by the angles $(\theta,\varphi)$, then the parameterization of $\xi_x$ and $\xi_y$ is
\begin{eqnarray}\label{eq:XiToSphericalCoordinates}
	\sin(\pi\frac{\xi_x}{L_x}) &=& \frac{a \Omega}{2 c}\sin(\theta) \cos(\varphi), \\
	\sin(\pi\frac{\xi_y}{L_y}) &=& \frac{b \Omega}{2 c}\sin(\theta) \sin(\varphi), \\
	\sin(\frac{1}{2}c p_z) &=& \frac{c \Omega}{2 c}\cos(\theta).
\end{eqnarray}

The first two equations give $\xi_x,\xi_y$ for  chosen $\theta, \varphi$ and $\Omega$, whereas  the third equation solves for $p_z$.

\section{Transfer and scattering matrices}\label{sec:TM}	
To go ahead and propagate from the IN  plane located at $n_z$ to the OUT at $n_z+L_z$, decompose the IN and OUT fields in the plane wave basis, (\ref{eq:PropagationS}), labeled  as $\ket{\psi_\text{pol,bf}^{\text{dir}}}$ and $\ket{\psi_\text{POL,BF}^{\text{DIR}}}$, respectively. The z-phases, $e^{\pm i p_{z}^{\text{IN}} n_{z}}$ and  $e^{\pm i p_{z}^{\text{OUT}} (n_{z}+L_z)}$, which are not included in the basis, are generated by the operators and so they do appear in the transfer matrix elements.
The transfer matrix elements are defined in (\ref{eq:DefinitionTransferElements}) as the factors that connect the coefficients of the $\ket{F^{\text{IN}}(n_\perp,n_z)}$ and $\ket{F^{\text{OUT}}(n_\perp,n_z+L_z)}$ fields decomposed in the reference medium plane wave basis
\begin{equation}\label{eq:DefinitionTransferElements}
	C_\text{POL,BF}^{\text{DIR}}=\sum \left(\large{T}^{\text{DIR};\text{dir}}_{\text{POL,BF};\text{pol,bf}}\right ) C_\text{pol,bf}^{\text{dir}}\;,
\end{equation}
the sum being over indices (dir,pol,bf) related to the IN parameters.
The transfer matrix elements $T$ in (\ref{eq:DefinitionTransferElements}) borrow, from the basis, all four index categories.
Building a matrix out of the $T$-coefficients of (\ref{eq:DefinitionTransferElements})  requires a map from the 2-dimensional Bloch-Floquet index $(M_x,M_y)$  to a 1-dimensional integer index $m=1,2,\cdots, L_x L_y$. We use the Rayleigh frequency of $(M_x,M_y)$ to both map and order the Bloch-Floquet index. If the Rayleigh frequency of $(M_x,M_y)$ is less than the Rayleigh frequency of $(M_x^{'},M_y^{'})$ we order the Bloch-Floquet indices as $(M_x,M_y)<(M_x^{'},M_y^{'})$. If two distinct Bloch-Floquet modes have the same Rayleigh frequency, the order used is the lexicographic order. Note that this ordering depends on $(\xi_x,\xi_y)$. The mode $(M_x,M_y)=(0,0)$ has $f_{\text{Rayleigh}}=0$ and it is mapped into $m=1$. The next  $f_{\text{Rayleigh}} > 0$ is mapped into $m=2$ and so on Table\ref{tab:Rayleigh} from Appendix I.

The process of building the transfer matrix out of its elements starts with  all $2\times 2$ matrices 
\begin{equation}\label{eq:TMMmn}
	T^{-,-}_{m,n}=	\left(
	\begin{array}{cc}
		T^{-,-}_{Sm,Sn}  &  T^{-,-}_{Sm,Pn}  \\
		T^{-,-}_{Pm,Sn}  &  T^{-,-}_{Pm,Pn}  \\
	\end{array}
	\right)
\end{equation} 
constructed for each pair $(m,n)$.
These matrices contain the information about the transfer, through the device, of the mode $n$ with a negative direction present at IN into the mode $m$, also with a negative direction but located at OUT. 

From these $ 2\times 2$-matrices, construct the transfer matrix $T^{-,-}$ by letting the Bloch-Floquet indices $m$ and $n$  run from 1 to $L_x L_y$
\begin{equation}\label{eq:TMM}
	T^{-,-}=	\left(
	\begin{array}{ccccc}
		T^{-,-}_{1,1}  & 	T^{-,-}_{1,2} & T^{-,-}_{1,3}&\cdots\\
		T^{-,-}_{2,1}  & 	T^{-,-}_{2,2} & T^{-,-}_{2,3}&\cdots\\
		T^{-,-}_{3,1}  & 	T^{-,-}_{3,2} & T^{-,-}_{3,3}&\cdots\\
		\cdots & \cdots &  \cdots & \cdots   \\
	\end{array}
	\right).
\end{equation}

Once  all four matrices $T^{-,-}, T^{-,+},	T^{+,-}, T^{+,+}$ are constructed in a similar manner, the complete transfer matrix appears as
\begin{equation}\label{eq:DeffTransfMatr}
	T(\xi_x,\xi_y)=	\left(
	\begin{array}{cc}
		T^{+,+}  & T^{+,-}  \\
		T^{-,+}  & T^{-,-}  \\
	\end{array}
	\right),
\end{equation}
where $(\xi_x,\xi_y)$ are explicitly written here to remind us of their role. 

The elements of the scattering matrix can now be obtained from the complete transfer matrix. The input and output have a different meaning for the scattering matrix than OUT and IN have for the transfer matrix. Instead of "DIR/dir", for the scattering matrix we use a position index, which is either $min$ or $max$ depending on the position on the z-axis of the field $\ket{F}$, at $z=n_z$ or at $z=n_z+L_z$, respectively. The polarization and the Bloch-Floquet  indices remain the same.

The well-known procedure to obtain the scattering matrix, \cite{pendry1994photonic},
\begin{equation}\label{eq:DeffScattMatr}
	t(\xi_x,\xi_y)=	\left(
	\begin{array}{cc}
		t^{max,min}  & t^{max,max}  \\
		t^{min,min}  & t^{min,max}  \\
	\end{array}
	\right),
\end{equation}
requires the inverse of the transfer matrix $T^{-,-}$
\begin{eqnarray}
	t^{min,max}&=&(T^{-,-})^{-1},\\
	t^{min,min}&=&-t^{min,max}T^{-,+},\\
	t^{max,max}&=&T^{+,-}t^{min,max},\\
	t^{max,min}&=&T^{+,+}+t^{max,max}T^{-,-}t^{min,min}.
\end{eqnarray}

\section{Propagation by path-operators}\label{sec:Paths}

The transfer matrix elements
(\ref{eq:DefinitionTransferElements}) can be written in terms of the propagation operator
\begin{equation}\label{eq:TransferV1}
	\begin{split}
		\text{\large{T}}^{\text{DIR};\;\text{dir}}_{\text{POL,BF};\;\text{pol,bf}}=&\|\scriptsize{\text{OUT}}\|^{-1}(L_xL_y)^{-1}\sum \mathop{}_{\mkern-5mu n_{\perp}}\\&
		\bra{\phi_\text{POL,BF}^{\text{DIR}}}\text{Operator}
		\ket{\psi_\text{pol,bf}^{\text{dir}}},
	\end{split}
\end{equation}	
where the normalization in the internal space and the Operator are
\begin{gather}\label{eq:NormOUT}
	\|\scriptsize{\text{OUT}}\|=\bra{p^{H,\text{OUT}}_\perp}\ket{p^{H,\text{OUT}}_\perp}\left(k_z^{H,\text{OUT}}+k_z^{E,\text{OUT}}\right),\\\label{eq:Operator}
\text{Operator}=O_H(n_z+L_z)O_E(n_z+ L_z-1)\\\nonumber
	\dots 
	O_H(n_z+1)O_E(n_z).
\end{gather} 

The common approach, also followed by \cite{pendry1994photonic}, is to place the resolution of identity between every pair of $O_HO_E$ operators in the product (\ref{eq:Operator}). Within this procedure, for a stack of laminae we need  to multiply a set of large-dimensional transfer matrices, one for each lamina. This implies that to grasp one desired transfer matrix element of the stack we actually need to compute all elements, which can be computationally taxing, at both levels, theoretical or numerical. Furthermore, the advantage of using a sequence of matrix products is diminished by numerical instabilities and, potentially, by approximating a large-dimensional transfer matrix by a smaller one. 

To avoid using  products of matrices  we develop a different procedure, which, to the best of our knowledge is new in the context of transfer matrices for non-homogeneous  permittivity and permeability materials. 

\begin{figure}[h]
	\includegraphics[scale=1.0]{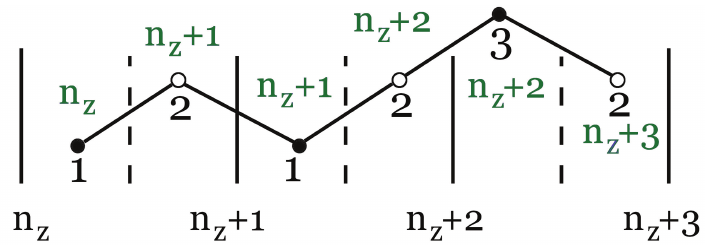}
	\caption{\label{fig:Flowchart} One possible path for a $L_z=3$ propagation. The path is IN121232OUT, labeled following the positive z-axis.}
\end{figure}

To exemplify, consider $\ket{\psi_\text{pol,bf}^{\text{dir}}}=\ket{p_\perp^{\text{IN}},S,+} e^{i p_\perp^{\text{IN}} n_\perp}$ at IN and $\bra{\phi_\text{POL,BF}^{\text{DIR}}}=\bra{p_{\perp}^{\text{OUT}},S,+} e^{-i p_\perp^{\text{OUT}} n_\perp}$ at OUT in (\ref{eq:TransferV1}).
There are several decisive moments in this manuscript. One is here, where we decide to expand the product operator as a sum of path-operators. Namely, since each $O_E$ and $O_H$ is a sum of there operators, (\ref{eq:SumEOperators}) and (\ref{eq:SumHOperators}), (\ref{eq:Operator}) becomes a sum of $3^{L_z+1}$ terms $O_H^{(j)}O_E^{(k)}\dots O_H^{(m)}O_E^{(n)}$, parameterized by a path $(\text{IN}n,m,\dots,k,j\text{OUT})$, where each index takes three values. Each path can be represented graphically, like in Fig.\ref{fig:Flowchart}, by associating a $ j \bullet$ and a $j \circ$ to $O_E^{(j)}$ and $O_H^{(j)}$, respectively.
For example, the path-operator $O_H^{(2)}O_E^{(3)}O_H^{(2)}O_E^{(1)}O_H^{(2)}O_E^{(1)}$ for the path IN121232OUT is represented in Fig.\ref{fig:Flowchart}

\section{Channels\label{sec:Channels}}
The transfer matrix element is thus a sum over path-operators
\begin{equation}\label{eq:RealSpacePropagationSminusPminus}
	\begin{split}
		\\&\|\scriptsize{\text{OUT}}\|^{-1}(L_xL_y)^{-1}\sum_{paths}
		\sum_{n_{\perp}}e^{-i p_\perp^{\text{OUT}} n_\perp}
		\\&
		\bra{p_{\perp}^{\text{OUT}},S,+}\text{path-operator}
		\ket{p_\perp^{\text{IN}},S,+} e^{i p_\perp^{\text{IN}} n_\perp}.
	\end{split}
\end{equation}
 We continue with the chosen example where S+ is placed at both  IN and OUT. However, the liberty of choosing any pair of polarization and direction at both IN and, independently, OUT, is manifestly visible in (\ref{eq:RealSpacePropagationSminusPminus}). 
 To proceed further, the tensor product decomposition of both the basis and the operators proves to be very useful. Each element of the  basis, (\ref{eq:pS+}), is a sum of two parts, the e-part and the h-part, that contains $\ket{e}$ and $\ket{h}$, respectively. The same kind of decomposition is valid for the path-operator, where there is an $I_E$-part and an $I_H$-part in  (\ref{eq:SumEOperators}) and (\ref{eq:SumHOperators}). Due to these decompositions, the  tensor product in $\bra{p_{\perp}^{\text{OUT}},S,+}\text{path-operator}
\ket{p_\perp^{\text{IN}},S,+}$ splits as a product of  two factors. The first factor is  of the form $\bra{e}I_H I_E I_2I_HI_1\cdots I_E\ket{h}$, and alike versions. It can take only two values, $0$ or $1$. For each non-zero path that survives this first-factor selection, we must compute the second factor, which contains the material properties. For this paper we confine ourselves to non-magnetic materials.
 
 The structure of the second factor is of the form $\bra{p_{\perp}^{H,\text{OUT}}}\ket{T_E}\Gamma_P\sigma_2\cdots\ket{p_{\perp}^{E,\text{IN}}}$, which, drawing on $\bra{T_E}\sigma_2\ket{T_H} e^{i p_{\perp}\cdot n_{\perp}}=0$, offers a second zero-selection procedure. Moreover, the second factor, for those paths that are overall non-zero, admits, as well, a pleasing and advantageous factorization into two terms. One term, call it the $\mathcal{Z}$-term, is a function of $(p_z^{\text{IN}},p_z^{\text{OUT}},\Omega)$ but not on the  materials parameters, whereas the other, call it the XY-term, depends on $p_{\perp}^{\text{IN}}$ and the deposited materials.
This factorization is crucial in simplifying the final result for the transfer matrix because, fortunately, the XY-term is common to a group of paths. 

We will call a channel a subset of paths, which have the same XY-term. The main significance is that the number of channels is much smaller than the number of non-zero paths, substantially reducing the final analytical expression for a transfer matrix element. Furthermore, the $\mathcal{Z}$-term of a channel, which is the sum of the $\mathcal{Z}$-terms of the paths that belong to that channel, simplifies into products of $\varphi_z$-dependent factors. 

The technique to separate a $\mathcal{Z}$-term from an XY-term is to forward-propagate the input phase $e^{i p_{\perp}^{\text{IN}}\cdot n_{\perp}}$ and backward-propagate the output phase  $e^{i p_{\perp}^{\text{OUT}}\cdot n_{\perp}}$, both through the chain of path-operator product. These propagations are carried through the path-operator by using the general relation

\begin{equation}\label{eq:PropagationRelation}
	\begin{split}
		\sum_{n_{\perp}}e^{-i p_{\perp}^{\text{OUT}}\cdot n_{\perp}} f_{\text{OUT}} (n_{\perp})\left(\ket{T_{E/H}}(f_{\text{IN}} (n_{\perp})e^{i p_{\perp}^{\text{IN}}\cdot n_{\perp}})\right)=\\
		\sum_{n_{\perp}}\left(-\ket{T_{H/E}}(e^{-i p_{\perp}^{\text{OUT}}\cdot n_{\perp}} f_{\text{OUT}} (n_{\perp}))\right)f_{\text{IN}} (n_{\perp})e^{i p_{\perp}^{\text{IN}}\cdot n_{\perp}}.
	\end{split}
\end{equation}	

For the specific aim of splitting-up the $\mathcal{Z}$ from the XY-factor, the forward-propagation proceeds as long as  $f_{\text{IN}} (n_{\perp})$ is a constant independent of $n_{\perp}$. With or without the use of (\ref{eq:PropagationRelation}), a trivial observation is that  the phase 
$e^{i p_{\perp}^{\text{IN}}}$ can be pushed through $\ket{T_E}$ leaving behind $\ket{p_E^{\text{IN}}}$.
However, (\ref{eq:PropagationRelation}) brings in clarity for the backward-propagation. In this case, i.e. for constant $f_{\text{OUT}} (n_{\perp})$, the phase 
$e^{-i p_{\perp}^{\text{OUT}}}$ is acted upon by  $-\ket{T_H}$, creating  $\ket{p_E^{\text{OUT}}}$, but at the output.
The splitting-up procedure stops when each phase meets its own $n_{\perp}$-dependent non-magnetic materials. The sub-chain of a path-operator that extends between the final propagation positions of $e^{-i p_{\perp}^{\text{OUT}}\cdot n_{\perp}}$ and $e^{i p_{\perp}^{\text{IN}}\cdot n_{\perp}}$ is the XY-operator, which defines a channel. Once the classification of channels is at hand, the complete solution for the transfer matrix elements is obtained by continuing the forward-propagation of the phase $e^{i p_{\perp}^{\text{IN}}}$, but now through the non-magnetic material-dependent XY-operators. We explain this below, solving the case of a structure composed of two nonhomogeneous non-magnetic laminae.

\section{Analytical solution for a bilaminar non-magnetic structure}\label{sec:Bilaminae}

It is not difficult to complete the path-operator calculation for a single lamina carrying an $(n_x,n_y)$-dependent non-magnetic materials. The next in line is a bilaminar structure, which gives rise to a greater range of scattering phenomena. Although its complete solution may seem arduous, the end result elucidates the role played by each propagation channel. Furthermore, the analytical solution combined with optimization procedures provides a method for designing practical devices, Section \ref{sec:Optim}. 

\begin{figure}[h]
	\includegraphics[scale=1.0]{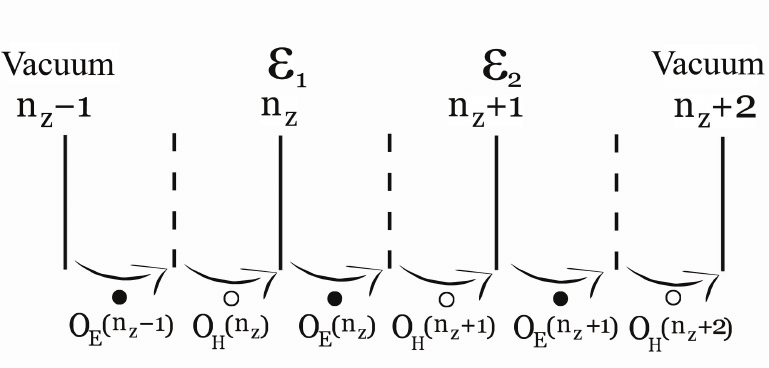}
	\caption{\label{fig:VacuumEEVacuum} Propagation from vacuum through two thin laminae, back into the vacuum. Planes 1 and 2, printed with nonhomogeneous $\epsilon_1(n_{\perp})$ and $\epsilon_2(n_{\perp})$,  form laminae 1 and 2, respectively. }
\end{figure}

In what follows we present the solution to the path-operator propagation for a device composed of two laminae with non-homogeneous and distinct non-magnetic materials  $\epsilon_1(n_x,n_y)\neq\epsilon_2(n_x,n_y)$, Fig.\ref{fig:VacuumEEVacuum}.

First we take an overall view of the results, followed by an in-depth study of  propagation through  path-operators.

From the total of $3^6=729$ paths only $188$ paths, counted for all possible polarization pairs, are non-zero. These paths are grouped into channels. The channels through which the plane waves propagate from IN at $n_z-1$ to OUT at $n_z+2$, depend on the polarization index, Tab.\ref{tab:Channels}. For example, the specific propagation from $S^{\text{IN}}$ into $S^{\text{OUT}}$ goes through 104 paths, the contributions from the rest of the paths, up to 188, being zero for this polarization pair. These 104 paths are grouped in 5 channels Tab.\ref{tab:Channels}. Channel 1 of $S^{OUT}S^{IN}$ contains 46 paths, Channel 2 comprises 22 and so on, until Channel 5, which is built on 20 paths. Tab.\ref{tab:Channels} contains, for each pair of IN-OUT polarization, the information for the number of channels and the number of paths per channel.

\begin{table}[h]
	\caption{\label{tab:Channels}Channels }
	\begin{ruledtabular}
		\begin{tabular}{ll}
			$Pol^{OUT}Pol^{IN}$ & \text{No. of paths per channel}\\
			\colrule
			\vspace{-1.5ex}\\
			$S^{OUT}S^{IN}$&\hspace{-3em}104=46+22+8+8+20\\
			$S^{OUT}P^{IN}$&\hspace{-3em}77=33+16+4+12+12\\
			$P^{OUT}S^{IN}$&\hspace{-3em}63=4+4+4+4+10+10+6+6+15\\
			$P^{OUT}P^{IN}$&\hspace{-3em}96=10+2+6+2+6+8+2+6+6+15+12+3+9+9\\
		\end{tabular}
	\end{ruledtabular}
	\footnotetext{The number of channels, from the second column, are: 5,5,9 and 14.  }
\end{table}

By way of example, to justify these results, and more, we will analyze three paths.
\begin{table}[h]
	\caption{\label{tab:Channel3Paths}Channel 3 SS }
	\begin{ruledtabular}
		\begin{tabular}{ll}
			paths & paths\\
			\colrule
			\vspace{-2.0ex}\\
			$\text{IN}221221\text{OUT}$&$\text{IN}321221\text{OUT}$\\
			$\text{IN}221222\text{OUT}$&$\text{IN}321222\text{OUT}$\\
			$\text{IN}221223\text{OUT}$&$\text{IN}321223\text{OUT}$\\
			$\text{IN}221233\text{OUT}$&$\text{IN}321233\text{OUT}$\\
		\end{tabular}
	\end{ruledtabular}
	\footnotetext{The common sequence 212 defines the XY-operator for this channel}
\end{table}

Two paths are selected form the 729-188 set of zero-paths to explain the source of their null contribution to propagation. 

The path  $\text{IN}321321\text{OUT}$ does not contribute at all to any polarization propagation because the operator product for this path,  $O_H^{(1)}O_E^{(2)}O_H^{(3)}O_E^{(1)}O_H^{(2)}O_E^{(3)}$, is of the form
$(I_H I_E\mathbbm{1}I_E I_H\mathbbm{1})\otimes(\text{operator} )$. The part $I_E\mathbbm{1}I_E=\ket{e}\bra{h}\mathbbm{1}\ket{e}\bra{h}$ is zero, since $\bra{h}\ket{e}=0$, which renders the paths inactive to any propagation.

Another path, $\text{IN}321121\text{OUT}$ is also zero, but for a different reason. Its path-operator $O_H^{(1)}O_E^{(2)}O_H^{(1)}O_E^{(1)}O_H^{(2)}O_E^{(3)}$ is of the form
$(I_H I_EI_HI_E I_H\mathbbm{1})\otimes(\text{operator})$. The first term of the tensor product, $(I_H I_E I_H I_E I_H\mathbbm{1})=\ket{h}\bra{e}$, is not zero. However, the second term of its tensor product contains $O_H^{(1)}O_E^{(1)}=\ket{T_H}\Gamma_{S}\bra{T_H}\sigma_2 (-c)\ket{T_E}(-c \Gamma_P)\bra{T_E}\sigma_2$. This brings into contact $\bra{T_H}$ with $\ket{T_E}$ acted by $\sigma_2$, which is zero

\begin{gather}
	\begin{matrix}
		\begin{bmatrix}
			T_x^E\\ \\T_y^E
		\end{bmatrix}
		\sigma_2
		\begin{bmatrix}
			T_x^E&T_y^E
		\end{bmatrix}
	\end{matrix}=i(T_y^ET_x^E-T_x^ET_y^E)=0,
\end{gather}
given that $T_y^E$ and $T_x^E$ commute. 
We are left with the propagation along 188 non-zero paths, which is exemplified next.   

\section{Channel 3 SS}\label{sec:Channel3SS}

Compute the contribution of the path $\text{IN}321221\text{OUT}$ to the propagation of $S^{\text{IN},+}$ into $S^{\text{OUT},+}$. This path is part of Channel 3 $S^{\text{OUT}}S^{\text{IN}}$, which contains 8 paths, Tab.\ref{tab:Channel3Paths}.
Its path-operator  is of the form
$\ket{h}\bra{e}\otimes(\text{operator})$. To evaluate its contribution to the transfer matrix element, the operator needs to act on both the input 
$\ket{p_\perp^{\text{IN}},S,+} e^{i p_\perp^{\text{IN}} n_\perp}$ 
and on the output 
$e^{-i p_\perp^{\text{OUT}} n_\perp}\bra{p_{\perp}^{\text{OUT}},S,+}$,
(\ref{eq:RealSpacePropagationSminusPminus}).
The first term $\ket{h}\bra{e}$ tells us that only the e-part of the input $\ket{p_\perp^{\text{IN}},S,+} $ is delivered, by this path, into the h-part of the output $\bra{p_{\perp}^{\text{OUT}},S,+}$.

Use, for a non-magnetic device, $\Gamma_{H}^{-1}=\Omega^2(i c)^{-1}$ and $\Gamma_{E}^{-1}=i \epsilon(n_{\perp})c^{-1}$ to arrive at

\begin{gather}\label{eq:RealSpacePropagationSminusPminusV3}
	\|\scriptsize{\text{OUT}}\|^{-1}(L_xL_y)^{-1}	
	\sum_{n_{\perp}}e^{-i p_\perp^{\text{OUT}}n_\perp}\\\nonumber ic^3\Omega^2\bra{p_{\perp}^{\text{OUT,H}}} \ket{T_H}\bra{T_H}\sigma_2\epsilon_2\ket{T_E}\epsilon_1^{-1}\bra{T_E}\epsilon_1\sigma_2\ket{p_\perp^{\text{IN,H}}}\\\nonumber e^{i p_\perp^{\text{IN}} n_\perp}.
\end{gather}
Factorize the summand into a $\mathcal{Z}$-factor and an XY-factor, by back-propagating $e^{-i p_\perp^{\text{OUT}}n_\perp}$ through $\ket{T_H}$ at OUT, (\ref{eq:PropagationRelation}), whereas at IN use   $\ket{T_H}e^{i p_\perp^{\text{IN}} n_\perp}=\ket{p_\perp^{\text{IN,H}}}e^{i p_\perp^{\text{IN}} n_\perp}$,

\begin{gather}\nonumber
	\|\scriptsize{\text{OUT}}\|^{-1}(L_xL_y)^{-1}\sum_{n_{\perp}}e^{-i p_\perp^{\text{OUT}}n_\perp} ic^3\Omega^2\bra{p_{\perp}^{\text{OUT,H}}} \ket{p_{\perp}^{\text{OUT,H}}}\\
	\bra{T_H}\sigma_2\epsilon_2\ket{T_E}\epsilon_1^{-1}\bra{T_E}\epsilon_1\sigma_2\ket{T_H} e^{i p_\perp^{\text{IN}} n_\perp}\label{eq:PathC3PropagationSplusSplusV2}.
\end{gather}
For this path, the $\mathcal{Z}$-factor is $ic^3\Omega^2\bra{p_{\perp}^{\text{OUT,H}}}\ket{p_{\perp}^{\text{OUT,H}}}$. For other paths, both variables $\varphi_z^{\text{OUT}}$ and $\varphi_z^{\text{IN}}$ are present in the result.  

The XY-operator, which is common to all 8 paths from $\text{channel}_3$-SS and thus  defines this channel, is

\begin{equation}\label{eq:Operator-XY-Channel3SSV7}
\begin{aligned}
\hspace{-1.294ex}\text{Operator-XY}_3^{SS}=\bra{T_H}\sigma_2\epsilon_2\ket{T_E}\epsilon_1^{-1}\bra{T_E}\epsilon_1\sigma_2\ket{T_H}
\end{aligned}.
\end{equation}

 In the general case, the action of an XY-operator on the input phase $e^{i p_\perp^{\text{IN}} n_\perp}$ delivers a function of $n_{\perp}$, $\chi(n_{\perp})$, which depends on $\epsilon_1(n_{\perp})$, $\epsilon_2(n_{\perp})$ and $p_{\perp}^{\text{IN}}$. This function,  $\chi(n_{\perp})$, contains the information from the propagation of the IN phase through the entire  non-magnetic media, all the way to the OUT. Here, for Channel 3 SS, 
\begin{equation}\label{eq:Chi3SS}
\begin{aligned}
\hspace{-1.294ex}\text{Operator-XY}_3^{SS}(e^{i p_\perp^{\text{IN}} n_\perp})=e^{i p_\perp^{\text{IN}} n_\perp} \chi_3^{S^{OUT},S^{IN}}(n_{\perp}).
\end{aligned}
\end{equation}	

Henceforth, to keep the formulas for $\chi(n_{\perp})$ simple,  we write interchangeably $p_x=p^{\text{IN}}_x$ and $p_y=p^{\text{IN}}_y$. However, to avoid confusion and to be able to distinguish between the IN and OUT phases, we will keep the index OUT for $p_{\perp}^{\text{OUT}}$.

From (\ref{eq:Chi3SS}) we deduce the dependence on $p_{\perp}^{\text{IN}}$ of the $\chi$-function.
Namely, $\chi_3^{S^{OUT},S^{IN}}(n_{\perp})=g_0+g_1\cos(a p_x)+g_2\cos(b p_y)+g_3\sin(a p_x)+g_4\sin(b p_y)+g_5\cos(a p_x-b p_y)+g_6\sin(a p_x-b p_y)$, where $g_j$, with $j=0\cdots6$, are functions of $n_{\perp}$ build exclusively on the non-magnetic materials deposited on the laminae. For example

\begin{equation}\label{eq:g6discrete}
	\begin{split}
		g_6= \frac{i\epsilon _1\left(n_x-1,n_y+1\right) \epsilon _2\left(n_x,n_y\right)}{a^2 b^2 \epsilon _1\left(n_x,n_y+1\right)}
		&\\
		-\frac{i\epsilon _1\left(n_x+1,n_y-1\right)
			\epsilon _2\left(n_x,n_y\right)}{a^2 b^2 \epsilon _1\left(n_x+1,n_y\right)}
		&\\
		+\frac{i\epsilon _1\left(n_x+1,n_y-1\right) \epsilon _2\left(n_x+1,n_y\right)}{a^2 b^2 \epsilon _1\left(n_x+1,n_y\right)}
		&\\-\frac{i\epsilon _1\left(n_x-1,n_y+1\right)
			\epsilon _2\left(n_x,n_y+1\right)}{a^2 b^2 \epsilon _1\left(n_x,n_y+1\right)}
		&\\
		+\frac{i\epsilon _2\left(n_x,n_y+1\right)}{a^2 b^2}-\frac{i\epsilon _2\left(n_x+1,n_y\right)}{a^2 b^2},
	\end{split}	
\end{equation}
where, for clarity, we used the shorthand notation $n_x-1=a n_x-a$, $n_y+1=b n_y+b$, and alike.
The periodic boundary conditions and the sum from (\ref{eq:PathC3PropagationSplusSplusV2}) bring in the Fourier transform of  $\chi(n_{\perp})$ over the variable $q_{\perp}=p_\perp^{\text{IN}}-p_\perp^{\text{OUT}}\equiv p_\perp-p_\perp^{\text{OUT}}$,
\begin{gather}\label{eq:DisctereFourierChi3}
		\hspace{-2.0ex}	\tilde{\chi}_3^{S^{OUT},S^{IN}}(q_{\perp}):=\\\nonumber
		(L_xL_y)^{-1}	\sum_{n_{\perp}}e^{i q_{\perp}\cdot n_\perp} \chi_3^{S^{OUT},S^{IN}}(n_{\perp}).
\end{gather}	

Since $\tilde{\chi}_3^{S^{OUT},S^{IN}}(q_{\perp})$ is common to all paths from Channel 3, the next focus is on the sum of all $\mathcal{Z}_{\text{paths}}$. It turns out that, for all channels, this sum factorizes into simple terms, each term being a function of either $\varphi _{z,\text{IN}}$ or $\varphi _{z,\text{OUT}}$, but not both. For the example under study

\begin{equation}\label{eq:ToalZ}
	\sum_{\substack{
			\text{paths from} \\
			\text{Channel 3}}}	\mathcal{Z}_{\text{path}}	=i c \Omega ^2 \varphi _{z,\text{IN}} \varphi _{z,\text{OUT}}.
\end{equation} 	

Finally, we arrive at the contribution of Channel 3 ${S^{OUT,+},S^{IN,+}}$, 
which, conveniently, is a product of $\tilde{\chi}_3$ from (\ref{eq:DisctereFourierChi3}) and the channel $\mathcal{Z}$-factor (\ref{eq:Z3SS}) 

\begin{equation}\label{eq:Z3SS}
	\mathcal{Z}_3^{\text{OUT\phantom{,}S+},\text{IN\phantom{,}S+}}=	\|{\scriptsize{\text{OUT}}}\|^{-1} i c \Omega ^2 \varphi _{z,\text{IN}} \varphi _{z,\text{OUT}}.
\end{equation}

The main result, for the IN S to OUT S propagation, is that the transfer matrix element is reduced from a sum of 729 paths to a sum of 5 channels

\begin{equation}\label{eq:TsumChannels}
	T_{Sm,Sn}^{+,+}=\sum_{\text{channel}=1}^5	\mathcal{Z}_{\text{channel}}^{\text{OUT\phantom{,}S+},\text{IN\phantom{,}S+}}\;\tilde{\chi}_{\text{channel}}^{S^{OUT},S^{IN}}(q_{\perp}).
\end{equation}	

The dependence on the Bloch-Floquet indices $m$ and $n$ is through both $p_{\perp}$ and $q_{\perp}$.

\section{$\mathcal{Z}$ and XY-factors for all Channels}\label{sec:ZandXYfactors}

\begin{figure}[h]
		\includegraphics[scale=1.0]{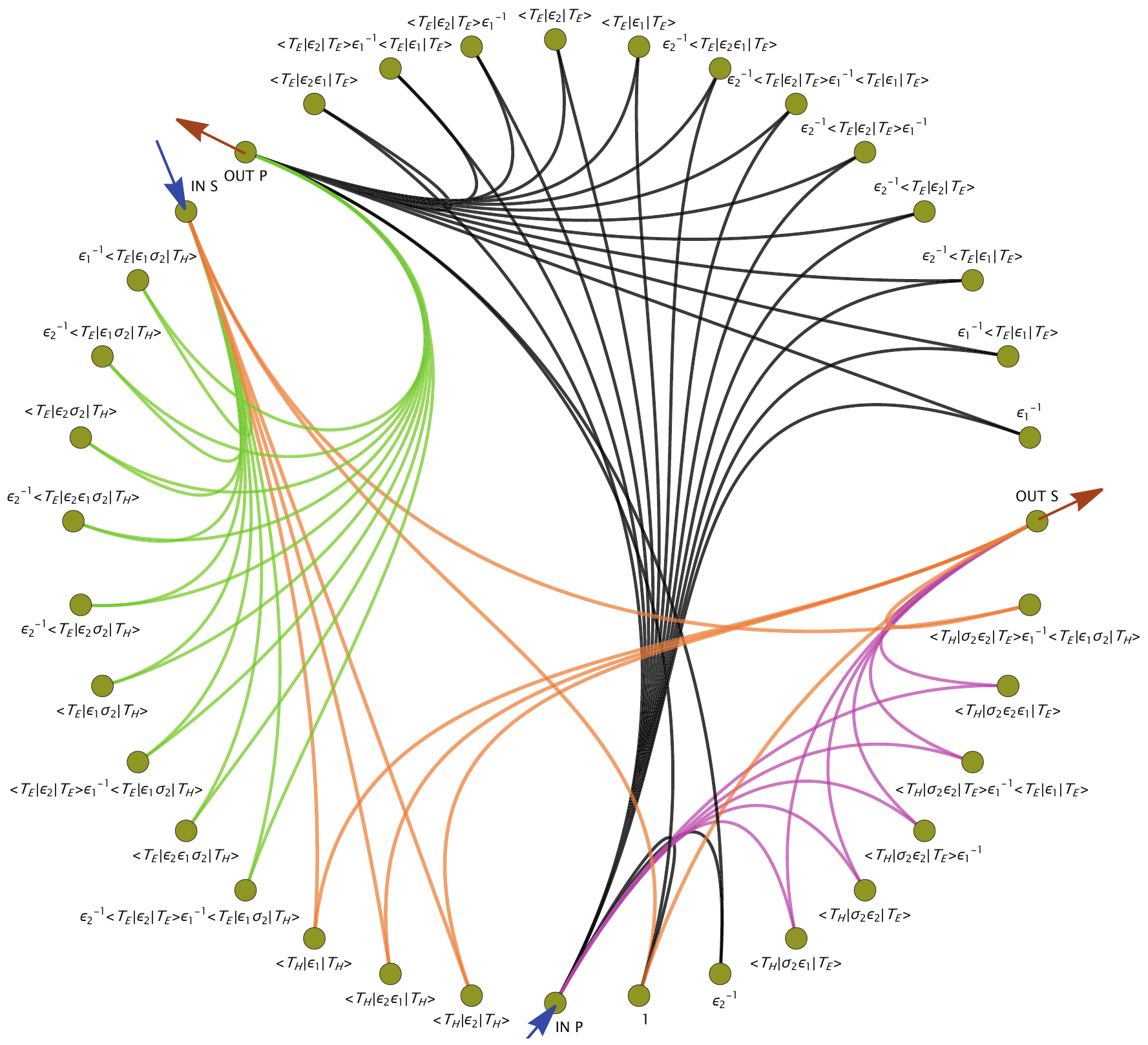}
		\caption{\label{fig:AllChannels} There are 32 channels, ferreted out from 729 paths, that transfer the input fields into the output. Not all 32 channels are open for every polarization pairs. To use this figure, chose an input, either InS or InP. Then follow a line that connects the input with a disk, which represents a channel. From the disk, follow one of the lines that exit that disk and ends on a chosen output.  The disks are labeled by the $(x,y)$-factor that defines the channel.}
	\end{figure}

The XY-channel-operators  for all four polarization combinations (OUT/IN,S/P) are represented in Fig.\ref{fig:AllChannels}. These operators are independent of the direction of propagation index $\pm$. The channels from IN S to OUT S were discussed above. There are 9 channels that connect  IN S polarization with  OUT P. All start with $\ket{T_E}$ and end in $\bra{T_H}$. They are colored green in Fig.\ref{fig:AllChannels}. Change the input into a P-polarization. There are 5 unique channels that transfer an IN P into an OUT S. They are colored purple in Fig.\ref{fig:AllChannels}. The last case  is the transfer of  IN P into  OUT P. There are 14 channels out of which 13 are unique to this polarization pair. They are colored black in Fig.\ref{fig:AllChannels}. The non-unique channel, shared with IN S to OUT S, has no $(x,y)$-dependence and so its $\chi$-function is $1$.

By contrast, the $\mathcal{Z}$-factors depend on the direction of propagation. Tab.\ref{tab:SSallmm} and Appendix B contain the information for all the other $\mathcal{Z}$-factors.

\begin{table}[h]
	\caption{\label{tab:SSallmm}Unnormalized $\mathcal{Z}$ factors for  OUT S$-$, IN S$-$}
	\begin{ruledtabular}
		\begin{tabular}{ll}
			$(\text{XY-operator})_{\text{No.}}$ &\hspace{-10ex} $\mathcal{Z} (\scriptsize{\text{OUT}}\;S-\;,\;\scriptsize{\text{IN}}\;S-)\;\|\scriptsize{\text{OUT}}\|$ \\
			\colrule\\[-1ex]
			$(1)_1$ &\hspace{-30ex}$ \frac{i \bra{p_{\perp}^{\text{OUT}}}\ket{p_{\perp}^{\text{OUT}}} \left(\left(\Omega ^2 \varphi
				_{z,\text{OUT}}+1\right){}^2-\varphi _{z,\text{OUT}}^2\right)}{c \varphi _{z,\text{OUT}}^4}  \delta _{\text{IN},\text{OUT}}$\\[2ex]
			$ ( \bra{T_H} \epsilon_1\ket{T_H})_2$ &\hspace{-7ex}$ -\frac{i \Omega ^2 \left(\Omega ^2 \varphi _{z,\text{OUT}}+1\right)}{c \varphi _{z,\text{IN}} \varphi
				_{z,\text{OUT}}^2} $\\[2ex]
			$ (\bra{T_H}\sigma_2 \epsilon_2\ket{T_{\text{E}}}\epsilon^{-1}_1\bra{T_{\text{E}}}\epsilon_1\sigma_2\ket{T_H})_3 $&\hspace{-5ex}$ -\frac{i c \Omega ^2}{\varphi _{z,\text{IN}} \varphi _{z,\text{OUT}}} $\\[2ex]
			$(\bra{T_H}\epsilon_2 \epsilon_1\ket{T_H})_4$ &\hspace{-5ex} $\frac{i \Omega ^4}{c \varphi _{z,\text{IN}} \varphi _{z,\text{OUT}}} $\\[2ex]
			$ (\bra{T_H}\epsilon_2\ket{T_H})_5$ &\hspace{-7ex}$ -\frac{i \Omega ^2 \left(\Omega ^2 \varphi _{z,\text{IN}}+1\right)}{c \varphi _{z,\text{IN}}^2 \varphi
				_{z,\text{OUT}}}$ \\	
		\end{tabular}
	\end{ruledtabular}
	\footnotetext{The channel number is the No. index of XY-operator. Here $1-2  \varphi _{z}+\Omega^2 \varphi _{z}+ \varphi _{z}^2=c^2 \varphi _{z}\bra{p_{\perp}}\ket{p_{\perp}}$ and $\delta _{\text{IN},\text{OUT}}$ is the Kronecker delta.}
\end{table}

The rule to obtain the $\mathcal{Z}$-factors for other directions is to change $\text{IN}-$ into $\text{IN}+$ and, at the same time, change
$\varphi_{\text{z,IN}}\rightarrow  \varphi_{\text{z,IN}}^{-1}$ and
$c \rightarrow  -c$. The same rule applies for $\text{OUT}-$ into $\text{OUT}+$.

However, this rule does not apply for Channel 1, for both  SS and PP, Appendix B,  Tables \ref{tab:SSchannel1}, \ref{tab:PPchannel1}. The reason is that, in vacuum, there are pairs of polarization and directions that cannot propagate into each other. For example, in vacuum,  $\text{IN S}+$ cannot propagate into  $\text{OUT S}-$. In contrast, Channel 1 is open in vacuum, so the rule does not apply to it.

At this point, the transfer matrix elements are  computed in closed analytical formulas for any non-magnetic bilaminar lattice structure. 
The solution is completely general being valid for elaborate non-magnetic material patterns that can be drawn on both laminae.

The decomposition of propagation by channels can be implemented also for three or more patterned laminae. The number of channels increases with the increase of the number of laminae, but the procedure remains the same.

Returning to the bilaminar structure, an avenue worth to explore is the continuum limit in the x and y dimensions. In this limit, the number of Bloch-Floquet modes become infinite, placing  weight to our method of directly seizing the transfer matrix elements by circumventing infinite-dimensional transfer matrices.
\phantom{XXXXXXXXXXXXXX}
\phantom{XXXXXXXXXXXXXX}
\phantom{XXXXXXXXXXXXXX}
\phantom{XXXXXXXXXXXXXX}
\phantom{XXXXXXXXXXXXXX}
\phantom{XXXXXXXXXXXXXX}
\phantom{XXXXXXXXXXXXXX}
\phantom{XXXXXXXXXXXXXX}
\phantom{XXXXXXXXXXXXXX}
\begin{table}[h]
	\caption{\label{tab:DielectricFunctions}Permittivity functions $(f_1,f_2)$}
	\begin{ruledtabular}
		\begin{tabular}{ll}
			$\text{Channel}$ &\hspace{-10ex} $f_1,f_2$ \\
			\colrule\\[-1ex]
			$\chi_3^{S^{OUT},S^{IN}}$ &\hspace{-30ex}$f_1=\ln(\epsilon),f_2=\epsilon$\\[2ex]
			$\chi_2^{P^{OUT},S^{IN}}$ &\hspace{-30ex}$f_1=\epsilon,f_2=\epsilon^{-1}$\\[2ex]
			$\chi_7^{P^{OUT},P^{IN}} $&\hspace{-30ex}$f_1=\epsilon^{-1},f_2=1$ and $f_1=\epsilon^{-1},f_2=\ln(\epsilon)$ \\
		\end{tabular}
	\end{ruledtabular}
	\footnotetext{Some channels, like $\chi_7$ in this table, operates with more than one pair of permittivity functions.}
\end{table}
\onecolumngrid
	
\begin{figure}[h]
		\includegraphics[scale=1.0]{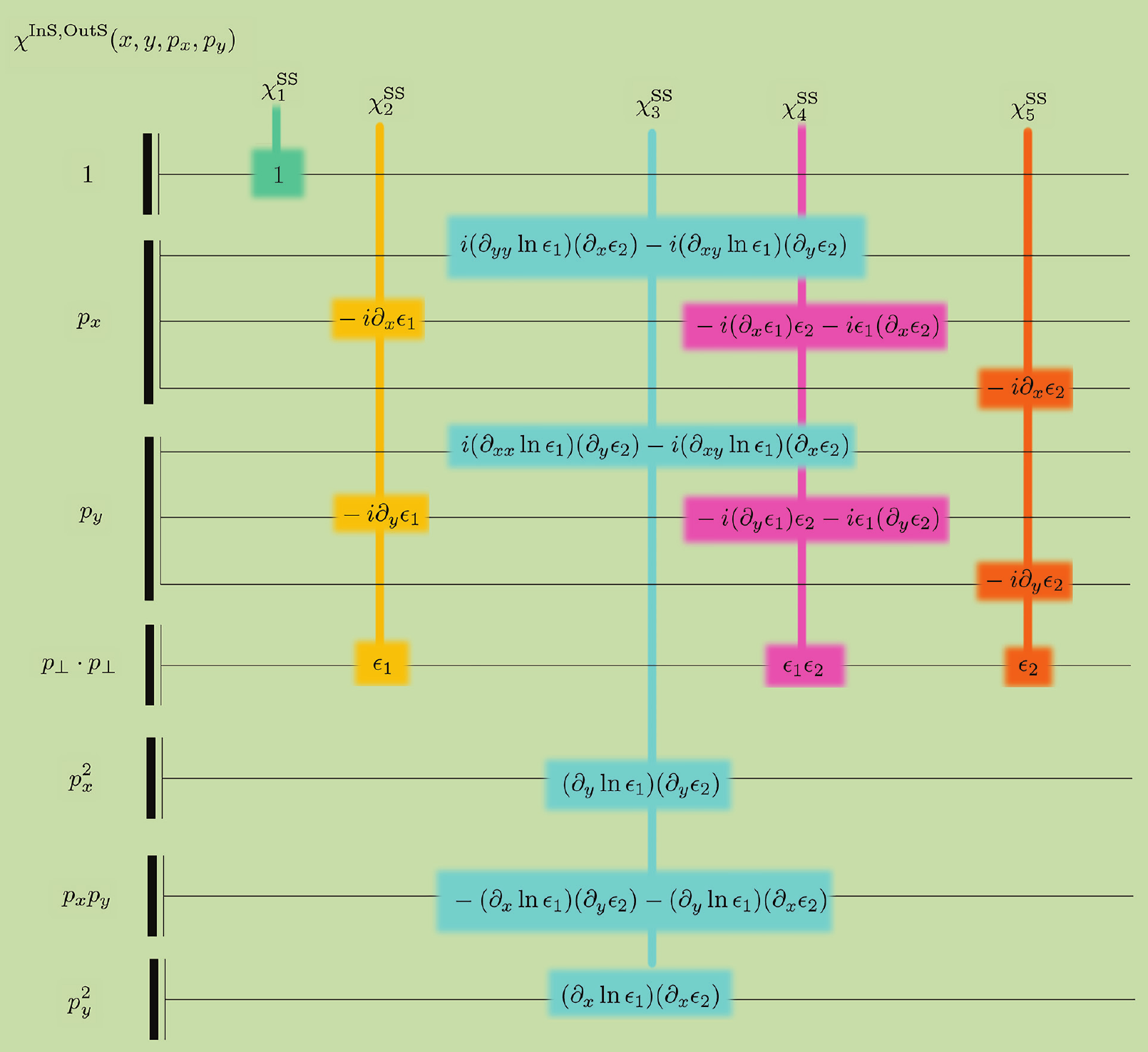}
		\caption{\label{fig:InSOutSPartitura.pdf}The IN S OUT S channels. There are 5 channels for the InS to OutS. The horizontal staves carry the polynomials  terms in $p_x$ and $p_y$. The vertical structure of the figure shows the composition of each channel as a product of the polynomial term in $p_x$ and $p_y$ to  its corresponding  permittivity function. Recall that $p_x$ and $p_y$ denote the input values, although, for simplicity, they do not carry the index IN. The third column corresponds to  formula (\ref{Continuum:3SS}). Appendix B contains the $\chi$-functions for all the other polarization pairs. }
\end{figure}
\begin{figure}[h]
		\includegraphics[width=1.0\textwidth]{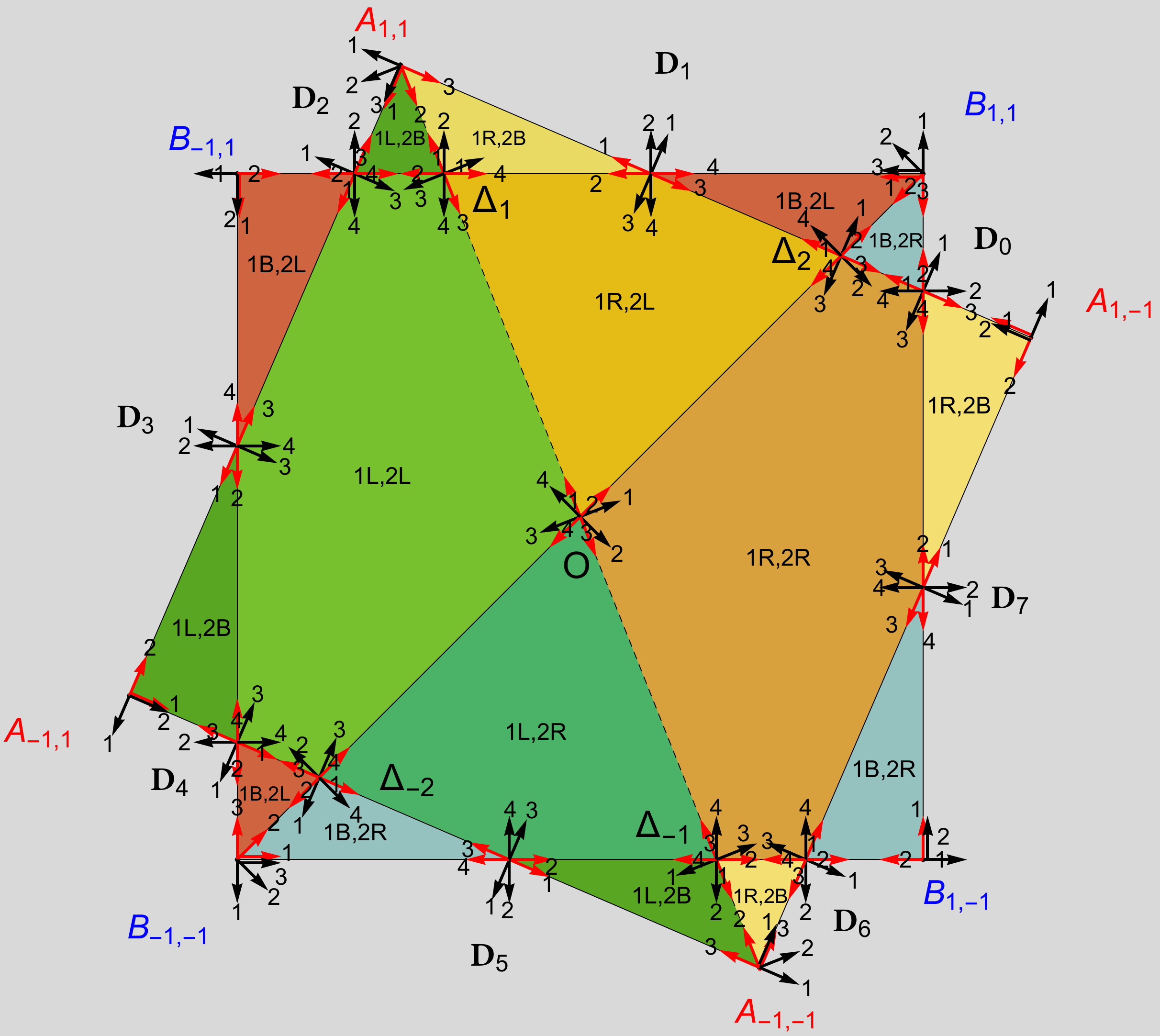}
		\caption{\label{fig:PiPiRig.pdf} The $\pi$-$\pi$-rig pattern.  To relate to Fig.\ref{fig:VacuumEEVacuum}, points denoted by A and B belong to plane 1 and 2, respectively. The $\hat{z}$-axis is coming out from the figure's page and $\hat{x}$ is oriented from $B_{-1,-1}$ to $B_{1,-1}$. The real points, of $V_1$ and $V_2$-type, are $A_{i,j}$  and $B_{i,j}$, $i=\pm 1,j=\pm 1$, respectively. The virtual points, $V_{12}$, at the intersection of one segment from lamina 1 and another from lamina 2, are $D_m, m=0,\cdots, 7$, $\Delta_n, n=\pm 1,\pm 2$ and the origin $O$. Here $\alpha >0$, so $A_{1,1}$ is rotated counterclockwise from its position on top of $B_{1,1}$ for $\alpha=0$. The black arrows represent the $\hat{\lambda}$-vectors, whereas the red arrows represent the $\hat{\mu}$-vectors. The permittivity $\epsilon_{1\text{L}}$ is printed on the left triangle $(A_{1,1},A_{-1,1},A_{-1,-1})$ whereas $\epsilon_{1\text{R}}$ on the right triangle  $(A_{1,-1},A_{1,1},A_{-1,-1})$, both on lamina 1. Same goes for lamina 2 where $\epsilon_{2\text{L}}$ and $\epsilon_{2\text{R}}$ labels the left and right triangles with respect to the diagonal $B_{-1,-1} B_{1,1}$. Outside the interior squares, and up to the boundaries of the unit cell, the background permittivity $\epsilon_{1\text{B}}$ is printed on lamina 1 and  $\epsilon_{2\text{B}}$ on lamina 2.  The color and label of each polygon express the decomposition of the  unit cell in polygons that carry a constant value of the product $\epsilon_1 \epsilon_2$. The unit cell corners, which are not represented, own their own $(\hat{\lambda},\hat{\mu})$-pairs.}
	\end{figure}
\twocolumngrid

\section{Continuum limit for $a,b\rightarrow 0$}\label{sec:Continuum}
Consider that the discrete lattice constants $a$ and $b$ are much smaller than $c$, the discrete lattice constant on the $z$-axis. In this case $n_{\perp}$ transforms into a continuum $r_{\perp}$ coordinate so that the unit cell length is  $l_x=\lim (a L_x)$ for $a\rightarrow 0$ and $L_x\rightarrow\infty$. Same goes for $l_y$. However, the $z$-axis remains discrete. In the continuum x,y-limit the discrete $\chi_3^{S^{OUT},S^{IN}}(n_{\perp})$, (\ref{eq:Chi3SS}), is a function of $(x,y)$
\begin{equation}\label{Continuum:3SS}
	\begin{split}
		&\chi_3^{\text{InS,OutS}}(x,y,p_x,p_y)=
		\\&
		p_x \left(
		i (\partial_{yy}\ln\epsilon_1)(\partial_x\epsilon_2)-i(\partial_{xy}\ln\epsilon_1)(\partial_y\epsilon_2)\right)+
		\\&
		p_y \left(
		i (\partial_{xx}\ln\epsilon_1)(\partial_{y}\epsilon_2)-i(\partial_{xy}\ln\epsilon_1)(\partial_x\epsilon_2)\right)+
		\\&
		p_x^2  (\partial_y\ln\epsilon_1)(\partial_y\epsilon_2)+
		\\&
		p_x p_y \left( -(\partial_x\ln\epsilon_1)(\partial_y\epsilon_2)-(\partial_y\ln\epsilon_1)(\partial_x\epsilon_2)\right)+
		\\&
		p_y^2 \left( (\partial_x\ln\epsilon_1)(\partial_x\epsilon_2)\right).
	\end{split}
\end{equation}
The dependence of (\ref{Continuum:3SS}) on $p_x$, $p_y$ is quadratic. Recall that $p_\perp=p_\perp^{\text{IN}}$. Each coefficient of this quadratic polynomial is a combination of partial derivatives of permittivities, Fig.\ref{fig:InSOutSPartitura.pdf}. We purposely placed these coefficients as a sum of terms, each being a product of a partial derivative of a function of $\epsilon_1$ into another one, which depends exclusively on $\epsilon_2$. This specific arrangement is crucial to get a closed formula for the Fourier integral for a polygonal tessellated  pattern, which will be a later subject to undertake. In addition to being technically useful, this arrangement has another facet. It reveals that the contribution of the permittivity constants to a channel is via pairs of functions $(f_1,f_2)$, where $f_1$ and $f_2$ acts only on $\epsilon_1$ and $\epsilon_2$, respectively. From (\ref{Continuum:3SS}) we get $f_1(\epsilon_1)= \ln(\epsilon_1)$ and $f_2(\epsilon_2)= \epsilon_2$.  These functions are easily gleaned, for each of the 33 channels, from Fig.\ref{fig:InSOutSPartitura.pdf} and Appendix B.  They are either $f(\epsilon)=\epsilon,\epsilon^{-1}, \ln(\epsilon)$ or the constant 1. 
All 33 channels, with the exception of the following 7 channels $\chi_{3,4}^{P^{OUT},S^{IN}}$,  $\chi_4^{S^{OUT},P^{IN}}$, and $\chi_{7,8,9,13}^{P^{OUT},P^{IN}}$, use one pair $(f_1,f_2)$. Channel  $\chi_{8}^{P^{OUT},P^{IN}}$ involves 3 pairs, whereas the other 6 from the exceptions use 2 pairs. Some examples are presented in Tab.\ref{tab:DielectricFunctions}. 
These channel-dependent permittivity functions are universal for a bilaminar structure, the distinction between devices with the same geometry being captured by the specific values of their arguments, $\epsilon_1$ and $\epsilon_2$. It is worth noting that, for a given IN-OUT polarization pair, the set of functions are uniquely specific to a channel. This property can thus be used to name a channel. For example, Channel 3 in Fig.\ref{fig:InSOutSPartitura.pdf} may bare the name SS$(\ln(x),x)$.

\section{c-scaling}\label{sec:Cscaling}

The transfer matrix elements get simpler once the influence of the discreteness  along the z-axis, $c$, is embodied into the scaling of the unit cell 
\begin{gather}\label{eq:xy-scaled}
	x = c x^{'},y=c y^{'}.
\end{gather}
The frequency was already c-scaled in (\ref{eq:DefinitionOmega}).
All parameters  become c-scaled $l_{x/y} = c l_{x/y}^{'}$,
$p_{x/y}= c^{-1}p_{x/y}^{'}$.  Scaling a $\chi$ channel-function is equivalent to multiplying it by a power of the scaling parameter $c$. For example, the c-scaling effect on (\ref{Continuum:3SS}) is
\begin{equation}\label{eq:c_ScalingForChi3SS}
	\chi_3^{\text{InS,OutS}}(x,y,p_x,p_y)=c^{-4}\chi_3^{\text{InS,OutS}}(x^{'},y^{'},p_x^{'},p_y^{'}).
\end{equation}

In the continuum $(x,y)$-plane the discrete Fourier transform becomes an integral, as $(a b)/((a L_x) (bL_y))\rightarrow (dx\;dy)/(l_x l_y)$.
In the scaled variables, the Fourier transform of the $\chi$ channel-function
\begin{gather}\label{eq:FromOperatorToFunctionInContinuum}
\hspace{-1ex}(l_x^{'}l_y^{'})^{-1}\int_{0}^{l_x^{'}}dx^{'} \int_{0}^{l_y^{'}}dy^{'}\; e^{i q_x^{'}  x^{'}} e^{i q_y^{'} y^{'}} \chi(x^{'},y^{'},p_x^{'},p_y^{'})
\end{gather}
is a function of $q_{\perp}^{'}=p_{\perp}^{'\text{IN}}-p_{\perp}^{'\text{OUT}}$ where
$p_{x/y}^{'\text{IN/OUT}}=2\pi\left( M_{x/y}^{\text{IN/OUT}}+\xi_{x/y}\right)(l_{x/y}^{'})^{-1}$
with $2\pi \xi_{x}/l_x^{'}$ and $2\pi \xi_{y}/l_y^{'}$ being equal to $\Omega\sin(\theta)\cos(\phi)$ and $\Omega\sin(\theta)\sin(\phi)$, respectively.

For $\mathcal{Z}$-factors, the c-scaling appears in the argument of $\varphi_z$
\begin{eqnarray}
	\varphi _{z,\text{IN}} &=& \varphi_z\left(1-2^{-1}\Omega ^2+2^{-1} \left(p_x^{'2}+p_y^{'2}\right)\right). 
\end{eqnarray}
For $\varphi _{z,\text{OUT}}$ the argument contains $p_{\perp}^{'\text{OUT}}$.
Below, we drop the prime notation for the c-scaled parameters, understanding that all variables are c-scaled.
The advantage of scaling is that the  $c$-monomial from  $\chi$ and its corresponding  $\mathcal{Z}$-factor cancel, rendering a c-independent formula for a channel.

\section{Polygonal tessellation patterns}\label{sec:Tessellation}

The transfer matrix elements, for the xy-continuous limit, are reduced at this point to the computation of 2-dimensional Fourier transforms of functions defined on the unit cell. To push forward the project of finding analytical closed-form formulas, we show that, in the general case, the Fourier transforms for all channels can be completely solved for polygonal subdomains. The approach and the notations for the general formulas are exemplified using a pattern,  Fig.\ref{fig:PiPiRig.pdf}, which we will call $\pi$-$\pi$-rig. The unit cell is a square with corners positioned at $(\pm\pi,\pm\pi)$, in length units chosen for a desired frequency band. Inside the unit cell sits an inner square. Fig.\ref{fig:PiPiRig.pdf} shows two inner squares, one located on plane 1, whereas the other on plane 2, both defined in Fig.\ref{fig:VacuumEEVacuum}. The corners of the unit cell are not visible in Fig.\ref{fig:PiPiRig.pdf} to save the space for the information pertaining to the inner squares. In the foreground, the square on plane 2 has its vertices labeled $B_{i,j}$, $i=\pm 1,j=\pm 1$ positioned at $(\pm \pi/2,\pm \pi/2)$. In the background, the square on plane 1, with vertices $A_{i,j}$, is rotated counterclockwise by an angle $\alpha$. For $\alpha=0$, the squares' vertices coincide. There are two kinds of points in Fig.\ref{fig:PiPiRig.pdf}. Points like $A_{i,j}$ and $B_{i,j}$ from plane 1 and plane 2, respectively, are called  $\text{\it{real}}$ . In contrast, we call $\text{\it{virtual}}$ those that appear as intersection points, like $D_1$, the origin O and $\Delta_1$. All points, real and virtual, have associated with them a set of unit vectors. Take for example the origin O. There are four segments that originate at O. One is the segment $\text{OA}_{1,1}$ which carries a unit vector denoted $\hat{\mu}_1$ in Fig.\ref{fig:PiPiRig.pdf}. Then  $\hat{\mu}_2$, $\hat{\mu}_3$ and  $\hat{\mu}_4$ are along  $\text{OB}_{1,1}$, $\text{OA}_{-1,-1}$ and  $\text{OB}_{-1,-1}$, respectively. All tails of these unit vectors sit on point O. From the same point O, originates another set of four unit vectors $\hat{\lambda}_i$, $i=1,\dots ,4$ in such a way that each pair $(\hat{\lambda}_i,\hat{\mu}_i)$ is orthogonal and each triplet $(\hat{\lambda}_i,\hat{\mu}_i, \hat{z})$ is right-oriented.
 The last geometrical parameter is related to the choice of a frequency band, say in GHz or THz, which scales the device's units to  mm or $\mu$m, respectively. Versus this unit of length we define a unitless thickness parameter $t$
\begin{equation}\label{eq:tParameter}
	t=[1\;\text{unit}](c \;[\text{unit}])^{-1}.
\end{equation}	
Although the frequency $\omega$ and the dimensions $l_x,l_y$ are fixed in the chosen units, the c-scaled parameters change with $t$. As the laminae get thinner the scaled unit cell $(l_x^{'}, l_y^{'})$ and the probing wavelength $\lambda^{'}$ get larger, $\lambda^{'}=t\frac{\lambda\;[\text{unit}]}{1\;[\text{unit}]}$, whereas the probing frequency $\Omega$ slides into lower regions,  
$\Omega=2 \pi \lambda^{'-1}$.  

The increase of the scaled unit cell repositions the coordinates of the vertices, but do not change the unit vectors $(\hat{\lambda},\hat{\mu})$. 

The coordinates of all points, real and virtual, together with the orientations of all unit vectors and a choice of the thickness parameter comprise the geometrical data for computing the Fourier transformation of any polygonal tessellation. The other data come from the jumps of the  permittivities across polygonal segments. For example, for point O in Fig.\ref{fig:PiPiRig.pdf} the relative permittivity pair $(\epsilon_{1\text{R}},\epsilon_{2\text{L}})$ appears towards to tip of $\hat{\lambda}_1$, i.e. on the right-side looking along $\hat{\mu}_1$. The pair $(\epsilon_{1\text{L}},\epsilon_{2\text{L}})$ sits towards the tail of $\hat{\lambda}_1$, i.e. on the left-side of $\hat{\mu}_1$. As we cross $\hat{\mu}_1$ walking in the direction of $\hat{\lambda}_1$, slightly above point O,  the change in the relative permittivities introduces  a jump
\begin{equation}\label{eq:JumpEpsilon}
(\Delta f_1f_2)_{\text{O},\hat{\mu}_1}= f_1(\epsilon_{1\text{R}})f_2(\epsilon_{2\text{L}})-f_1(\epsilon_{1\text{L}})f_2(\epsilon_{2\text{L}}),
\end{equation}	
where $(f_1,f_2)$ is a channel-specific pair of permittivity functions as in Tab.\ref{tab:DielectricFunctions}.


The Fourier transforms for the $\chi$ channel-functions include both the geometrical configuration and the permittivities jumps.
 In (\ref{eq:ResultIntegralC3SS}) we show, as an example, the result of the Fourier transform, $\mathcal{F}(\chi)$, of  the third channel for OutS,InS and for the specific case $q_{\perp}\cdot\hat{\mu}\neq 0$ for all $\hat{\mu}$,
\begin{gather}\nonumber
\sum_{(V_{12},\hat{\mu})}e^{i (q_{\perp}\cdot V_{12})_2} (p_{\perp}\cdot \hat{\mu}_1)(p_{\perp}\cdot \hat{\mu}_2) 
\left( \frac{q_{\perp}\cdot \hat{\mu}_2}{\hat{\mu}_1\cdot \hat{\lambda}_2}\right)\left( \frac{q_{\perp}\cdot \hat{\mu}_1}{\hat{\mu}_2\cdot \hat{\lambda}_1}\right)\\\nonumber
\hspace{-10em}	\Delta (\epsilon_2\ln\epsilon_1)_{\hat{\mu}}\frac{W\cdot\hat{\lambda}}{W\cdot q_{\perp}} \frac{1}{q_{\perp}\cdot\hat{\mu}}\;-\\\nonumber
\sum_{(V_{12},\hat{\mu})}e^{i (q_{\perp}\cdot V_{12})_3} (p_{\perp}\cdot \hat{\mu}_1)(\hat{\mu}_1\cdot \hat{\lambda}_2) 
\left( \frac{q_{\perp}\cdot \hat{\mu}_2}{\hat{\mu}_1\cdot \hat{\lambda}_2}\right)^2\left( \frac{q_{\perp}\cdot \hat{\mu}_1}{\hat{\mu}_2\cdot \hat{\lambda}_1}\right)\\\label{eq:ResultIntegralC3SS}
\hspace{-10em}	\Delta (\epsilon_2\ln\epsilon_1)_{\hat{\mu}}\frac{W\cdot\hat{\lambda}}{W\cdot q_{\perp}} \frac{1}{q_{\perp}\cdot\hat{\mu}}.		
\end{gather}	
For each virtual vertex $V_{12}$ from Fig.\ref{fig:PiPiRig.pdf}, the sum covers all four $\hat{\mu}'\text{s}$ associated with it. The indices 1 and 2 refer to the line from plane 1 and 2, respectively. In this formula we are free to choose one pair $(\hat{\mu}_1,\hat{\mu}_2)$ of unit vectors along line 1 and line 2, out of all four possible choices of the pairs  exiting $V_{12}$. The result is independent on the choice of the pair because the terms are invariant to the transformation $\hat{\mu}_1\rightarrow -\hat{\mu}_1$ or $\hat{\mu}_2\rightarrow -\hat{\mu}_2$.

The Fourier transform in (\ref{eq:ResultIntegralC3SS}) depends on $p_{\perp}$, i.e.  $p_{\perp}^{\text{IN}}$, because the channel's $\chi$-function (\ref{Continuum:3SS}) depends on it. In addition, the transform is naturally a function of the  $q_{\perp}$, (\ref{eq:FromOperatorToFunctionInContinuum}).  The first sum in (\ref{eq:ResultIntegralC3SS}) displays the quadratic dependence on $p_{\perp}$ as a dot product, invariant to rotations of the $(x,y)$ coordinate system. An invariant form also appears for the second sum, which contains the linear dependence on  $p_{\perp}$. In this formula, the variables $q_{\perp},p_{\perp}$ and $V_{12}$ are c-scaled and depend on the thickness parameter $t$. As we announced before, we do not carry the primed symbol, (\ref{eq:xy-scaled}), anymore.
In the next section we lay down the Fourier transform results, with their proofs left to Appendices D to H.

\section{Fourier transforms for the $\chi-\text{channel-functions}$}\label{sec:FourierChi}

 We encounter four types of integrals on the unit cell, based on the number of derivatives $m^1=(m_x^{1},m_y^{1})$ and $m^2=(m_x^{2},m_y^{2})$ being zero or not

\begin{enumerate}
	\item[(00).] $\iint dx dy (f_1)(f_2) e^{q_{\perp}\cdot r_{\perp}}$,\;$m^1=0,m^2=0$,
	\item[(10).] $\iint dx dy (\partial_{m^1}f_1)(f_2) e^{q_{\perp}\cdot r_{\perp}}$,\; $m^1\neq 0, m^2= 0$,
	\item[(01).] $\iint dx dy (f_1) (\partial_{m^2}f_2) e^{q_{\perp}\cdot r_{\perp}}$,\; $m^1= 0, m^2\neq 0$,
	\item[(11).] $\iint dx dy (\partial_{m^1}f_1)(\partial_{m^2}f_2) e^{q_{\perp}\cdot r_{\perp}}$,\; $m^1\neq 0, m^2\neq 0$.
\end{enumerate}

There are three cases for (00), which depend on the relation between the geometry of the device, expressed by the set of vectors $\hat{\mu}$ and the mode-scattering vector $q_{\perp}$. For case  $(1)$, $q_{\perp}\cdot\hat{\mu}\neq0$.
Case $(2)$ consists of $q_{\perp}\cdot\hat{\mu}=0 \; \text{and}\; q_{\perp}\neq 0$, which means that $q_{\perp}=\pm\hat{\lambda}\abs{q_{\perp}}$. Finally for $(3)$ $q_{\perp}=0$.

The formulas for (00), from (\ref{Eq81}) and (\ref{StokesFinalResultqZEROTHEOREM}), are written on purpose as a sum of terms contributed by each pair $(V,\hat{\mu})$. This is akin to a potential theory where only the ends of the curve contribute. In addition, also reminiscent of a potential theory,  gauge vectors appear in the solution. They are the same for all pairs $(V,\hat{\mu})$ and their arbitrariness helps eliminating all singularities that may appear in the denominators. The gauge vectors are denoted as $W$ and $\mathcal{D}$, and can be arbitrarily chosen with the provision that 
$W\cdot q_{\perp}\neq 0$ and $\mathcal{D}\cdot \hat{\mu}\neq 0$.
The jump in $\epsilon$, as the vector $\hat{\mu}$ is traversed along $\hat{\lambda}$, is captured in $(\Delta f_1f_2)_{\hat{\mu}}$,
as exemplified in (\ref{eq:JumpEpsilon}).
 
Outside the unit cell we place a fictitious material with $\epsilon=0$ to get, for the (00) integrals, the correct terms generated by the unit cell boundaries. 

The (00) integration formula for a polygonal tessellation of the unit cell, for cases (1) and (2), is
\begin{gather}\label{Eq81}
		\iint\limits_\text{ } dxdy (f_1f_2) e^{i q_{\perp}\cdot r_{\perp}}=\\\nonumber
		\;\sum_{V,\hat{\mu}}e^{i q_{\perp}\cdot V}\frac{1}{q_{\perp}\cdot \hat{\mu}}\;\frac{W\cdot\hat{\lambda}}{W\cdot q_{\perp}} (-\Delta(f_1f_2))_{\hat{\mu}}\;+\\
		\;\sum_{V,\hat{\mu}}e^{i q_{\perp}\cdot V}(i \hat{\mu}\cdot V)\;\frac{W\cdot\hat{\lambda}}{W\cdot q_{\perp}} (-\Delta(f_1f_2))_{\hat{\mu}}\;.\nonumber
\end{gather}	
The first and the second sum applies for case(1) and (2), respectively.
For (00) in case (3) we need to use two gauge vectors. The $W$ is the same as before, whereas $\mathcal{D}$ is specific for this case and comes from taking the limit $s\rightarrow 0$ for $q_{\perp}=s \mathcal{D}$. This formula,as proven in Appendix E, is independent on the choice of $\mathcal{D}$ 
\begin{gather}\label{StokesFinalResultqZEROTHEOREM}
	\iint\limits_\text{ } dxdy (f_1f_2) =\\ \nonumber
	\sum_{V,\hat{\mu}}\frac{1}{2} (i \mathcal{D}\cdot V)^2 \frac{1}{\mathcal{D}\cdot \hat{\mu}}\;\frac{W\cdot\hat{\lambda}}{W\cdot \mathcal{D}} (-\Delta (f_1f_2))_{\hat{\mu}}\;. 
\end{gather}
We intentionally drafted the (00) Fourier transforms as a sum of contributions from vertices. This format can be handed on to the next integrals, (01), (1,0) and (1,1), in a consistent manner. The (00) integral appear in other works \cite{lee1983fourier},\cite{wuttke2017form}, but as a sum over the edges of the polygon. 

For the formulas that follow, we will list only the contribution of a pair $(V,\hat{\mu})$, the summation being understood.

\section{ Fourier transforms for product of derivatives of type (01), (10) and (11)}\label{sec:Fourier11}

The result for the (00) case is now used to generate the integrals for which a derivative acts on at least one function. The procedure implies two steps. The first step is to compute 
\begin{equation}\label{eq:IntAndTranslation}
	\begin{split}
		\iint\limits_\mathbb{ } dxdy f_1(r_{\perp}+\delta r_1) f_2(r_{\perp}+\delta r_2) e^{i q_{\perp}\cdot r_{\perp}}=&\\
		\text{Int}(\delta x_1,\delta y_1,\delta x_2,\delta y_2)
	\end{split} 
\end{equation}
as a function of two independent translations, $\delta r_1$ and $\delta r_2$, on plane 1 and plane 2, respectively. These translations change the position of a vertex V 
\begin{equation}
	V_{\text{translated}}=V-\vv{\tau}_{x}^1\delta x_1-\vv{\tau}_{y}^1\delta y_1-\vv{\tau}_{x}^2\delta x_2-\vv{\tau}_{y}^2\delta y_2.
\end{equation}	
The four vectors $\vv{\tau}$ depend on the vertex $V$ being subjected to a free or constraint translation, which will be defined subsequently. If the vertex is associated with a free translation, we have $\vv{\tau}_x^1=\vv{\tau}_x^2=\hat{x}$	and $\vv{\tau}_y^1=\vv{\tau}_y^2=\hat{y}$. 
A vertex at the intersection of two lines from different planes,  subjected to a constraint translation, moves to
\begin{equation}\label{eq:V12Translation}
	(V_{12})_{\text{translated}}=V_{12}-\frac{\delta r_1\cdot \hat{\lambda}_1}{\hat{\mu}_2\cdot\hat{\lambda}_1}\hat{\mu}_2-\frac{\delta r_2\cdot \hat{\lambda}_2}{\hat{\mu}_1\cdot\hat{\lambda}_2}\hat{\mu}_1,
\end{equation}
where $\hat{\mu}_1$ and $\hat{\mu}_2$ represent the directions of the two intersecting lines situated on plane 1 and 2, respectively. For constraint translations, the vectors $\vv{\tau}^1$ and $\vv{\tau}^2$ can be extracted from (\ref{eq:V12Translation}).
In what follows we associate vertices  $V_1$  and $V_2$ that are exclusively situated either in plane 1 or 2, respectively, to a free translation. For a  $V_1$-type vertex $\vv{\tau}^2=0$ and, in reverse, $\vv{\tau}^1=0$ for a  $V_2$-type.
Virtual vertices $V_{12}$ are associated to a constraint translation.
The sets $V_1$, $V_2$ and $V_{12}$ are exemplified in Fig.\ref{fig:PiPiRig.pdf}.

For the second step, once the function $\text{Int}(\delta x_1,\delta y_1,\delta x_2,\delta y_2)$ is at hand, we compute the rest of the integrals by taking its derivatives
\begin{gather}\label{Taylor:fgIntroduction}
	\iint\limits_\mathbb{ } dxdy \left(\partial_x^{m_{x}^1}\partial_y^{m_{y}^1}f_1(r_{\perp})\right)\left(\partial_x^{m_{x}^2}\partial_y^{m_{y}^2}f_2(r_{\perp})\right) e^{i q_{\perp}\cdot r_{\perp}}\\ \nonumber
=\partial_{\delta x_1}^{m_{x}^1}\partial_{\delta y_1}^{m_{y}^1}\partial_{\delta x_2}^{m_{x}^2}\partial_{\delta y_2}^{m_{y}^2}\text{Int}
\end{gather}
 at $\delta x_j,\delta y_j=0$, $j=1,2$. The results are listed below and the details follow in Appendices D to H.

Unlike for the (00) integrals, here the unit cell boundaries do not generate any terms, because the derivative  of a permittivity function is zero across the unit cell boundaries.

For $m^1+m^2\neq0$ and case (1), $q_{\perp}\cdot\hat{\mu}\neq0$, the contribution of $(V,\hat{\mu})$  is 
\begin{equation}\label{Taylor:fg}
	\begin{split}
		e^{i (q_{\perp}\cdot V) }\;
		(-i q_{\perp}\cdot \vv{\tau}_x^1)^{m_{x}^1}
		(-i q_{\perp}\cdot\vv{\tau}_y^1)^{m_{y}^1}
		&\\
		(-i q_{\perp}\cdot \vv{\tau}_x^2)^{m_{x}^2}
		(-i q_{\perp}\cdot\vv{\tau}_y^2)^{m_{y}^2}&\\
		\frac{1}{q_{\perp}\cdot\hat{\mu}}\;	\frac{W\cdot\hat{\lambda}}{W\cdot q_{\perp}}(-\Delta(f_1f_2)_{\hat{\mu}}).
	\end{split}
\end{equation}
For case (2), $q_{\perp}\cdot\hat{\mu}=0$, the contribution of $(V,\hat{\mu})$ is
\begin{gather}\label{Taylor:fgCase2}
	\frac{d}{ds}\Bigr|_{s=0} e^{i (q_{\perp}+s \hat{\mu})\cdot V) }\;
	(-i (q_{\perp}+s \hat{\mu})\cdot \vv{\tau}_x^1)^{m_{x}^1}
	\\ \nonumber
	(-i (q_{\perp}+s \hat{\mu})\cdot\vv{\tau}_y^1)^{m_{y}^1}				
	(-i (q_{\perp}+s \hat{\mu})\cdot \vv{\tau}_x^2)^{m_{x}^2}
	\\ \nonumber
	(-i (q_{\perp}+s \hat{\mu})\cdot\vv{\tau}_y^2)^{m_{y}^2}
	\frac{W\cdot\hat{\lambda}}{W\cdot q_{\perp}}(-\Delta(f_1f_2)_{\hat{\mu}}).
\end{gather}
For case (3), $q_{\perp}=0$, the contribution of $(V,\hat{\mu})$ is
\begin{gather}\label{Taylor:fgCase3}
	\frac{1}{2}	\frac{d^2}{ds^2}\Bigr|_{s=0} e^{i (s\mathcal{D} \cdot V) }s^{m_{x}^1+m_{y}^1+m_{x}^2+m_{2}^2}\\\nonumber
	(-i \mathcal{D}\cdot \vv{\tau}_x^1)^{m_{x}^1}
	(-i \mathcal{D}\cdot\vv{\tau}_y^1)^{m_{y}^1}				
	(-i \mathcal{D}\cdot \vv{\tau}_x^2)^{m_{x}^2}
	\\ \nonumber
	(-i \mathcal{D}\cdot\vv{\tau}_y^2)^{m_{y}^2}
	\frac{W\cdot\hat{\lambda}}{W\cdot  \mathcal{D}}(-\Delta(f_1f_2)_{\hat{\mu}}).
\end{gather}

The transfer matrix for a polygonal tessellated bilaminar non-magnetic structure, continuous in the x and y directions, is completely solved at this point.  We are moving to test the theoretical results by comparing our closed analytical formulas for the $\pi$-$\pi$-rig structure, to numerical results obtained from a high-performance analysis software package \cite{studios2008cst}. Complex symbolic computations were handled by \cite{Mathematica}.

\section{Evanescent resonant modes by optimization design}\label{sec:Optim}

The intent is to design-optimize a dielectric structure from Fig.\ref{fig:PiPiRig.pdf} such that a subset of evanescent Bloch-Floquet modes $(M_x,M_y)\neq0$ acquire large scattering matrix elements at a specified frequency. Such an excited resonant mode propagates along the plane $(x,y)$ at a frequency smaller than its Rayleigh frequency. Specifically, in the frequency range under study, $37$ GHz to  $50$ GHz, the optimization function is based on four modes. Two modes, $(0,0)$ and $(-1,0)$, propagate, and two other modes, $(0,\pm 1)$, are evanescent. The goal is to find a set of dielectric constants for which the modes $(0,\pm 1)$ resonate.  To excite the resonant modes, the plane wave $(0,0)$ coming from z-$min$ lands at angles $\theta=17.5^{\circ}$ and $\varphi=0^{\circ}$  on the face A of the $\pi$-$\pi$-rig oriented at $\alpha=45^{\circ}$. The thickness of each of the two laminae is
$c=0.25\, \mathtt{mm}$.
The design-objective-function, constructed from the analytical formulas for the transfer matrix elements, selects a point in the 6-dimensional dielectric space $\epsilon _{1\text{B}}= 4.99$, $\epsilon _{2\text{B}}= 2.96$, $\epsilon _{1\text{L}}= 7.60$,
$\epsilon _{2\text{L}} = 5.66$, $\epsilon _{1\text{R}}=5.93$ and $\epsilon _{2\text{R}}= 1.63$. 

Technologically, this point may not be easily implemented, although we constrained the dielectric constants to be less than 12. Instead, we focus  on studying the predictions  put forward by the analytic solution and compare them against numerical results from a high-performance software package. 

These predictions are grouped in three categories. The first category, inspired by topological photonics, is related to robustness of the resonant modes with respect to the change of the dielectric constants, quantified as a map from the real to integer numbers.
The second is based on resonant frequency identification, whereas the third is focused on high Q-factors and the use of complex frequency plane to estimate the Fano-Lorentz spectral line shape for the resonant modes. In what follows we confine ourselves to evanescent S-polarized resonant waves excited by the $(0,0)$ S-polarized incoming wave. Only the main results are presented below, with specific details on the design optimization being summarized in  Appendix I. 

\section{A  map from the real to integer numbers}\label{sec:Map} 

The resonances in the scattering matrix elements appear as zeros of the determinant of the $T^{-,-}$ matrix, (\ref{eq:TMM}). The minimal $T^{-,-}$ matrix can be reduced to just the matrix element $T^{-,-}_{Sm,Sm}$, the zeros of which give the first approximation for the resonant frequencies. To see that, take a  $2\times 2$ transfer matrix  built only from the set $T^{\pm,\pm}_{Sm,Sm}$ and look at the resonance for the mode $Sm$ as a singularity for which  (\ref{eq:DefinitionTransferElements}) has a non-zero solution if all coefficients but (DIR,POL,BF)=(+,S,m) and  (dir,pol,bf)=($-$,S,m) are zero. It turns out that this simple approximation has predictive power.
As the dielectric constants change their values, the  zeroes of $T^{-,-}_{Sm,Sm}$ appear or disappear, creating or destroying a resonance. We study this phenomenon for variable $\epsilon_{1\text{B}}=\epsilon_{2\text{B}}$, running from 1 to $\infty$, with the rest of the dielectric constants kept constant at the optimized values. 

\begin{figure}[h]
	\includegraphics[scale=1.0]{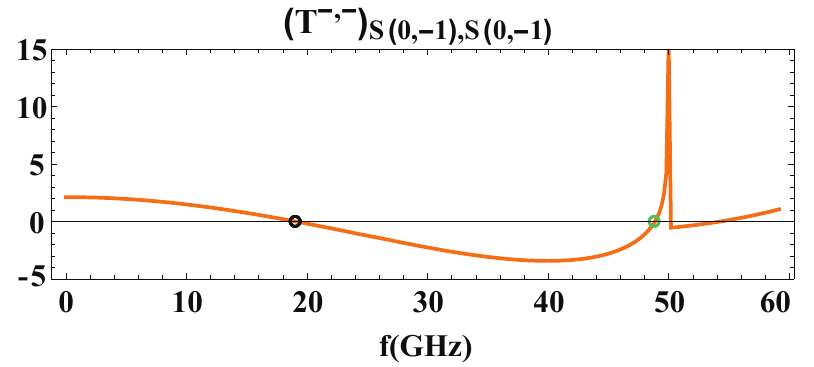}
	\caption{\label{fig:TSSMode3} The theoretical matrix element $T^{-,-}_{S(0,-1),(S(0,-1)}$ as a function of frequency. The zeroes at $19.9$ GHz and $48.86$ GHz have a negative and a positive slope, respectively. The Rayleigh frequency for the mode $(0,-1)$ is at $50.03$ GHz, where a pole singularity is present. The dielectric constants are $\epsilon_{1\text{B}}=\epsilon_{2\text{B}}=35.$  }
\end{figure}

\begin{figure}[h]
	\includegraphics[scale=1.0]{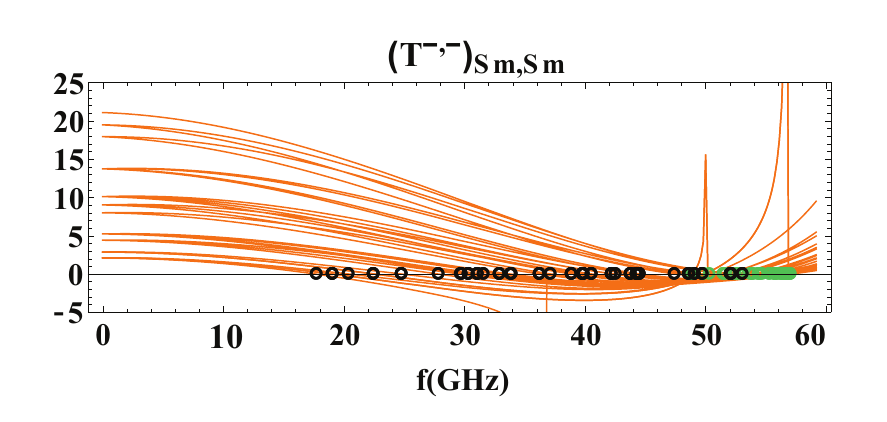}
	\caption{\label{fig:ZeroesTss} All matrix elements $T^{-,-}_{S m,S m}$ that have at least one zero on the real frequency axis. The zeroes with a negative slope are marked as empty black circles, whereas the zeroes with a positive slope are in green. There are three pole singularities in the range of frequency up to $60$ GHz, for $(-1,0), (0,\pm 1)$ and $(-1,\pm 1)$ at $36.69, 50.03$ and $56.71$ GHz, respectively.}
\end{figure}

Hence, map $\epsilon_{1\text{B}}=\epsilon_{1\text{B}}$ to the number of modes $m=(M_x,M_y)$ for which the diagonal transfer matrix element $T^{-,-}_{S m,S m}$, Appendix C, has at least one zero on the frequency axis in the interval $\left[0,60\right]$ GHz,
\begin{gather}
	\epsilon \in \mathbb{R} \mapsto \text{number of modes with zeroes}.
\end{gather}	
The analytic formulas position the zeros as in Fig.(\ref{fig:TSSMode3}) and Fig.(\ref{fig:ZeroesTss})

\begin{figure}[h]
	\includegraphics[scale=1.0]{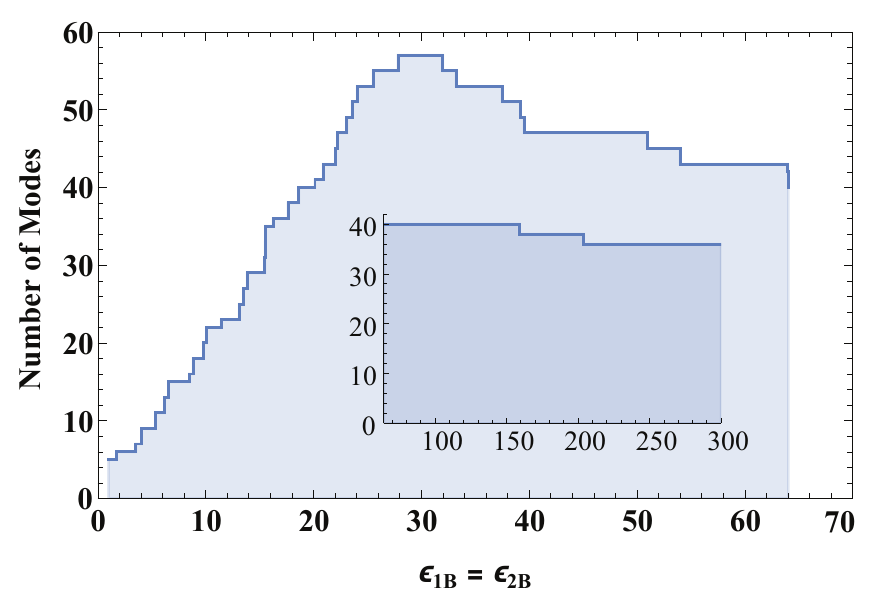}
	\caption{\label{fig:NumberOfModes} The topological map of the number of modes $m=(M_x,M_y)$ for which the corresponding diagonal transfer matrix element $T^{-,-}_{S m,S m}$ has a zero on the real frequency axis in the interval $\left[0,60\right]$ GHz. The continuous variable $\epsilon_{1\text{B}}=\epsilon_{2\text{B}}$  runs from 1 to $\infty$. }
\end{figure}

\begin{figure}[h]
	\includegraphics[scale=1.0]{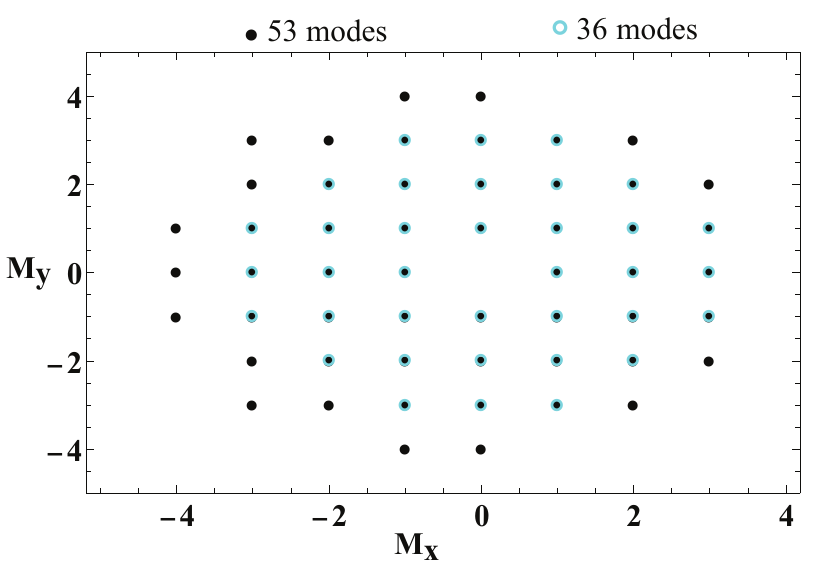}
	\caption{\label{fig:Modes53AND36} Modes $m=(M_x,M_y)$. There are  53 modes for  $\epsilon_{1\text{B}}=\epsilon_{2\text{B}}=35$ and 36 modes for  $\infty$ , respectively. }
\end{figure}

As $\epsilon_{1\text{B}}=\epsilon_{2\text{B}}$ increases from $1$ to $\infty$, Fig.\ref{fig:NumberOfModes}, the number of modes for which $T^{-,-}_{S m,S m}$ crosses the real frequency axis  increases in discontinuous steps, then decreases as the dielectric constant gets above 30. It reaches a plateau of 36 modes as $\epsilon \rightarrow \infty$. Fig.\ref{fig:Modes53AND36} shows the position in the $(M_x,M_y)$ plane of the 53  and 36 modes for $\epsilon_{1\text{B}}=\epsilon_{2\text{B}}=35$ and $\infty$, respectively. Two adjacent steps in  Fig.\ref{fig:NumberOfModes} have two distinct topological invariants, i.e. distinct  number of resonant modes. The boundaries of these steps  are not uniformly distributed on the horizontal dielectric axis. Knowing the distribution of these boundary values may play a significant role in assembling compound devices, which are built by bringing together two structures with distinct topological invariants.
Numerical simulations confirm that 5 modes resonate for $\epsilon_{1\text{B}}=\epsilon_{2\text{B}}=1$ and 53 for $\epsilon_{1\text{B}}=\epsilon_{2\text{B}}=35$. 

\begin{figure}[h]
	\includegraphics[scale=1.0]{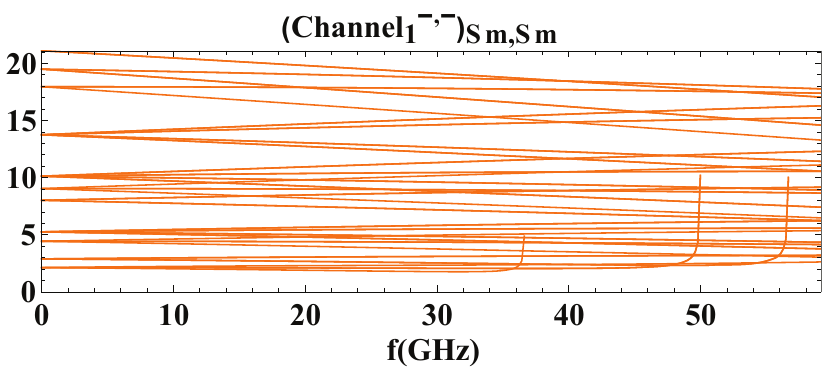}
	\caption{\label{fig:Channel1For53Modes} Channel 1 for all 53 matrix elements $T^{-,-}_{S m,S m}$ from Fig.\ref{fig:Modes53AND36}. Channel 1 is independent on  $\epsilon$.}
\end{figure}

\begin{figure}[h]
	\includegraphics[scale=1.00]{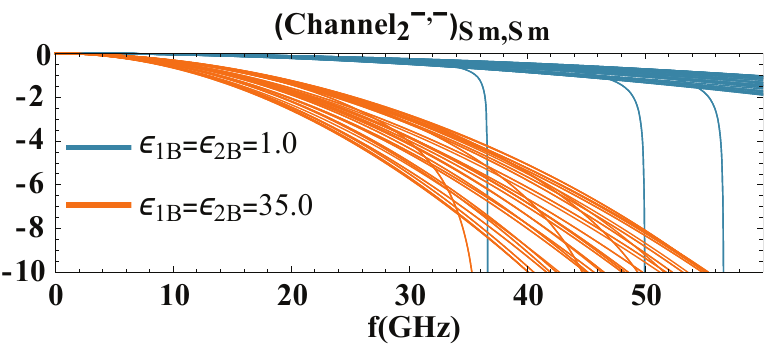}
	\caption{\label{fig:Channel2ForEpsilon35AndEpsilon1} Channel 2 for all 53 matrix elements $T^{-,-}_{S m,S m}$  from Fig.\ref{fig:Modes53AND36},  for two values of  $\epsilon_{1\text{B}}=\epsilon_{2\text{B}}.$}
\end{figure}

\begin{figure}[h]
	\includegraphics[scale=1.0]{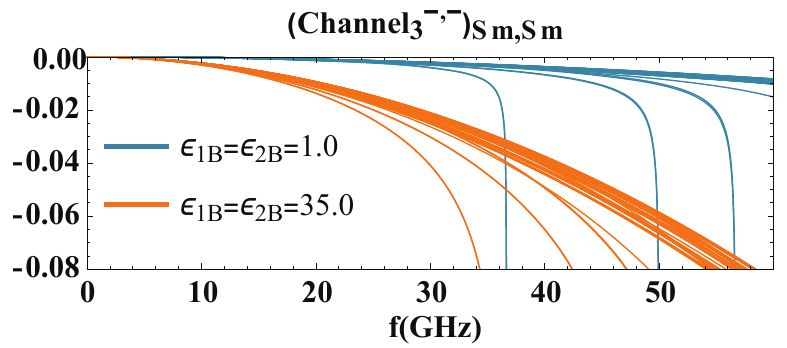}
	\caption{\label{fig:Channel3ForEpsilon35AndEpsilon1} Channel 3 for all 53 matrix elements $T^{-,-}_{S m,S m}$  from Fig.\ref{fig:Modes53AND36}, for two values of  $\epsilon_{1\text{B}}=\epsilon_{2\text{B}}.$}
\end{figure}

\begin{figure}[h]
	\includegraphics[scale=1.00]{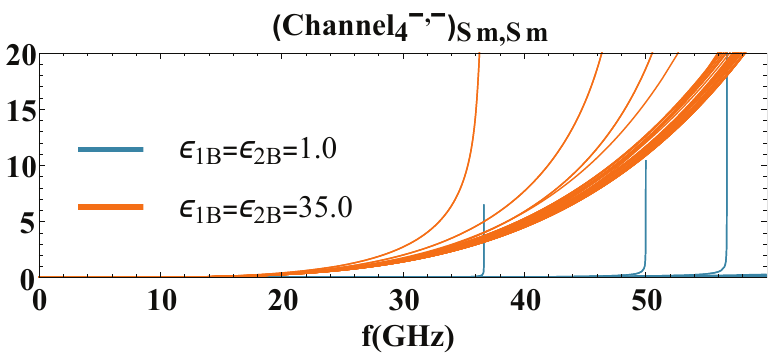}
	\caption{\label{fig:Channel4ForEpsilon35AndEpsilon1} Channel 4 for all 53 matrix elements $T^{-,-}_{S m,S m}$  from Fig.\ref{fig:Modes53AND36}, for two values of  $\epsilon_{1\text{B}}=\epsilon_{2\text{B}}.$}
\end{figure}

\begin{figure}[h]
	\includegraphics[scale=1.0]{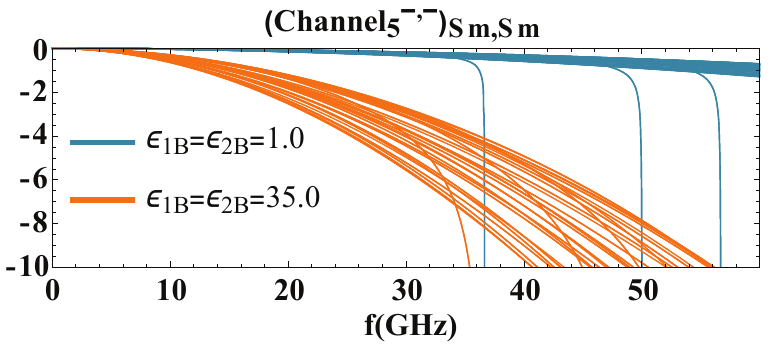}
	\caption{\label{fig:Channel5ForEpsilon35AndEpsilon1} Channel 5 for all 53 matrix elements $T^{-,-}_{S m,S m}$   from Fig.\ref{fig:Modes53AND36}, for two values of  $\epsilon_{1\text{B}}=\epsilon_{2\text{B}}.$}
\end{figure}

One step deeper into the analysis of the zeros is provided by  propagation along channels. Channel 1 SS does not depend on the dielectric constants and  is positive for any frequency and direction $p_{\perp}$, Fig.\ref{fig:Channel1For53Modes}. Some other channels have to be negative for $T^{-,-}_{S m,S m}$ to cross the frequency axis. Interestingly, each channel keeps a constant sign  as a function of frequency, namely $(1,+),(2,-),(3,-),(4,+)$ and $(5,-)$, noticeable in  Fig.\ref{fig:Channel2ForEpsilon35AndEpsilon1} to Fig.\ref{fig:Channel5ForEpsilon35AndEpsilon1}.

From Fig.\ref{fig:Channel3ForEpsilon35AndEpsilon1} we see that Channel 3 is two orders of magnitude smaller than the other ones, so it does not lead to creation of destruction of a zero, but contributes to its location on the frequency axis.
 As  $\epsilon_{1\text{B}}=\epsilon_{2\text{B}}$ increases the negative contribution starts to dominate the positive one and two phenomena appear: the zeros are pushed into lower frequencies and a variable number of modes start to cross the frequency axis. This explains the increase in the number of resonant modes as $\epsilon_{1\text{B}}=\epsilon_{2\text{B}}$ increases.


\subsection{Zeroes of the diagonal matrix elements of $T^{-,-}$}

A distinctive advantage of analytical formulas over the numerical approaches is in the study of infinite limits
\begin{equation}
	\epsilon_{1\text{B}} =\epsilon_{2\text{B}} \rightarrow \infty.
\end{equation}	

In this limit, the zeros of the transfer matrix elements
$T^{-,-}_{S m,S m}(\Omega)=0$, appear at small $\Omega\cong 0$. So, approximate
$1+\Omega ^2 \varphi _{z,\text{OUT}}\cong1$ to get, from Appendix C,
\begin{equation}\label{eq:Channe1LApprox}
	\begin{split}
		(T^{-,-}_{S m,S m})_{\text{Channel 1}}\cong\frac{1}{ \varphi _{z,\text{OUT}}^3 },
	\end{split}
\end{equation}
\begin{equation}\label{eq:Channel2Approx}\nonumber
\begin{split}
\hspace{-1ex}(T^{-,-}_{S m,S m})_{\text{Channel 2}}\cong-\frac{\Omega ^2 }{8 \varphi _{z,\text{OUT}}^2 \left(1-\varphi _{z,\text{OUT}}^2\right)}\left( 6 \epsilon _{1\text{B}}\right),
\end{split}
\end{equation}
\begin{equation}\label{eq:Channel3Approx}\nonumber
	\begin{split}
	(T^{-,-}_{S m,S m})_{\text{Channel 3}}\cong0,
	\end{split}
\end{equation}
\begin{equation}\label{eq:Channel4Approx}\nonumber
\begin{split}
\hspace{-1ex}(T^{-,-}_{S m,S m})_{\text{Channel 4}}\cong\frac{\Omega ^4 }{ \varphi _{z,\text{OUT}} \left(1-\varphi _{z,\text{OUT}}^2\right)}\left(  \frac{\epsilon_{1\text{B}} \epsilon_{2\text{B}}}{\sqrt{2}}\right),
\end{split}
\end{equation}
\begin{equation}\label{eq:Channel5Approx}\nonumber
\begin{split}
\hspace{-1ex}(T^{-,-}_{S m,S m})_{\text{Channel 5}}\cong-\frac{\Omega ^2 }{8 \varphi _{z,\text{OUT}}^2 \left(1-\varphi _{z,\text{OUT}}^2\right)}\left( 6 \epsilon _{2\text{B}}\right).
\end{split}
\end{equation}
The equation
$(T^{-,-}_{S m,S m})=0$, under the above approximations and using the notation $\epsilon_{1\text{B}}=\epsilon_{2\text{B}}=\epsilon$, turns into
\begin{equation}
	-\frac{3}{2}\Omega ^2\varphi _{z,\text{OUT}}\epsilon+\frac{1}{\sqrt{2}}\Omega ^4\varphi _{z,\text{OUT}}^2\epsilon^2+1-\varphi _{z,\text{OUT}}^2=0.
\end{equation}
As $\epsilon\rightarrow \infty$, the zero of $T^{-,-}_{S m,S m}(\Omega)=0$ tends to zero in such a way that	$\Omega^2\epsilon\rightarrow \text{finite number}$. In terms of  this finite number, which we call X, the equation becomes  
\begin{equation}{\label{Eq:Modes}}
	-\frac{3}{2}X \varphi _{z,\text{OUT}}+\frac{1}{\sqrt{2}}X ^2\varphi _{z,\text{OUT}}^2+1-\varphi _{z,\text{OUT}}^2=0.
\end{equation}
The requirement for $X$ to be positive imposes a constrain on mode numbers $M_x$ and $M_y$, for which the diagonal transfer matrix elements posses a zero on the real frequency axis, in the limit  $\epsilon_{1\text{B}} =\epsilon_{2\text{B}} \rightarrow \infty$
\begin{equation}
	\varphi_z \left(\frac{1}{2} \left(\frac{M_x^2}{t^2}+\frac{M_y^2}{t^2}\right)+1\right) \geqslant \frac{1}{4} \sqrt{16-9 \sqrt{2}}.
\end{equation}
This inequality is valid for 36 modes, which are visible in Fig.\ref{fig:Modes53AND36}.

\section{Analytical prediction of the resonant frequency}\label{sec:ResonantFreq}

\subsection{Thinner the lamina higher the precision}\label{subsec:A}	

From counting the number of zeros for many modes, we change the focus, Table\ref{tab:ex}, to predict the position of the zero, $\Omega_{\text{theory}}$, of $T^{-,-}_{S m,S m}$ for the optimized design  modes, $(0,\pm 1)$.
Both modes have the same zero at $\varphi=0^{\circ}$. The resonant frequency depends on the thickness of the structure's lamina, $c$. All the other parameters are kept constant.

\begin{table}[h]
	\caption{\label{tab:ex}Variable layer width.}\label{Table:LayerWidth}
	\begin{ruledtabular}
		\begin{tabular}{llllll}
			c [mm] & t & f[GHz] & $\Omega \, \mathtt{num}$ & $\Omega \, \mathtt{theory}$ & $\Omega \,\mathtt{\% -error}$\\
			\colrule\\[-1ex]
			0.100 & 10 & 47.540 & 0.0996365 & 0.0993646 & 0.27\\
			0.200 & 5 & 43.376 & 0.181819 & 0.180396 & 0.78\\
			0.250 & 4 & 41.667 & 0.218319 & 0.215294 & 1.39\\
			0.275 & 4/11 & 40.693 & 0.234537 & 0.231701 & 1.21\\
			0.300 & 10/3 & 39.908 & 0.250923 & 0.247551 & 1.34\\
		\end{tabular}
	\end{ruledtabular}
\end{table}
Notice that $\Omega \,\mathtt{\%}\text{-error}=100\frac{\abs{\Omega\, \mathtt{num}-\Omega\, \mathtt{ theory}}}{\Omega\, \mathtt{num}}$, which measures the distance to the numerical simulation result $\Omega \, \mathtt{num}$, \cite{studios2008cst}, decreases with the decrease of the width c. This is to be expected because as $c\rightarrow 0$ the discrete approximation of Maxwell's equations improves.

\subsection{The influence of the rotation angle $\alpha$ on the resonant frequency}

\begin{figure}[h]
	\includegraphics[scale=1.0]{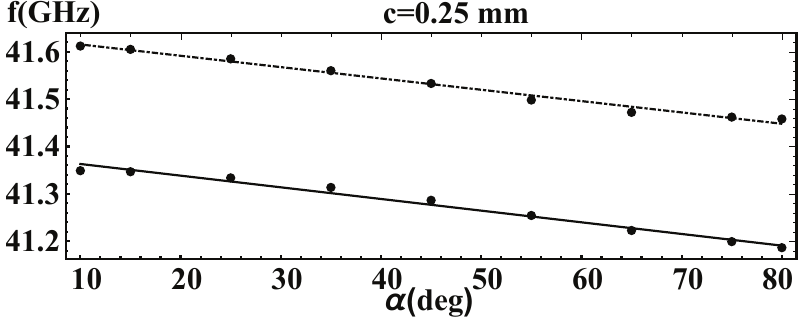}
	\caption{\label{fig:SlopeALPHAc0p25} The resonant frequency of the modes $S(0,-1)$ and $S(0,1)$ as a function of  $\alpha.$ The dotted-dashed line corresponds to the numerical data, whereas the continuum line to the analytical transfer matrix.}
\end{figure}

\begin{figure}[h]
	\includegraphics[scale=1.0]{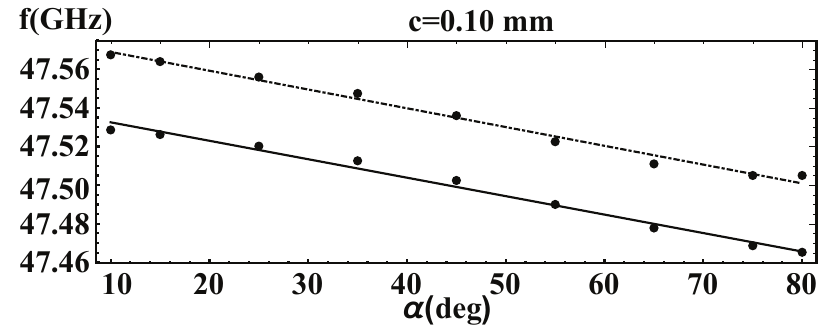}
	\caption{\label{fig:SlopeALPHAc0p10} The resonant frequency of the modes $S(0,-1)$ and $S(0,1)$ as a function of  $\alpha.$ The dotted-dashed line corresponds to the numerical data, whereas the continuum line to the analytical transfer matrix.}
\end{figure}

Out of the 5 channels for transfer of the SS-polarization, only Channels 3 and 4 depend on the angle $\alpha$. 
As the angle $\alpha$ is varied, the resonant frequency for the degenerate modes $(0,-1)$ and $(0,1)$ changes in the vicinity of the frequencies determined at $\alpha =45^{\circ}$, Figs.\ref{fig:SlopeALPHAc0p25},\ref{fig:SlopeALPHAc0p10}.  Although the relative changes are small, with an estimated linear slope of $-0.96 \text{MHz/deg}$ and $-2.46 \text{MHz/deg}$, respectively, the slopes are close to the ones from the numerically simulated data.
The systematic shift between the theoretical and numerical values, described for $\alpha=45^\circ$ in Tab.\ref{Table:LayerWidth}, is the same for all angles $\alpha$.

The study of the wedge effect, that comes out for $0^{\circ}<\alpha<10^{\circ}$ or for $80^{\circ}<\alpha<90^{\circ}$, will be covered in a different manuscript.

The transfer matrix elements depend on the logarithm of the permittivities, so we need to test the formulas for a metal, which exhibits  complex permitivitty. 
We take $\epsilon _{\text{2B}}= \text{ Johnson-Chrysti gold   }$ and find a local minimum of an objective function in the 5-dimensional space of the other dielectric constants.
The frequency range is now placed in the infrared at $f=286.2$ THz. 
The optimization process to find the five unknown dielectric constants is similar to that for the all-dielectric case. This time, though, the objective function is based on one propagative mode $(0,0)$ and one evanescent mode  $(-1,0)$, both P-polarized. The position and shape of the theoretical resonance were confirmed by numerical simulations performed via \cite{studios2008cst}. 
\section{A high Q-factor bilaminar structure}\label{sec:HighQ}
\begin{figure}[h]
	\includegraphics[scale=1.00]{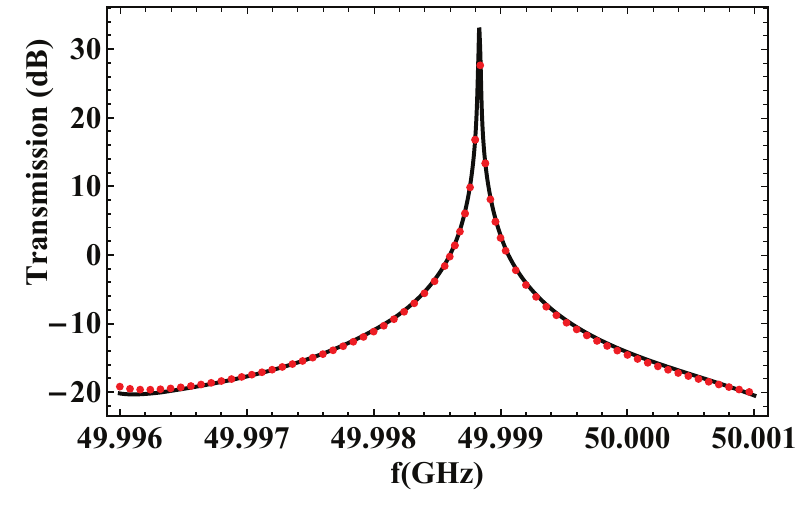}
	\caption{\label{fig:ResonanceEpsilonHighQ} A resonance with a high Q-factor. The Q-factor at -3 dB is $Q=4.8\times 10^6$. The black curve represents the theoretical transmission from S(0,0) at z-$min$ to S(0,1) at z-$max$. The red points were numerically simulated via \cite{studios2008cst}. All  parameters are equal to the optimized values except for  $c=0.01$mm and $\epsilon_{2\text{R}}=1.0$.}
\end{figure}

To study the theoretical prediction for the amplitude we choose a series of  resonances with large Q-factors. The first has $Q=4.8\times 10^6$ at $-3\;\text{dB}$, Fig.\ref{fig:ResonanceEpsilonHighQ}. The theoretical values were computed using a $20\times 20$ transfer matrix. Two systematic shifts, $\text{f}_{\text{shift}}$ and $\text{t}_{\text{shift}}$, are needed to superpose the theoretical formula $20\log_{10}\abs{t^{max,min}_{S(0,1),S(0,0)}(f-\text{f}_{\text{shift}})}+t_{\text{shift}}$ on numerical simulations. For Fig.\ref{fig:ResonanceEpsilonHighQ},  $\text{f}_{\text{shift}}=3.2\;10^{-5}\text{ GHz}$ and $t_{\text{shift}}=-3.9\text{ dB}$. The lamina being thin, $c=0.01\text{mm}$, the shifts are small with respect to the frequency and amplitude at resonance.

\begin{figure}[h]
	\includegraphics[scale=1.0]{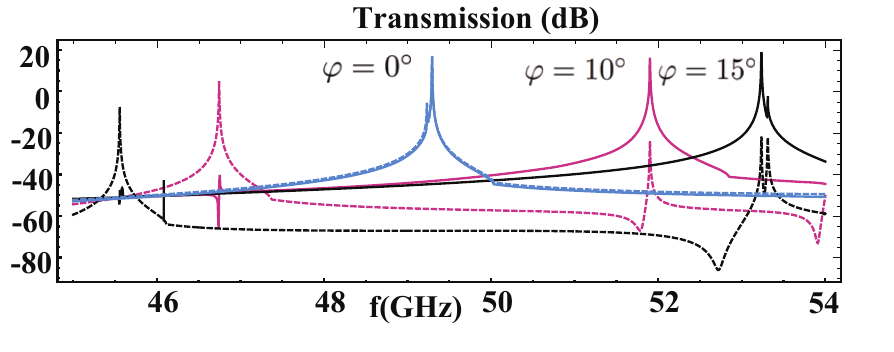}
	\caption{\label{fig:PHI0AND10AND15MinMaxCROPPED}Resonances with  a high Q-factor for variable incident angle $\varphi$. The parameters are $c=0.05$ mm and $\varphi={0^{\circ},10^{\circ}, \text{and} 15^{\circ}}$. All the other parameters are placed at the optimized values. The continuous and dashed curves represent the theoretical transmission from S(0,0) at z-$min$ to S(0,1) and S(0,-1) at z-$max$, respectively.  For $\varphi={0^{\circ}},{10^{\circ}}$ and $\varphi={15^{\circ}}$, the resonant frequency, in GHz, and the Q-factors are  $(49.30,3.6\times 10^4),(51.90,5.2\;10^3)$ and $(53.24,1.3\;10^4)$, respectively.}
\end{figure}

Variable $\varphi$ offers another venue to study the amplitude around a high Q-factor resonance. The modes $(0,-1)$ and $(0,1)$ are degenerate for $\varphi=0^{\circ}$, having the same Rayleigh and resonant frequency. As the angle $\varphi$ slides away from $0^{\circ}$, the scattering amplitude of the mode $S(0,-1)$  decreases, whereas the scattering amplitude for the mode $S(0,1)$ remains excited at high dB levels, Fig.\ref{fig:PHI0AND10AND15MinMaxCROPPED}.
The same trend was confirmed by numerical simulation. Table\ref{tab:Mode01PhiShifts} contains additional data for the comparison between theory and numerical simulations for the high Q-factors.

\begin{table}[h]\label{TablePHIshifts}
	\caption{\label{tab:Mode01PhiShifts}Variable $c$ for $\varphi={15^{\circ}}$. Scattering matrix element for input S(0,0) at z-$min$ output S(0,1) at z-$max$. Frequency shifts are measured in GHz and the amplitude of shifts of the Scattering matrix elements in dB.}
	\begin{ruledtabular}
		\begin{tabular}{lllll}
			$c$[mm]  & $\text{f}_{\text{shift}}$ &  $t_{\text{shift}}$& $Q$\\
			\colrule\\[-1ex]
			0.25 & 3.2 $10^{-1}$ & 9.81 & 1.1 $10^3$  \\
			0.10 & 5.4 $10^{-2}$ & 5.67 & 2.6 $10^3$   \\
			0.05 & 9.4 $10^{-3}$ & 2.85 & 1.3 $10^4$   \\
			0.01 & 9.0 $10^{-5}$ & -3.78 & 1.2 $10^6$  
		\end{tabular}
	\end{ruledtabular}
\end{table} 

\section{Complex-frequency plane and the weighted Fano-Lorentz line shape}\label{sec:ComplexOmega}

\begin{figure}[h]
	\includegraphics[scale=1.0]{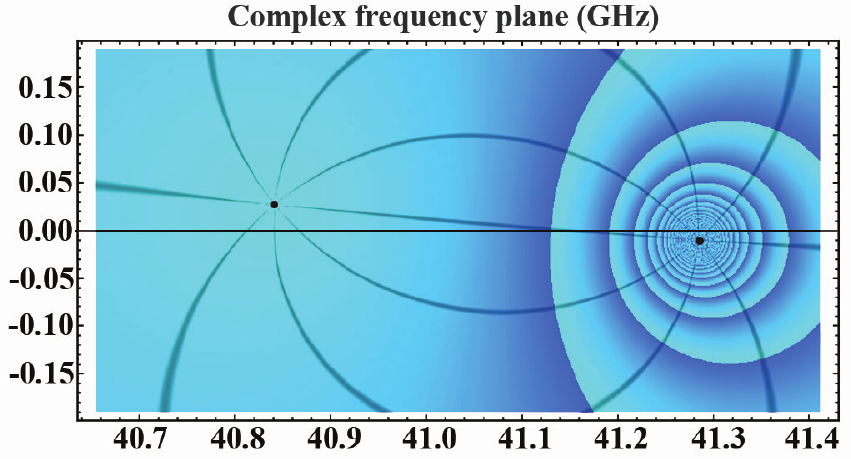}
	\caption{\label{fig:ComplexPlaneFrequency} Representation, in the complex frequency plane, of the scattering matrix element Out S(0,1) at z-$min$, In S(0,0) at z-$max$. The closed curves around the pole 41.286$-$i 0.009, show the absolute value of the scattering matrix as it tends to infinity. The argument of the scattering matrix element is constant on the curves that connect the zero at 40.843 + 0.027 i with the pole. The real resonant frequency is located above the pole, at 41.286  GHz. }
\end{figure}

Moving beyond predicting the position of the resonant frequency and its amplitude, the complete analytical solution for transfer matrix provides a simple but meaningful approximation for the resonant line shape
\begin{gather}\nonumber
	t^{min,max}_{S(0,1),S(0,0)}=\frac{\text{Cofactor}_{S(0,0),S(0,1)}(T^{-,-})}{\text{Det}(T^{-,-})}\\\label{eq:ZeroPole}
	\cong \rho e^{i\Phi}\frac{f-f_{zero}}{f-f_{pole}},
\end{gather}
where the zero and the pole are located in the complex frequency plane, both in the vicinity of  the real resonant frequency, \cite{avrutsky2013linear}. The meromorphic approximation (\ref{eq:ZeroPole}) has a natural structure of dipoles with the monopoles placed at $f_{zero}$ and $f_{pole}$, Fig.\ref{fig:ComplexPlaneFrequency}. 

To estimate the pole and the zero, we use $\varphi_z(\zeta)$ already defined for complex values of $\zeta$. Here we take advantage of the continuity at the real $\zeta$-axis of $\varphi_z(\zeta)$ coming from below,  $\text{Im}(\zeta)\leqslant 0$. This translates, for our case  of $\varphi=0^{\circ}$ and $M_x=0$, into the continuity from above, $\text{Im}(\Omega)\geqslant 0$,  of the cofactor and the determinant at the real $\Omega$-axis. Both functions of $\Omega$ are thus extended into the lower half of complex frequencies plane from the values they take above and on the real frequency axis, Appendix J. 
After this analytic continuation, the zero-pole model, extracted from a  $20\times 20$ matrix $t^{min,max}$ , is
\begin{equation}
	t^{min,max}_{S(0,1),S(0,0)}=0.053\; e^{-2.732 i}\;\frac{f-f_{\text{zero}}}{f-f_{\text{pole}}}\; ,
\end{equation}
with frequency, in GHz, given by
\begin{eqnarray}\nonumber
	f_{\text{zero}}=40.843+ 0.027 i \\\nonumber
	f_{\text{pole}}=41.286- 0.009 i
\end{eqnarray}
Following \cite{avrutsky2013linear}, we express the scattering line shape $\abs{t^{min,max}_{S(0,1),S(0,0)}}^2$ in different forms. The first is 
\begin{gather}
	\abs{t^{min,max}_{S(0,1),S(0,0)}}^2(x)=\rho^2\frac{(\delta+x)^2+\gamma^2}{1+x^2}\\\nonumber
	=2.81\times 10^{-3}\frac{(-47.47+x)^2+8.55}{1+x^2},
\end{gather}
where $x=-\frac{f-\text{Re}(f_{\text{pole}})}{\text{Im}(f_{\text{pole}})}$, $\delta=-\frac{\text{Re}(f_{\text{pole}}-f_{\text{zero}})}{\text{Im}(f_{\text{pole}})}$, and $\gamma=\frac{\text{Im}(f_{\text{zero}})}{\text{Im}(f_{\text{pole}})}.$
The second form is connected with the presence of a continuum background
\begin{equation}
	\abs{t^{min,max}_{S(0,1),S(0,0)}}^2(x)=A_0+F_0\frac{(\nu+x)^2}{1+x^2}.
\end{equation}
We found two solutions for the parameters $A_0,F_0$ and $\nu$. Namely,
$A_0=1.06\times 10^{-5}$, $F_0=2.8\times 10^{-3}$ and $\nu=-47.65$, or
$A_0=6.37$, $F_0=-6.36$ and $\nu=2.09\times 10^{-2}.$ This form is thus ambiguous as is also noted in  \cite{avrutsky2013linear}.

The third, and the final format, presents the resonant line shape as a weighted sum of a Fano and Lorentz line shapes
\begin{equation}\label{FanoLorentz}
\hspace{-0.20ex}\abs{t^{min,max}_{S(0,1),S(0,0)}}^2(x)=A \Bigr(\eta\frac{(\delta+x)^2}{1+x^2}+(1-\eta)\frac{1}{1+x^2}\Bigl).
\end{equation}
where the weight parameter is $\eta=1/(1+\gamma^2)=0.105$ and $A=\rho^2 (1+\gamma^2)=0.027.$

The Fano line shape, which is the first term in (\ref{FanoLorentz}), contributes 10.5\text{\%} to the resonance line shape.
\section{Conclusion }\label{sec:Conclusions}
To sum up, we show how to derive the transfer matrix using  Pendry-MacKinnon discrete Maxwell's equations. Space is an orthogonal lattice with lattice constants $a,b$ and $c$ on the $x,y$ and $z$ directions, respectively. The z-axis is singled-out as the transfer axis. A finite number of parallel discrete laminae of thickness $c$ are  placed perpendicular along the z-axis. Each lamina can carry any non-magnetic, frequency-dependent pattern, independent from each other. The method we introduce computes, in an analytical closed form, the transfer matrix elements  for all Bloch-Floquet modes. The computation starts with a large number of path-operators that transfer the field from the first to the last lamina. Fortunately, many of these path-operators come up with a zero contribution and the results of the paths that contribute can be further grouped together in simple terms, which we call channels. Each channel has the interesting attribute of being a product of two terms that bear distinctive features. One term, called the channel's $\chi$-function, depends on the permittivities, whereas the other term, called the channel's $\mathcal{Z}$-factor, does not. We computed the $\chi$ and the $\mathcal{Z}$-factors for a bilaminar device for all pairs of input output polarization. 

To exemplify the building up of a transfer matrix element consider, in continuum case of $a,b\to 0$, the P polarization at the input and the S at the output. The method for the discrete case is similar to the continuum, except that the $\chi$ functions are discrete like in (\ref{eq:g6discrete}). For this (P,S) pair of polarization the transfer matrix element is a sum over 5 channels, having their $\chi$-functions displayed in Fig.\ref{fig:InPOutSPartitura.pdf}. For Channel 2, as an example, the  $\chi$-function is a polynomial in  $(p_x,p_y)$ given by $p_y (-\partial_x\epsilon_2)+p_x \partial_y\epsilon_2$, for any permittivity pattern. The transfer matrix elements depend on the Fourier transforms of the  $\chi$-functions, which contain all Bloch-Floquet modes through the variable $q_{\perp}=p_{\perp}-p_{\perp}^{\text{OUT}}$. Next, the choice of a pair of  $\pm$ propagation direction decides the specific form of the $\mathcal{Z}$-factor. For example, if we are looking for the matrix element $T^{-,-}$, then use Table \ref{tab:SPall} and divide its entries by $\|\text{OUT}\|$  to normalize the $\mathcal{Z}$-factor. 

The rule to obtain the $\mathcal{Z}$-factors for other directions  is to change $\text{IN}-$ into $\text{IN}+$ and, at the same time, change
$\varphi_{\text{z,IN}}\rightarrow  \varphi_{\text{z,IN}}^{-1}$, and $c\rightarrow -c$, before moving to the c-scaled variables. The same rule applies for $\text{OUT}-$ into $\text{OUT}+$. However, this rule does not apply for Channel 1, for both  SS and PP, for which all unnormalized  $\mathcal{Z}$-factors are tabulated in Appendix B Tables \ref{tab:SSchannel1} to \ref{tab:SPall}. Finally, the matrix element is a sum over the product of the Fourier transform of  $\chi$ and the $\mathcal{Z}$-factor for  all channels that belong to the selected  polarization pair.

A  valuable way of proceeding further is to extend to range of applications. One possibility is to exploit Fig.\ref{fig:NumberOfModes} and bring together two devices with distinct number of resonant modes, and examine the propagation along different channels at the interface. Other applications may be connected to metasurfaces \cite{barbuto2021metasurfaces}, energy harvesting \cite{zhou2021metamaterials}, \cite{atwater2010plasmonics},\cite{eteng2021review}, leaky-wave theory \cite{eteng2021review} and frequency-selective surfaces \cite{ferreira2018review}.

\section{Acknowledgment}
 One of us, O.L., wishes to thank the Fulbright Commission for a Fulbright Scholarship grant and the host Victor Lopez-Richard from Universidade Federal de S\~{a}o Carlos, SP, Brazil, where part of this work was done. In addition, O.L. thanks his colleague Mariama Rebello de Sousa Dias for many conversations during the course of this research. Both of us wish to thank Dr. Andrei Silaghi, with Continental Automotive Romania SRL, for facilitating the access to numerical simulation using CST microwave studio \cite{studios2008cst}.

\vspace{1.0em}
\textbf{Appendix A: A summary of symbols with explanatory notes}

\textbf{1.}The discrete position vectors are $\vv{n}=(a n_x,b n_y,c n_z)$ and $n_{\perp}=(a n_x,b n_y)$. In the continuous, $a,b\to 0$, the position  $n_{\perp}$ becomes $r_{\perp}=(x,y)$.

\textbf{2.} The vector $p_\perp=(p_x,p_y)$ appear in the formulation of the plane wave basis, $\ket{\Psi_{S/P,p_\perp}^{\pm}} e^{\pm i p_z n_z c} e^{- i \omega t}$ where $\ket{\Psi_{S/P,p_\perp}^{\pm}}=\ket{p_\perp,S/P,\pm}e^{i p_\perp n_\perp}$, (\ref{eq:PropagationS}). To distinguish between two Bloch-Floquet modes, one at the input IN and the other one at the output OUT, we write $p_{\perp}^{\text{IN}}$ and $p_{\perp}^{\text{OUT}}$, respectively. To keep the final formulas for the transfer matrix elements simple, we use $p_{\perp}$ instead of $p_{\perp}^{\text{IN}}$, as is the case in Fig.\ref{fig:InSOutSPartitura.pdf}.

\textbf{3.} Connection between $p_x$ and $k_x^E = (i a)^{-1}(e^{i a p_x}-1)$, (\ref{eq:KxE}).

\textbf{4.} The unitless frequency $\Omega$ is defined by $\Omega v_0=\omega c$, (\ref{eq:DefinitionOmega})

\textbf{5.} The $p_z$ component appears in $\varphi _{z}=e^{i c p_z}$. $\varphi_z(\zeta)=\zeta-\sqrt{\abs{\zeta^2-1}}e^{\frac{i}{2}\big(\text{arg}(\zeta-1)+\text{arg}(\zeta+1)\big)}$, (\ref{eq:phizWitharg}). The $\text{arg}(\zeta)$ takes values on $[-\pi,\pi)$ in such a way  that $\text{arg}(\zeta)=-\pi$ if $\text{Im}(\zeta)=0$ and $\zeta<0$. The argument $\zeta$ is defined in (\ref{eq:PhizEquation2}).

\textbf{6.} Periodic boundary conditions introduce the modes $(M_x,M_y)$ and the direction $(\xi_x,\xi_y)$ of the mode $(0,0)$ through  (\ref{eq:MXMY}) and (\ref{eq:MXMY2}).

\textbf{7.} Transfer matrix element $T^{-,-}_{m,n}$ are organized in (\ref{eq:TMMmn}). 

\textbf{8.} All XY-operators are listed in Fig.\ref{fig:AllChannels}.

\textbf{9.} An example of a discrete $\chi$-function, (\ref{eq:g6discrete}). 

\textbf{10.} Discrete Fourier transform of a $\chi$-function, (\ref{eq:DisctereFourierChi3}).  

\textbf{11.} The Fourier transform goes from space $n_{\perp}$ to the mode variable $q_{\perp}=p_\perp^{\text{IN}}-p_\perp^{\text{OUT}}\equiv p_\perp-p_\perp^{\text{OUT}}$.

\textbf{12.} An example of a $\mathcal{Z}$-factor, (\ref{eq:Z3SS}). 

\textbf{13.} Transfer matrix element as a sum over channels, (\ref{eq:TsumChannels}).

\textbf{14.} The $\mathcal{Z}$-factors are tabulated in Table \ref{tab:SSallmm} and \ref{tab:SSchannel1} to \ref{tab:SPall}. 

\textbf{15.} Rules to obtain all $\mathcal{Z}$-factors for all directions are presented in  Appendix B. 

\textbf{16.} Some examples of  permittivity functions for the continuum $(x,y)$-plane are in Table \ref{tab:DielectricFunctions}. 

\begin{figure}[h]
	\includegraphics[width=1.0\textwidth]{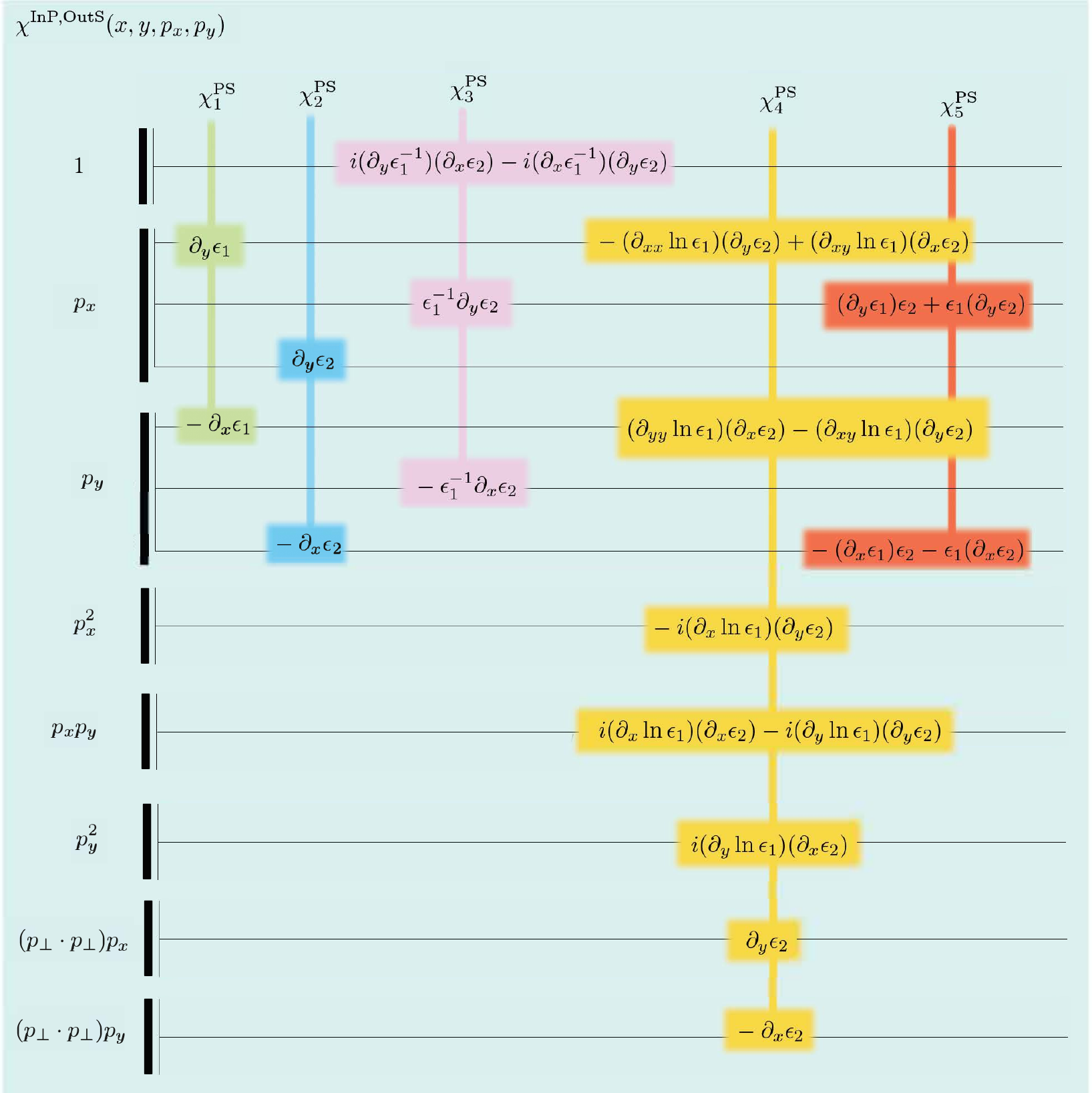}
	\caption{\label{fig:InPOutSPartitura.pdf}The $\chi$-functions for the  In P Out S channels. From the total of  5 channels, the more elaborate is Channel 4, which is cubic in $(p_x,p_y)$.}
\end{figure}

\begin{figure}[h]
\includegraphics[width=1.0\textwidth]{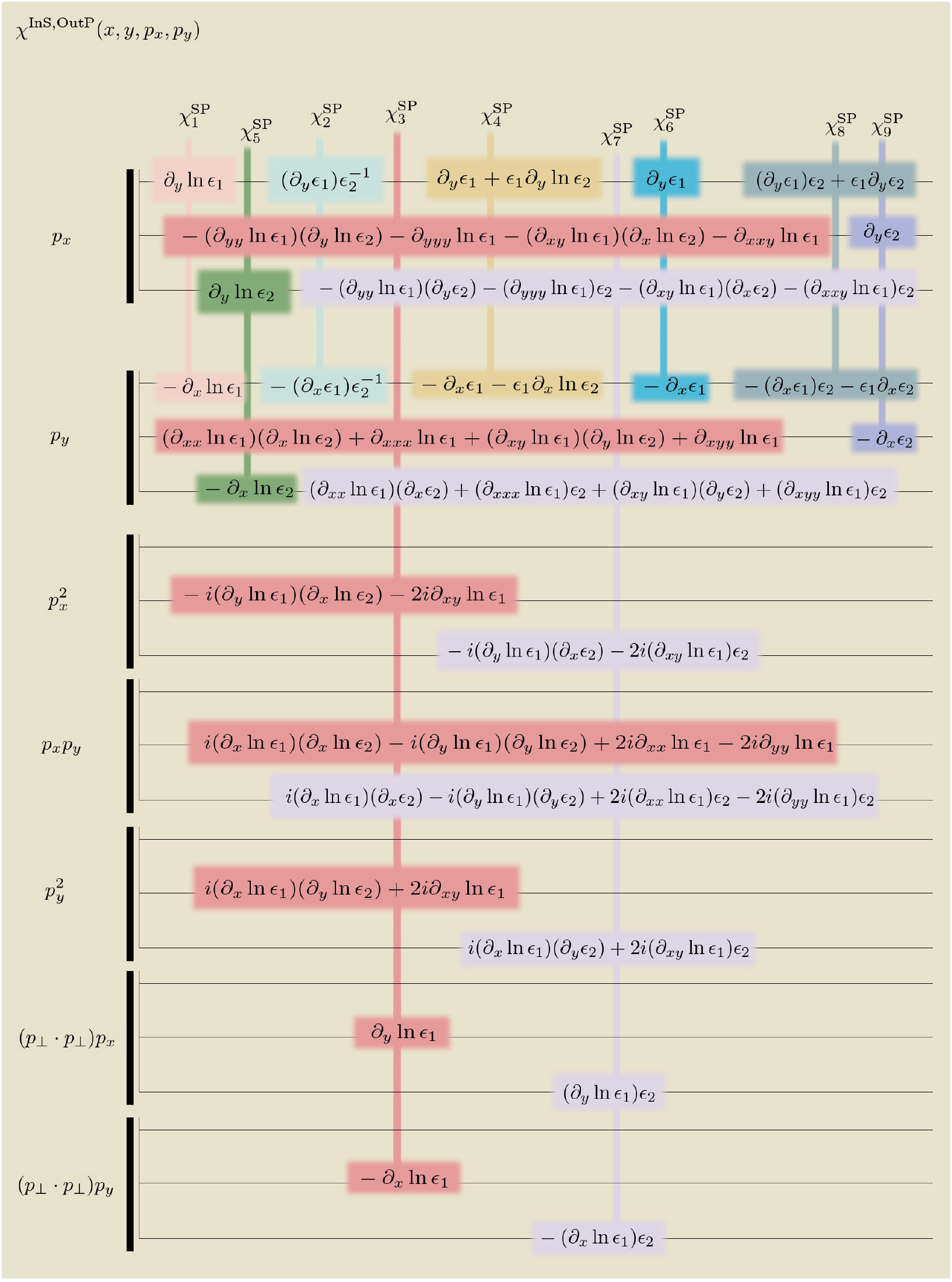}
\caption{\label{fig:InSOutPPartitura.pdf}The $\chi$-functions for the  In S Out P channels. From the total of  9 channels, the more elaborate are Channel 3 and 7, which are cubic in $(p_x,p_y)$.}
\end{figure}

\begin{figure}[h]
\includegraphics[width=1.0\textwidth]{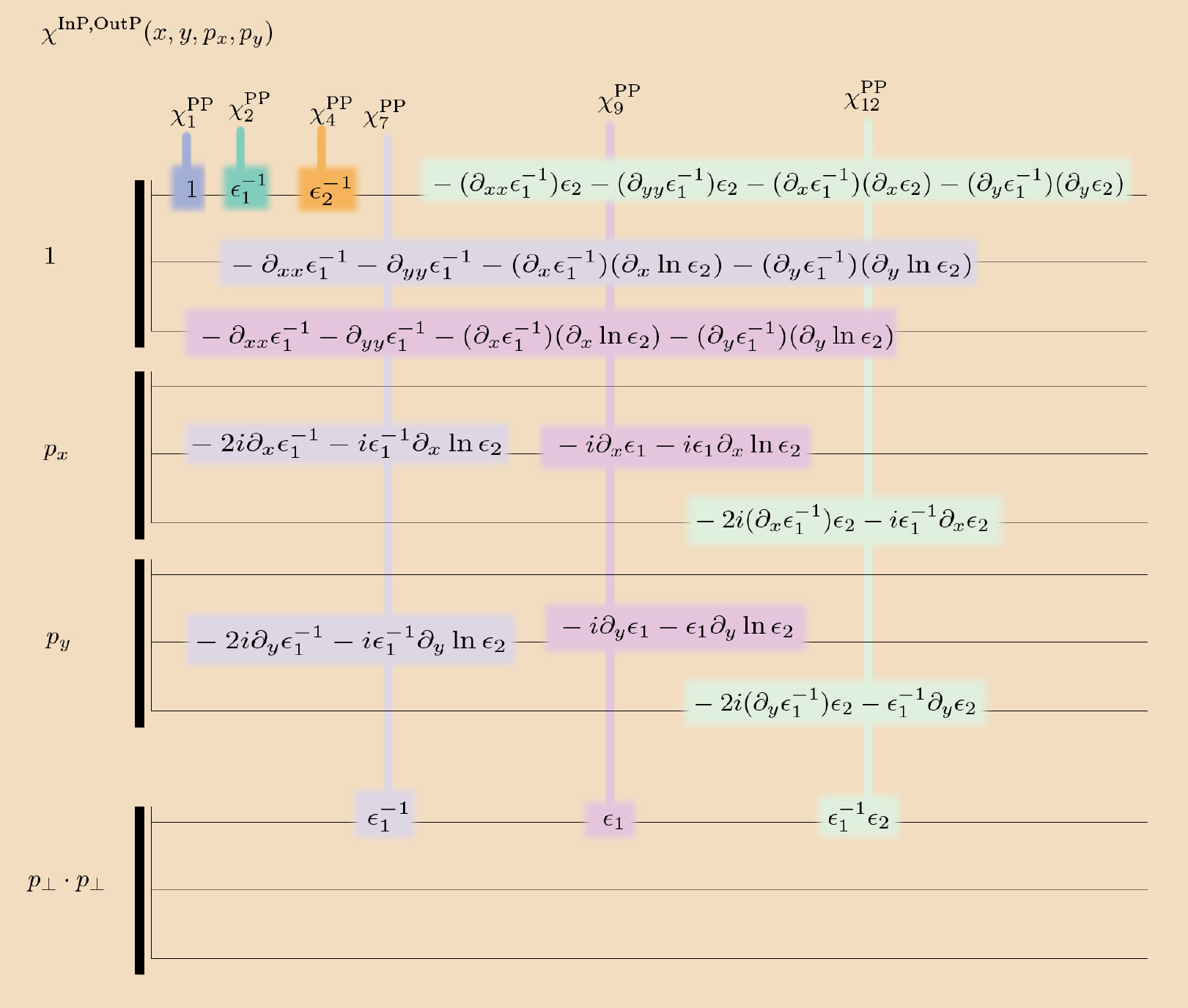}
\caption{\label{fig:InPOutPPartitura1.pdf}The $\chi$-functions for the  In P Out P channels. This figure contains 6 channels  out of the total of 14, which have a nonzero free term as a polynomial in $p_x$ and $p_y$.}
\end{figure}

\begin{figure}[h]
\includegraphics[width=1.0\textwidth]{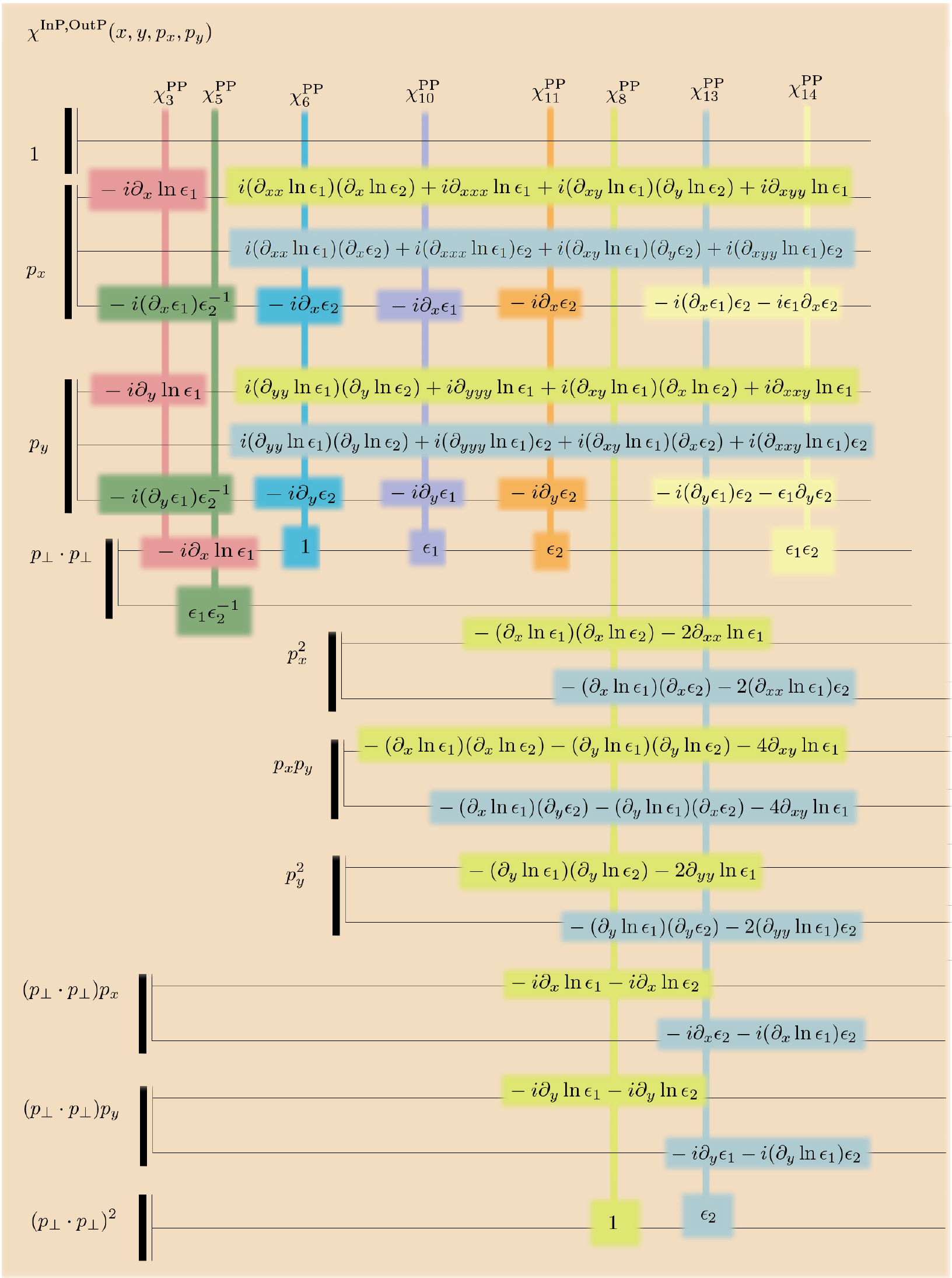}
\caption{\label{fig:InPOutPPartitura2.pdf}The $\chi$-functions for the  In P Out P channels. This figure contains 8 channels  out of the total of 14, which have no free term as a polynomial in $p_x$ and $p_y$.}
\end{figure}

\textbf{17.} For the continuum $(x,y)$ case, the $\chi$-functions, the permittivity functions and the polynomial dependence on $p_{\perp}\equiv p_{\perp}^{\text{IN}}$ are found in (\ref{Continuum:3SS}), Fig.\ref{fig:InSOutSPartitura.pdf} and Fig.\ref{fig:InPOutSPartitura.pdf} to \ref{fig:InPOutPPartitura2.pdf}

\textbf{18.} Through  $c$-scaling,  $x = c x^{'},y=c y^{'}$ from (\ref{eq:xy-scaled}), the lattice constant $c$ simplifies in the transfer matrix element.

\textbf{19.} Choose a unit of length in accordance with the frequency band under study. Use the thickness parameter $t=[1\;\text{unit}](c \;[\text{unit}])^{-1}$, (\ref{eq:tParameter}), to write the transfer matrix element in terms of only unitless parameters.

\textbf{20.} Each vertex for any polygonal tessellation pattern carry a pair of orthogonal unit vectors $(\hat{\lambda},\hat{\mu})$. An example is the $\pi$-$\pi$-rig from Fig.\ref{fig:PiPiRig.pdf}. 

\textbf{21.} The jumps in the relative permittivities are measured across $\hat{\mu}$ in the direction of $\hat{\lambda}$, (\ref{eq:JumpEpsilon}). 

\textbf{22.} The Fourier transforms of a tessellated pattern, in continuum, from the position space vector $r_{\perp}$ to the mode vector $q_{\perp}$, are built on $(\hat{\lambda},\hat{\mu})$, the permittivities jumps and two constant gauge vector $W$ and $\mathcal{D}$.

\onecolumngrid
\twocolumngrid

\begin{table*}
	\begin{flushleft}
		\textbf{Appendix B: The $\mathcal{Z}$-factors and the $\chi$-functions} 
\vspace{1em}

 For all tables, $\bra{p_{\perp}}\ket{p_{\perp}} $ can be written in terms of $\varphi _{z}$ through $\bra{p_{\perp}}\ket{p_{\perp}} c^2\varphi _{z}=1-2  \varphi _{z}+\Omega^2 \varphi _{z}+ \varphi _{z}^2$.
 
 The subscript No. of the XY-operator represents the Channel number.
 
 The $\mathcal{Z}$-factors are valid for both discrete and continuous $(x,y)$-plane, since they are independent of permittivities.
 
 The $\chi$-functions from Fig.\ref{fig:InPOutSPartitura.pdf} to Fig.\ref{fig:InPOutPPartitura2.pdf} are valid for a continuous $(x,y)$-plane.

	\end{flushleft}	
	\centering
	\begin{minipage}[t]{\columnwidth}
		\caption{\label{tab:SSchannel1}Unnormalized $\mathcal{Z}_1$ factor for SS $\text{Channel}_1$}
		\begin{ruledtabular}
			\begin{tabular}{ll}
				$\text{OutS,InS}$ &\hspace{10ex} $\mathcal{Z}\text{-channel}_1\;\|\text{OUT}\|$ \\
				\colrule\\[-1ex]
				$ -,- $&\hspace{-5ex} $\frac{i \bra{p_{\perp}^{\text{OUT}}}\ket{p_{\perp}^{\text{OUT}}} \left(\left(\Omega ^2 \varphi _{z,\text{OUT}}+1\right){}^2-\varphi _{z,\text{OUT}}^2\right)}{c \varphi _{z,\text{OUT}}^4} \delta _{\text{IN},\text{OUT}}$ \\[2ex]
				$+,-$ &\hspace{-5ex} $-\frac{i \Omega ^2 \bra{p_{\perp}^{\text{OUT}}}\ket{p_{\perp}^{\text{OUT}}} \left(\Omega ^2 \varphi
					_{z,\text{OUT}}+\varphi _{z,\text{OUT}}^2+1\right)}{c \varphi _{z,\text{OUT}}}  \delta _{\text{IN},\text{OUT}}$\\[2ex]
				$-,+$ &\hspace{-5ex} $\frac{i \Omega ^2 \bra{p_{\perp}^{\text{OUT}}}\ket{p_{\perp}^{\text{OUT}}} \left(\Omega ^2 \varphi
					_{z,\text{OUT}}+\varphi _{z,\text{OUT}}^2+1\right)}{c \varphi _{z,\text{OUT}}} \delta _{\text{IN},\text{OUT}} $\\[2ex]
				$+,+$ &\hspace{-7ex} $-\frac{i \bra{p_{\perp}^{\text{OUT}}}\ket{p_{\perp}^{\text{OUT}}} \varphi _{z,\text{OUT}}^2 \left(\left(\varphi _{z,\text{OUT}}+\Omega ^2\right){}^2-1\right)}{c} \delta _{\text{IN},\text{OUT}}$ \\
			\end{tabular}
		\end{ruledtabular}
\vspace{1em}

\begin{flushleft}	
 The rule to obtain the $\mathcal{Z}$-factors for other directions  is to change $\text{IN}-$ into $\text{IN}+$ or $\text{OUT}-$ into $\text{OUT}+$ and, at the same time, change
$\varphi_{\text{z,IN}}\rightarrow  \varphi_{\text{z,IN}}^{-1}$ or $\varphi_{\text{z,OUT}}\rightarrow  \varphi_{\text{z,OUT}}^{-1}$, and $c\rightarrow -c$. However, this rule does not apply for Channel 1, for both  SS and PP, for which all unnormalized  $\mathcal{Z}$-factors are tabulated in Tables \ref{tab:SSchannel1} and \ref{tab:PPchannel1}.
Notice that the lattice constant  $c$  is present in these tables to enable the transformation $c \rightarrow -c$. The same constant simplifies from the transfer matrix elements after the variables are $c$-scaled. To normalize the $\mathcal{Z}$-factors, divide the second column by $\|\scriptsize{\text{OUT}}\|$, which, by formula (\ref{eq:NormOUT}), can be written in terms of $\varphi_{\text{z,OUT}}$.
\end{flushleft}			
	\end{minipage}\hfill
	\begin{minipage}[t]{\columnwidth}
		\caption{\label{tab:PSall}Unnormalized $\mathcal{Z}$ factor for  OUT P-, IN S-}
		\begin{ruledtabular}
			\begin{tabular}{ll}
				$(\text{XY-operator})_{\text{No.}}$ &\hspace{-10ex} $\mathcal{Z} \text{(OUT,P-)(IN,S-)}\;\|\text{OUT}\|$ \\
				\colrule\\[-1ex]
				$(\epsilon^{-1}_1\bra{T_{\text{E}}
				} \epsilon_1\sigma_2\ket{T_H})_1$& $-\frac{c \bra{p_{\perp}^{\text{OUT}}}\ket{p_{\perp}^{\text{OUT}}}}{\varphi _{z,\text{IN}} \varphi _{z,\text{OUT}}} $\\[2ex]
				$( \epsilon^{-1}_2\bra{T_{\text{E}}} \epsilon_1\sigma_2\ket{T_H})_2 $&$ -\frac{c \bra{p_{\perp}^{\text{OUT}}}\ket{p_{\perp}^{\text{OUT}}}}{\varphi _{z,\text{IN}} \varphi _{z,\text{OUT}}} $\\[2ex]
				$(\epsilon^{-1}_2\bra{T_{\text{E}}} \epsilon_2\ket{T_{\text{E}}}\epsilon^{-1}_1\bra{T_{\text{E}}}\epsilon_1\sigma_2\ket{T_H})_3 $&$ -\frac{c^3 \bra{p_{\perp}^{\text{OUT}}}\ket{p_{\perp}^{\text{OUT}}}}{\varphi _{z,\text{IN}} \varphi _{z,\text{OUT}}}$ \\[2ex]
				$( \epsilon^{-1}_2\bra{T_{\text{E}}}\epsilon_2 \epsilon_1\sigma_2\ket{T_H})_4 $& $\frac{c \Omega ^2 \bra{p_{\perp}^{\text{OUT}}}\ket{p_{\perp}^{\text{OUT}}}}{\varphi _{z,\text{IN}} \varphi _{z,\text{OUT}}} $\\[2ex]
				$(\epsilon^{-1}_2\bra{T_{\text{E}}}\epsilon_2\sigma_2\ket{T_H})_5 $&\hspace{-12ex} $-\frac{c \bra{p_{\perp}^{\text{OUT}}}\ket{p_{\perp}^{\text{OUT}}} \left(\Omega ^2 \varphi _{z,\text{IN}}+1\right)}{\varphi _{z,\text{IN}}^2 \varphi
					_{z,\text{OUT}}} $\\[2ex]
				$(\bra{T_E}\epsilon_1\sigma_2\ket{T_H})_6 $& $\frac{\varphi _{z,\text{OUT}}+2 \Omega ^2-1}{c \varphi _{z,\text{IN}} \varphi _{z,\text{OUT}}}$ \\[2ex]
				$( \bra{T_E} \epsilon_2\ket{T_{\text{E}}}\epsilon^{-1}_1\bra{T_{\text{E}}}\epsilon_1\sigma_2\ket{T_H})_7 $&\hspace{-2ex} $\frac{c \left(\varphi _{z,\text{OUT}}+\Omega ^2-1\right)}{\varphi _{z,\text{IN}} \varphi _{z,\text{OUT}}} $\\[2ex]
				$(\bra{T_E}\epsilon_2 \epsilon_1\sigma_2\ket{T_H})_8 $&\hspace{-5ex} $-\frac{\Omega ^2 \left(\varphi _{z,\text{OUT}}+\Omega ^2-1\right)}{c \varphi _{z,\text{IN}} \varphi
					_{z,\text{OUT}}}$ \\[2ex]
				$(\bra{T_E}\epsilon_2\sigma_2\ket{T_H})_9 $&\hspace{-12ex} $\frac{\left(\Omega ^2 \varphi _{z,\text{IN}}+1\right) \left(\varphi _{z,\text{OUT}}+\Omega ^2-1\right)}{c \varphi
					_{z,\text{IN}}^2 \varphi _{z,\text{OUT}}}$ \\	
			\end{tabular}
		\end{ruledtabular}
	\end{minipage}
	\vspace{-17.90em}
	\begin{minipage}[t]{\columnwidth}
		\caption{\label{tab:PPchannel1}Unnormalized $\mathcal{Z}_1$ factor for PP $\text{Channel}_1$}
		\begin{ruledtabular}
			\begin{tabular}{ll}
				$\text{OutP,InP}$ &\hspace{10ex} $\mathcal{Z}\text{-channel}_1\;\|\text{OUT}\|$ \\
				\colrule\\[-1ex]
				$ -,- $ & $-\frac{i \bra{p_{\perp}^{\text{OUT}}}\ket{p_{\perp}^{\text{OUT}}}  \left(2 \Omega ^2 \varphi
					_{z,\text{OUT}}+\varphi _{z,\text{OUT}}^2-1\right)}{c \varphi _{z,\text{OUT}}^2}\delta _{\text{IN},\text{OUT}}$ \\[2ex]
				$ +,-$ &$ \frac{2 i \Omega ^2 \bra{p_{\perp}^{\text{OUT}}}\ket{p_{\perp}^{\text{OUT}}}  \varphi _{z,\text{OUT}}}{c}\delta _{\text{IN},\text{OUT}} $\\[2ex]
				$ -,+$ &$ -\frac{2 i \Omega ^2 \bra{p_{\perp}^{\text{OUT}}}\ket{p_{\perp}^{\text{OUT}}} }{c \varphi _{z,\text{OUT}}}\delta _{\text{IN},\text{OUT}}$ \\[2ex]
				$ +,+$ &$ -\frac{i \bra{p_{\perp}^{\text{OUT}}}\ket{p_{\perp}^{\text{OUT}}}  \left(-2 \Omega ^2 \varphi
					_{z,\text{OUT}}+\varphi _{z,\text{OUT}}^2-1\right)}{c}\delta _{\text{IN},\text{OUT}} $\\
			\end{tabular}
		\end{ruledtabular}
	\end{minipage}\hfill
	\begin{minipage}[t]{\columnwidth}
		\caption{\label{tab:PPall} Unnormalized $\mathcal{Z}$ factor for  OUT P-, IN P-}
		\begin{ruledtabular}
			\begin{tabular}{ll}
				$(\text{XY-operator})_{\text{No.}}$ &\hspace{-10ex} $ \mathcal{Z} \text{(OUT,P-)(IN,P-)}\;\|\text{OUT}\|$ \\
				\colrule\\[-1ex]
				$(1)_1$ &\hspace{-25ex}$ -\frac{i \bra{p_{\perp}^{\text{OUT}}}\ket{p_{\perp}^{\text{OUT}}} \left(2 \Omega ^2 \varphi _{z,\text{OUT}}+\varphi _{z,\text{OUT}}^2-1\right)
				}{c \varphi _{z,\text{OUT}}^2}\delta _{\text{IN},\text{OUT}}$ \\[2ex]
				$(\epsilon^{-1}_1)_2$ &\hspace{-10ex}$ \frac{i c \bra{p_{\perp}^{\text{IN}}}\ket{p_{\perp}^{\text{IN}}} \bra{p_{\perp}^{\text{OUT}}}\ket{p_{\perp}^{\text{OUT}}}}{\varphi _{z,\text{OUT}}}$ \\[2ex]
				$(\epsilon^{-1}_1\bra{T_{\text{E}}}\epsilon_1\ket{T_E})_3$ &\hspace{-10ex}$ -\frac{i c \bra{p_{\perp}^{\text{OUT}}}\ket{p_{\perp}^{\text{OUT}}} \left(\varphi _{z,\text{IN}}-1\right)}{\varphi _{z,\text{IN}} \varphi
					_{z,\text{OUT}}}$ \\[2ex]
				$(\epsilon^{-1}_2)_4$ &\hspace{-10ex} $\frac{i c \bra{p_{\perp}^{\text{IN}}}\ket{p_{\perp}^{\text{IN}}} \bra{p_{\perp}^{\text{OUT}}}\ket{p_{\perp}^{\text{OUT}}}}{\varphi _{z,\text{OUT}}}$ \\[2ex]
				$(\epsilon^{-1}_2\bra{T_{\text{E}}}\epsilon_1\ket{T_E})_5$ &\hspace{-10ex}$ -\frac{i c \bra{p_{\perp}^{\text{OUT}}}\ket{p_{\perp}^{\text{OUT}}} \left(\varphi _{z,\text{IN}}-1\right)}{\varphi _{z,\text{IN}} \varphi
					_{z,\text{OUT}}}$ \\[2ex]
				$(\epsilon^{-1}_2\bra{T_{\text{E}}}\epsilon_2\ket{T_E})_6$ &\hspace{-15ex}$ -\frac{i c \bra{p_{\perp}^{\text{OUT}}}\ket{p_{\perp}^{\text{OUT}}} \left(\Omega ^2 \varphi _{z,\text{IN}}+\varphi _{z,\text{IN}}-1\right)}{\varphi
					_{z,\text{IN}} \varphi _{z,\text{OUT}}} $\\[2ex]
				$(\epsilon^{-1}_2\bra{T_{\text{E}}}\epsilon_2\ket{T_{\text{E}}}\epsilon^{-1}_1)_7$ &\hspace{-10ex}$ \frac{i c^3 \bra{p_{\perp}^{\text{IN}}}\ket{p_{\perp}^{\text{IN}}} \bra{p_{\perp}^{\text{OUT}}}\ket{p_{\perp}^{\text{OUT}}}}{\varphi _{z,\text{OUT}}}$ \\[2ex]
				$ (\epsilon^{-1}_2\bra{T_{\text{E}}}\epsilon_2\ket{T_{\text{E}}}\epsilon^{-1}_1\bra{T_{\text{E}}}\epsilon_1\ket{T_E})_8$ &\hspace{-1ex}$ -\frac{i c^3 \bra{p_{\perp}^{\text{OUT}}}\ket{p_{\perp}^{\text{OUT}}} \left(\varphi _{z,\text{IN}}-1\right)}{\varphi _{z,\text{IN}} \varphi
					_{z,\text{OUT}}}$ \\[2ex]
				$(\epsilon^{-1}_2\bra{T_{\text{E}}}\epsilon_2\epsilon_1\ket{T_E})_9$ &\hspace{-10ex}$ \frac{i c \Omega ^2 \bra{p_{\perp}^{\text{OUT}}}\ket{p_{\perp}^{\text{OUT}}} \left(\varphi _{z,\text{IN}}-1\right)}{\varphi _{z,\text{IN}} \varphi
					_{z,\text{OUT}}} $\\[2ex]
				$(\bra{T_E}\epsilon_1\ket{T_E})_{10}$ &\hspace{-10ex} $\frac{i \left(\varphi _{z,\text{IN}}-1\right) \left(\varphi _{z,\text{OUT}}+2 \Omega ^2-1\right)}{c \varphi
					_{z,\text{IN}} \varphi _{z,\text{OUT}}}$ \\[2ex]
				$(\bra{T_E}\epsilon_2\ket{T_E})_{11}$ &\hspace{-10ex}$ \frac{i \left(\Omega ^2 \varphi _{z,\text{IN}}+\varphi _{z,\text{IN}}-1\right) \left(\varphi
					_{z,\text{OUT}}+\Omega ^2-1\right)}{c \varphi _{z,\text{IN}} \varphi _{z,\text{OUT}}}$ \\[2ex]
				$(\bra{T_E}\epsilon_2\ket{T_{\text{E}}}\epsilon^{-1}_1)_{12}$ &\hspace{-10ex}$ -\frac{i c \bra{p_{\perp}^{\text{IN}}}\ket{p_{\perp}^{\text{IN}}} \left(\varphi _{z,\text{OUT}}+\Omega ^2-1\right)}{\varphi _{z,\text{OUT}}}$ \\[2ex]
				$(\bra{T_E}\epsilon_2\ket{T_{\text{E}}}\epsilon^{-1}_1\bra{T_{\text{E}}}\epsilon_1\ket{T_E})_{13}$ &\hspace{-2ex}$ \frac{i c \left(\varphi _{z,\text{IN}}-1\right) \left(\varphi _{z,\text{OUT}}+\Omega ^2-1\right)}{\varphi
					_{z,\text{IN}} \varphi _{z,\text{OUT}}}$ \\[2ex]
				$(\bra{T_E}\epsilon_2\epsilon_1\ket{T_E})_{14}$ &\hspace{-10ex}$ -\frac{i \Omega ^2 \left(\varphi _{z,\text{IN}}-1\right) \left(\varphi _{z,\text{OUT}}+\Omega ^2-1\right)}{c
					\varphi _{z,\text{IN}} \varphi _{z,\text{OUT}}}$ \\
			\end{tabular}
		\end{ruledtabular}
	\end{minipage}
	\begin{minipage}[t]{\columnwidth}	
		\caption{\label{tab:SPall}Unnormalized $\mathcal{Z}$ factor for  OUT S-, IN P-}
		\begin{ruledtabular}
			\begin{tabular}{ll}
				$(\text{XY-operator})_{\text{No.}}$ &\hspace{-10ex} $\mathcal{Z} \text{(OUT,S-)(IN,P-)}\;\|\text{OUT}\|$ \\
				\colrule\\[-1ex]
				$ (\bra{T_H}\sigma_2 \epsilon_1\ket{T_E})_1$ &$ \frac{\Omega ^2 \left(\varphi _{z,\text{IN}}-1\right) \left(\Omega ^2 \varphi _{z,\text{OUT}}+1\right)}{c \varphi
					_{z,\text{IN}} \varphi _{z,\text{OUT}}^2} $\\[2ex]
				$(\bra{T_H}\sigma_2 \epsilon_2\ket{T_E})_2$ & $\frac{\Omega ^2 \left(\Omega ^2 \varphi _{z,\text{IN}}+\varphi _{z,\text{IN}}-1\right)}{c \varphi _{z,\text{IN}}
					\varphi _{z,\text{OUT}}}$ \\[2ex]
				$(\bra{T_H}\sigma_2\epsilon_2\ket{T_{\text{E}}}\epsilon^{-1}_1)_3 $&$ -\frac{c \Omega ^2 \bra{p_{\perp}^{\text{IN}}}\ket{p_{\perp}^{\text{IN}}}}{\varphi _{z,\text{OUT}}} $\\[2ex]
				$ (\bra{T_H}\sigma_2 \epsilon_2\ket{T_{\text{E}}}\epsilon^{-1}_1\bra{T_{\text{E}}}\epsilon_1\ket{T_E})_4$ &$ \frac{c \Omega ^2 \left(\varphi _{z,\text{IN}}-1\right)}{\varphi _{z,\text{IN}} \varphi _{z,\text{OUT}}} $\\[2ex]
				$(\bra{T_H}\sigma_2 \epsilon_2\epsilon_1\ket{T_E})_5$ &$ -\frac{\Omega ^4 \left(\varphi _{z,\text{IN}}-1\right)}{c \varphi _{z,\text{IN}} \varphi _{z,\text{OUT}}}$ \\	
			\end{tabular}
		\end{ruledtabular}
	\end{minipage}
	\begin{minipage}[t]{\columnwidth}
		\centering 
		\phantom{	\begin{tabular}{ll}
				\hline a6 & b6 \\ \hline
		\end{tabular}}
	\end{minipage}
\end{table*}
\clearpage
\onecolumngrid

\clearpage
\appendix{\bf{Appendix C: The diagonal $T^{-,-}$ matrix elements ($q_{\perp}=0$) for the $\pi$-$\pi$-rig  and $a,b\to 0$.}}



\begin{gather}\label{eq:u}
	(T^{-,-}_{S m,S m})_{\text{Channel 1}}=
	-\delta _{\text{IN},\text{OUT}}\frac{\left((\Omega ^2 \varphi _{z,\text{OUT}}+1)^2-(\varphi _{z,\text{OUT}})^2\right) }{\left(\varphi _{z,\text{OUT}}^2-1\right) \varphi _{z,\text{OUT}}^3 },
\end{gather}


\begin{gather}\label{eq:u}
	(T^{-,-}_{S m,S m})_{\text{Channel 2}}=\left( 6 \epsilon _{\text{1B}}+\epsilon _{\text{1L}}+\epsilon _{\text{1R}}\right)
	\frac{\Omega ^2 \left(\Omega ^2 \varphi _{z,\text{OUT}}+1\right)}{8 \varphi _{z,\text{IN}} \left(\varphi _{z,\text{OUT}}^2-1\right) \varphi _{z,\text{OUT}}
	},
\end{gather}

\begin{gather}\label{eq:u}
	\hspace{-65ex}	(T^{-,-}_{S m,S m})_{\text{Channel 3}}=\\\nonumber
	\frac{\Omega^2}{4 \pi ^2 t^2 \left(p_x^2+p_y^2\right) \varphi _{z,\text{IN}} \left(\varphi _{z,\text{OUT}}^2-1\right)}\\\nonumber
	\Biggl(\Biggr.
	p_x^2\Biggl(
	\csc (\alpha ) \sec (\alpha ) \epsilon _{\text{2B}} \left(2 \ln \left(\epsilon _{\text{1B}}\right)-\ln \left(\epsilon _{\text{1L}}\right)-\ln \left(\epsilon
	_{\text{1R}}\right)\right)\\\nonumber
	+\epsilon _{\text{2L}} \left(-\csc (\alpha ) \sec (\alpha ) \log \left(\epsilon _{\text{1B}}\right)+\frac{3 \csc (\alpha ) \log \left(\epsilon _{\text{1L}}\right)}{2
		(\sin (\alpha )+\cos (\alpha ))}-\frac{(\cot (\alpha )-2) \csc (\alpha ) \sec (\alpha ) \log \left(\epsilon _{\text{1R}}\right)}{2 (\cot (\alpha )+1)}\right)\\\nonumber
	+\epsilon _{\text{2R}} \left(-\csc (\alpha ) \sec (\alpha ) \log \left(\epsilon _{\text{1B}}\right)-\frac{(\cot (\alpha )-2) \csc (\alpha ) \sec (\alpha ) \log
		\left(\epsilon _{\text{1L}}\right)}{2 (\cot (\alpha )+1)}+\frac{3 \csc (\alpha ) \log \left(\epsilon _{\text{1R}}\right)}{2 (\sin (\alpha )+\cos (\alpha
		))}\right)
	\Bigl.	\Biggr)\\\nonumber
	+p_x p_y\frac{ \csc (\alpha ) }{\sin (\alpha )+\cos (\alpha )}\left(\epsilon _{\text{2L}}-\epsilon _{\text{2R}}\right) \left(\log \left(\epsilon _{\text{1L}}\right)-\log \left(\epsilon
	_{\text{1R}}\right)\right)\\\nonumber
	+p_y^2 \Biggl( \Biggr.
	\csc (\alpha ) \sec (\alpha ) \epsilon _{\text{2B}} \left(2 \log \left(\epsilon _{\text{1B}}\right)-\log \left(\epsilon _{\text{1L}}\right)-\log \left(\epsilon
	_{\text{1R}}\right)\right)\\\nonumber
	+\epsilon _{\text{2L}} \left(-\csc (\alpha ) \sec (\alpha ) \log \left(\epsilon _{\text{1B}}\right)+\frac{(3 \cot (\alpha )+2) \csc (\alpha ) \sec (\alpha ) \log
		\left(\epsilon _{\text{1L}}\right)}{2 (\cot (\alpha )+1)}-\frac{\csc (\alpha ) \log \left(\epsilon _{\text{1R}}\right)}{2 (\sin (\alpha )+\cos (\alpha ))}\right)\\\nonumber
	+\epsilon _{\text{2R}} \left(-\csc (\alpha ) \sec (\alpha ) \log \left(\epsilon _{\text{1B}}\right)-\frac{\csc (\alpha ) \log \left(\epsilon _{\text{1L}}\right)}{2
		(\sin (\alpha )+\cos (\alpha ))}+\frac{(3 \cot (\alpha )+2) \csc (\alpha ) \sec (\alpha ) \log \left(\epsilon _{\text{1R}}\right)}{2 (\cot (\alpha )+1)}\right)
	\Biggl. \Biggr)
	\Biggl.	\Biggr )
	\Bigr. ,
\end{gather}

\begin{gather}\label{eq:u}
	(T^{-,-}_{S m,S m})_{\text{Channel 4}}=-\frac{\Omega^4}{128\, \varphi _{z,\text{IN}} \left(\varphi _{z,\text{OUT}}^2-1\right) }\\\nonumber
	\Bigl(\Bigr.
	64 \epsilon _{\text{1B}} \epsilon _{\text{2B}}+16 \epsilon _{\text{1B}} \epsilon _{\text{2L}}+16 \epsilon _{\text{1B}} \epsilon _{\text{2R}}+16 \epsilon _{\text{2B}}
	\epsilon _{\text{1L}}+16 \epsilon _{\text{2B}} \epsilon _{\text{1R}}+\frac{4 (\cos (\alpha )-\sin (\alpha )) \left(\epsilon _{\text{1L}}-\epsilon _{\text{1R}}\right) \left(\epsilon _{\text{2L}}-\epsilon _{\text{2R}}\right)}{\sin (\alpha
		)+\cos (\alpha )}
	\\\nonumber
	\csc ^2\left(\frac{1}{4} (2 \alpha +\pi )\right) \sec (\alpha ) (2 \sin (\alpha )+\cos (2 \alpha )+3) \left(\epsilon _{\text{1B}}-\epsilon _{\text{1R}}\right)
	\left(\epsilon _{\text{2B}}-\epsilon _{\text{2L}}\right)
	\\\nonumber
	-4 \cot (\alpha ) \cot ^2\left(\frac{1}{4} (2 \alpha +\pi )\right) \left(\epsilon _{\text{1B}}-\epsilon
	_{\text{1L}}\right) \left(\epsilon _{\text{2B}}-\epsilon _{\text{2L}}\right)
	\\\nonumber
	+\csc ^2\left(\frac{1}{4} (2 \alpha +\pi )\right) \sec (\alpha ) (2 \sin (\alpha )+\cos (2 \alpha )+3) \left(\epsilon _{\text{1B}}-\epsilon _{\text{1L}}\right)
	\left(\epsilon _{\text{2B}}-\epsilon _{\text{2R}}\right)
	\\\nonumber
	-4 \cot (\alpha ) \cot ^2\left(\frac{1}{4} (2 \alpha +\pi )\right) \left(\epsilon _{\text{1B}}-\epsilon _{\text{1R}}\right) \left(\epsilon _{\text{2B}}-\epsilon
	_{\text{2R}}\right)
	\\\nonumber
	4 \left(2 \tan \left(\frac{\alpha }{2}\right)+\cot (\alpha )\right) \left(\left(\epsilon _{\text{1B}}-\epsilon _{\text{1L}}\right) \left(\epsilon _{\text{2B}}-\epsilon
	_{\text{2L}}\right)+\left(\epsilon _{\text{1B}}-\epsilon _{\text{1R}}\right) \left(\epsilon _{\text{2B}}-\epsilon _{\text{2R}}\right)\right)
	\\\nonumber
	-4 \tan ^2\left(\frac{\alpha }{2}\right) \tan (\alpha ) \left(\left(\epsilon _{\text{1B}}-\epsilon _{\text{1L}}\right) \left(\epsilon _{\text{2B}}-\epsilon
	_{\text{2L}}\right)+\left(\epsilon _{\text{1B}}-\epsilon _{\text{1R}}\right) \left(\epsilon _{\text{2B}}-\epsilon _{\text{2R}}\right)\right)
	\Bigl)\Bigr. ,
\end{gather}


\begin{gather}\label{eq:u}
	(T^{-,-}_{S m,S m})_{\text{Channel 5}}=
	\left (6 \epsilon _{\text{2B}}+\epsilon _{\text{2L}}+\epsilon _{\text{2R}}\right)	\frac{\Omega ^2 \left(\Omega ^2 \varphi _{z,\text{IN}}+1\right)}{8 \varphi _{z,\text{IN}}^2 \left(\varphi _{z,\text{OUT}}-1\right) \left(\varphi
		_{z,\text{OUT}}+1\right)}\;.
\end{gather}
\clearpage
\twocolumngrid

\appendix{\bf{Appendix D: Fourier transform of a constant function on a polygon. Case $q_{\perp}\neq 0$}}

Start with the Stokes theorem applied for a polygon in the (x,y)-plane, \cite{wuttke2017form},
\begin{equation}\label{Stokes}
	\iint\limits_\text{polygon} (\nabla \times \vv{F})\cdot \vv{dA}=\oint\limits_{\text{right-oriented boundary}}\vv{F}\cdot\vv{ds},
\end{equation}		
where $\vv{dA}=\hat{z} dx dy$. Take the filed $\vv{F}$ to be of a special case $\vv{F}(x,y)=(\hat{z}\times W) e^{q_{\perp}\cdot r_{\perp}}$, with $W=(W_x,W_y)$ a constant 2-dimensional field.
Use the right-oriented pair $(\hat{\mu}_j,\hat{\lambda}_j)$ to compute, along each segment of the polygon, the right-hand  integral from (\ref{Stokes}). The integral over each segment gives different results for $q_{\perp}\cdot \hat{\mu}_j \neq 0$ or $q_{\perp}\cdot \hat{\mu}_j = 0$. At this stage in the computations we notice two features
(i) the sum over segments is written in terms of sums over vertices,
(ii) we can drop the requirement of using a specific orientation of the polygon by using both $\hat{\mu}_j$ and $-\hat{\mu}_j$ for the $j^{\text{th}}$-segment. To accomplish this, if  $\hat{\mu}_j$ is associated with the vertex $V_j$, then associate $-\hat{\mu}_j$ with the vertex $V_{j+1}$.
The result is a vertex-wise formula, with the sum running over the index $k$ that labels the vertices of the polygon 
\begin{equation}\label{Stokes4}
	\begin{split}
		\iint\limits_\text{polygon} dx dy e^{i q_{\perp}\cdot r_{\perp}}=&\sum_{k\in \text{vertices}}\pm\frac{W\cdot \hat{\lambda}_k}{ W\cdot q_{\perp}}\; \frac{1}{ q_{\perp}\cdot \hat{\mu}_k}\;e^{i q_{\perp}\cdot V_{k}}
	\end{split}.
\end{equation}	

The $\pm$ is the last remnant of the need of an orientation in the Stokes's theorem. To incorporate this $\pm$ into an unoriented polygon, place a permittivity $\epsilon$ inside the polygon, $\epsilon=0$ outside and define the change in $\epsilon$ like in (\ref{eq:JumpEpsilon}).

The need for an orientation in the Stokes' theorem is exchange with the need to use a right-oriented pair $(\hat{\mu},\hat{\lambda})$. This, in turn, imposes the jump to change sign when $\hat{\mu}$ changes sign,
which makes the $\pm$ in (\ref{Stokes4}) to become part of $(\Delta\epsilon)_{\hat{\mu}}$.
We thus arrived at the formula  (\ref{Eq81}).

Note that the four corners of the unit cell contribute to the (00)-type Fourier integrals. For Channel 1, which is permittivity-independent, the corners contribute with jumps  $(\Delta(f_1f_2))_{\hat{\mu}}$ that are equal to $\pm 1$, given that outside the unit cell we work with $\epsilon=0$.
\vspace{1em}
\appendix{\bf{Appendix E: Fourier transform of a constant function on a polygon. Case $q_{\perp}= 0$}}

This integral is the area of the polygon, for which a simple answer is given by the sum of the areas of triangles, each triangle producing a term $\frac{1}{2}\hat{z}\cdot (V_{j+1}\times V_j).$ The disadvantage for us is that the vertices $V_{j+1}$ and $V_J$ are not separated, as they are in the formula for $q_{\perp}\neq 0$.
Start from (\ref{Eq81}) and take the limit of $q_{\perp}\rightarrow 0$ along a selected direction $\mathcal{D}$
\begin{equation}
	q_{\perp} = s \mathcal{D}\rightarrow 0,\;\text{as}\; s\rightarrow 0.
\end{equation}
We will show that the final result is independent of the choice of the direction $\mathcal{D}.$
With this choice of $q_{\perp}$ and $\mathcal{D}\cdot\hat{\mu}\neq 0$ for all $\hat{\mu}$, formula  (\ref{Eq81}) becomes
\begin{gather}\label{FormulaD1}
	\;\sum_{\text{vertices}}e^{i \mathcal{D}\cdot V s}\sum_{\hat{\mu}_\text{vertex}}\frac{W\cdot\hat{\lambda}}{W\cdot \mathcal{D}} \frac{1}{s}(-\Delta\epsilon)_{\hat{\mu}} \frac{1}{\mathcal{D}\cdot\hat{\mu}}\frac{1}{s}\;.
\end{gather}

In (\ref{FormulaD1}) there is a second-order singularity, $s^{-2}$, which we need to eliminate. To do just that, expand the exponential in Taylor series and examine the terms that belong to  $\hat{\mu}$ and $-\hat{\mu}$ oriented along the same edge and notice that the second-order singularity is eliminated. However, a first-order singularity $s^{-1}$ came into being. This singularity goes away too once we sum up its contribution along a closed loop.
Once the singularity $s^{-1}$ is eliminated, we are left, in the limit $s\rightarrow 0$, with one term per segment
\begin{equation}
	\frac{W\cdot\hat{\lambda}_j}{W\cdot \mathcal{D}}  \frac{1}{\mathcal{D}\cdot\hat{\mu}_j}(-\Delta\epsilon)_{\hat{\mu}_j}\frac{1}{2}\left((i \mathcal{D}\cdot V_j)^2-(i \mathcal{D}\cdot V_{j+1})^2\right).
\end{equation}

This term can be placed back into a format that contains two $\hat{\mu}$ vectors per each edge. This is the result (\ref{StokesFinalResultqZEROTHEOREM}).

We  prove that the result thus obtained is independent on the choice of $W$ and $\mathcal{D}$ by showing that
(\ref{StokesFinalResultqZEROTHEOREM}) is equal to sum of areas of triangles.
Call the vertices of the j-edge by $V_{j}$ and $V_{j+1}$. Placed at $V_j$ and $V_{j+1}$ are $(\hat{\lambda}_j,\hat{\mu}_j)$ and $(\hat{\lambda}_{j+1},\hat{\mu}_{j+1})=(-\hat{\lambda}_j,-\hat{\mu}_j)$, respectively. 
Along the computation we use  $V_{j+1}-V_j=\hat{\mu}_j l_j$, where $l_j$ is the length of the edge. Change the dot product into a cross product,  $W\cdot \hat{\lambda}=(\hat{z}\times W)\cdot \hat{\mu}$, to prove the equality

\begin{equation}\label{SecondProofZeroq}
	\begin{split}
		\frac{W\cdot\hat{\lambda}_j}{W\cdot \mathcal{D}}  \frac{1}{\mathcal{D}\cdot\hat{\mu}_j}(-\Delta\epsilon)_{\hat{\mu}_j}\frac{1}{2}(i \mathcal{D}\cdot V_j)^2+&\\
		\frac{W\cdot\hat{\lambda}_{j+1}}{W\cdot \mathcal{D}}  \frac{1}{\mathcal{D}\cdot\hat{\mu}_{j+1}}(-\Delta\epsilon)_{\hat{\mu}_{j+1}}\frac{1}{2}(i \mathcal{D}\cdot V_{j+1})^2=&\\
		(-\Delta\epsilon)_{\hat{\mu}_j}	\frac{1}{W\cdot \mathcal{D}}  \frac{1}{2}(\hat{z}\times W)\cdot &\\ \left(\frac{V_{j+1}-V_j}{l_j}( \mathcal{D}\cdot V_j)+\frac{V_{j+1}-V_j}{l_j}( \mathcal{D}\cdot V_{j+1})\right)l_j\; .
	\end{split}
\end{equation}

There are four terms in the last line of (\ref{SecondProofZeroq}). Two out of these four sum up to zero when we go around a closed loop
keeping in mind that $(-\Delta\epsilon)_{\hat{\mu}_j}$ is the same for all $\hat{\mu}_j$ because we are working for a polygon with one kind of permittivity inside and $\epsilon=0$ outside.

We are left with the sum of the other two terms, which, by the Binet-Cauchy vector relation, is transformed into a sum of  oriented areas of triangles $O V_j V_{j+1}$, where $O$ is some arbitrary origin.

\vspace{1em}
\appendix{\bf{Appendix F: The integral (00) on a polygonal tessellation of the plane}  }
In the tessellated plane, for a selected vertex $V$ and a pair $(\hat{\lambda},\hat{\mu})$ associated with this vertex, there are exactly two contributions from the two polygons  that have a common edge along $\hat{\mu}$.
 Two permittivity jumps along $\hat{\mu}$ originate from these two distinct polygons, each polygon being surrounded by a medium with $\epsilon=0$. The sum of these polygonal jumps is actually equal to $(-\Delta(f_1f_2))_{\hat{\mu}}$ but on the tessellated structure.
We arrived at the formula for the polygonal tessellation (\ref{Eq81}).
\vspace{1em}
\appendix{\bf{Appendix G: Translation of a virtual point}}\label{section:TranslationV12}

A virtual point appears at the intersection of a line from plane 1 and another line from plane 2, which are described by $r_1=V_{12}+\mu_1\hat{\mu}_1$ and $r_2=V_{12}+\mu_2 \hat{\mu}_2$, respectively. Parameters $\mu_1$ and $\mu_2$ run on the real axis.
Apply a translation in plane 1, 
$\delta r_1=(\delta x_1,\delta y_1)$, and plane 2,
$\delta r_2=(\delta x_2,\delta y_2)$, respectively.
Written in terms of a new set of  parameters, $\mu_1^{'}$ and $\mu_2^{'}$, the translated lines are 
\begin{gather}
	\begin{split}
		(r_1)_{\text{translated}}=&V_{12}-(\delta r_1\cdot \hat{\lambda}_1)\hat{\lambda}_1+\mu_1^{'}\hat{\mu}_1,\\
		(r_2)_{\text{translated}}=&V_{12}-(\delta r_2\cdot \hat{\lambda}_2)\hat{\lambda}_2+\mu_2^{'}\hat{\mu}_2.
	\end{split}	
\end{gather}
To compute the position of the translated intersection, $(V_{12})_{\text{translated}}$, solve for $\mu_1^{'}$ and $\mu_2^{'}$ such that $(r_1)_{\text{translated}}=(r_2)_{\text{translated}}$.
Use $\hat{\mu}_2\cdot\hat{\lambda}_1=-\hat{\mu}_1\cdot\hat{\lambda}_2$ and get at the final result (\ref{eq:V12Translation}).
\vspace{1em}
\appendix{\bf{Appendix H: Fourier transforms for product of derivatives of type (01), (10) and (11)}}

%
We expand on (\ref{Taylor:fgCase2}) and (\ref{Taylor:fgCase3}).
In order to simplify the computation of $\partial_{\delta x_1}$ and others of this kind, for case (2), (i.e. $q_{\perp}\cdot \hat{\mu}= 0$ and $q_{\perp}\neq 0$), compound the product of the two vertex-dependent terms in one exponential
\begin{equation}
	e^{i q_{\perp}\cdot V}(i \hat{\mu}\cdot V)=\frac{d}{d s}\left( e^{i q_{\perp}\cdot V+s i \hat{\mu}\cdot V}\right)|_{s=0}\;.
\end{equation}	
For case (3), $q_{\perp}=0$, in a similar way as above, we get
\begin{equation}
	\frac{1}{2} (i \mathcal{D}\cdot V)^2=\frac{1}{2} \left(\frac{d^2}{ds^2}\right)_{s= 0}e^{i (\mathcal{D}\cdot V)s }.
\end{equation}	

To group together the results for case (1) and (3), we found it useful to define a  function  $e^{i (q_{\perp}\cdot V)_m }$, which for  $q_{\perp}\neq0$ it is equal to  $e^{i q_{\perp}\cdot V }$. For  $q_{\perp}=0$ and for any natural number $m$ it is $\frac{1}{2} \left(\frac{d^2}{ds^2}\right)_{s= 0}\left(e^{i (\mathcal{D}\cdot V)s }s^{m}\right)$. So
\begin{equation}
	e^{i (q_{\perp}\cdot V)_m }=\begin{dcases}
		e^{i (q_{\perp}\cdot V) } &\text{for all}\; m,\text{if}\; q_{\perp}\neq 0,\\
		\frac{1}{2}(i \mathcal{D}\cdot V)^2 &\text{if}\; m=0,\text{and}\; q_{\perp}= 0,\\
		i \mathcal{D}\cdot V &\text{if}\; m=1,\text{and}\; q_{\perp}= 0,\\
		1&\text{if}\; m=2,\text{and}\; q_{\perp}= 0,\\
		0&\text{if}\; m\geqslant3,\text{and}\; q_{\perp}= 0.\\
	\end{dcases}
\end{equation}
The function, for $q_{\perp}=0$, is equal to the term number  $2-m$,  from the Taylor expansion.

With the help of this function, the contribution to the type (10)-integral, Section \ref{sec:FourierChi},
 from a pair from plane 1, i.e. $(V_1,\hat{\mu})$, is 
\begin{equation}\label{Taylor:fg}
	\begin{split}
	e^{i (\mathcal{D}\cdot V_1)_{m_{x}^1+m_{y}^1} } (-i\mathcal{D}_x )^{m_{x}^1} (-i\mathcal{D}_y )^{m_{y}^1}\\
		\frac{W\cdot\hat{\lambda}}
		{W\cdot \mathcal{D}}(-\Delta(f_1f_2)_{\hat{\mu}})\frac{1}{\mathcal{D}\cdot\hat{\mu}},
	\end{split}
\end{equation}
where $\mathcal{D}$ is arbitrary, subject to the condition that $\mathcal{D}\cdot\hat{\mu}\neq0$ for all $\hat{\mu}$ in plane 1.
The contribution to the type (01)-integral from all real points from plane 2, i.e. $V_2$, is a mirror image. 

The contribution to the type (11)-integral from a virtual point, $V_{12}$, under constrained translation has a different aspect, being
\begin{equation}\label{Taylor:fg}
	\begin{split}
		\sum_{\text{all four}\; \hat{\mu}^{\text{'}}\text{s}}e^{i (\mathcal{D}\cdot V_{12})_{m_{x}^1+m_{y}^1+m_{x}^2+m_{y}^2} }&\\
		(\lambda_{1x})^{m_{x}^1}	(\lambda_{1y})^{m_{y}^1}
		(\lambda_{2x})^{m_{x}^2}	(\lambda_{2y})^{m_{y}^2}
		&\\	\left(i\frac{\mathcal{D}\cdot\hat{\mu}_2}{\hat{\mu}_1\cdot\hat{\lambda}_2} \right)^{m_{x}^1+m_{y}^1}
		\left(i\frac{\mathcal{D}\cdot\hat{\mu}_1}{\hat{\mu}_2\cdot\hat{\lambda}_1} \right)^{m_{x}^2+m_{y}^2}&\\
		\frac{W\cdot\hat{\lambda}}{W\cdot \mathcal{D}}(-\Delta(f_1f_2)_{\hat{\mu}})\frac{1}{\mathcal{D}\cdot\hat{\mu}}\; .
	\end{split}
\end{equation}
To avoid carrying over a minus sign and to arrive at the above formula, we use $\hat{\mu}_2\cdot\hat{\lambda}_1=-\hat{\mu}_1\cdot\hat{\lambda}_2$.
The sum, for a fixed virtual point $V_{12}$, is over all four $\hat{\mu}^{\text{'}}\text{s}$, which have the tail located at $V_{12}$.

The above sum, if needed, can be place in a form which is independent of the gauge vector $W$. For that, first notice that 
\begin{equation}\label{eq:CommonFactor}
	\begin{split}
		(\lambda_{1x})^{m_{x}^1}	(\lambda_{1y})^{m_{y}^1}
		(\lambda_{2x})^{m_{x}^2}	(\lambda_{2y})^{m_{y}^2}
		\\	\left(i\frac{\mathcal{D}\cdot\hat{\mu}_2}{\hat{\mu}_1\cdot\hat{\lambda}_2} \right)^{m_{x}^1+m_{y}^1}
		\left(i\frac{\mathcal{D}\cdot\hat{\mu}_1}{\hat{\mu}_2\cdot\hat{\lambda}_1} \right)^{m_{x}^2+m_{y}^2}
	\end{split}
\end{equation}
is the same for all four $\hat{\mu}^{\text{'}}\text{s}$, which have the tail located at $V_{12}$. 

Second, use the relation
\begin{equation}\label{eq:SumFourMus}
	\begin{split}
		\sum_{\text{all four}\; \hat{\mu}^{\text{'}}\text{s}} \frac{W\cdot\hat{\lambda}}{W\cdot \mathcal{D}}(-\Delta(f_1f_2)_{\hat{\mu}})\frac{1}{\mathcal{D}\cdot\hat{\mu}}=\\
		\frac{\abs{\hat{\lambda}_1\cdot\hat{\mu}_2}}{(\mathcal{D}\cdot\hat{\mu}_1)(\mathcal{D}\cdot\hat{\mu}_2)}
		(\Delta f)_{\hat{\mu}_1}(\Delta g)_{\hat{\mu}_2}
	\end{split}
\end{equation}	
valid for any two choices of $\hat{\mu}_1$ and $\hat{\mu}_2$,  which have the tail located at $V_{12}$ and the tip on line 1 and line 2, respectively. 

The contribution of a virtual point can be thus brought into a W-independent form
\begin{equation}\label{Taylor:fg}
	\begin{split}
		\sum_{V_{12},\text{two}\; \hat{\mu}^{\text{'}}\text{s}}e^{i (\mathcal{D}\cdot V_{12})_{m_{x}^1+m_{y}^1+m_{x}^2+m_{y}^2} }
		&\\	\left(i\frac{\mathcal{D}\cdot\hat{\mu}_2}{\hat{\mu}_1\cdot\hat{\lambda}_2}\lambda_{1x} \right)^{m_{x}^1}
		\left(i\frac{\mathcal{D}\cdot\hat{\mu}_2}{\hat{\mu}_1\cdot\hat{\lambda}_2}\lambda_{1y} \right)^{m_{y}^1}
		&\\	\left(i\frac{\mathcal{D}\cdot\hat{\mu}_1}{\hat{\mu}_2\cdot\hat{\lambda}_1}\lambda_{2x} \right)^{m_{x}^2}
		\left(i\frac{\mathcal{D}\cdot\hat{\mu}_1}{\hat{\mu}_2\cdot\hat{\lambda}_1}\lambda_{2y} \right)^{m_{y}^2}&\\
		\frac{\abs{\hat{\lambda}_1\cdot\hat{\mu}_2}}{(\mathcal{D}\cdot\hat{\mu}_1)(\mathcal{D}\cdot\hat{\mu}_2)}
		(\Delta f)_{\hat{\mu}_1}(\Delta g)_{\hat{\mu}_2}\; .
	\end{split}
\end{equation}

To show (\ref{eq:SumFourMus}), choose  two out of four vectors that have the tail on the virtual point and direction on the line in plane 1 and plane 2, and call them $\hat{\mu}_1$ and $\hat{\mu}_2$, respectively. The other two vectors are $-\hat{\mu}_1$ and $-\hat{\mu}_2$.
Use 
\begin{equation}
	\begin{split}
		-\Delta(f_1f_2)_{\hat{\mu}_1}-\Delta(f_1f_2)_{-\hat{\mu}_1}=\\
		-(-\Delta(f_1f_2)_{\hat{\mu}_2}-\Delta(f_1f_2)_{-\hat{\mu}_2})
	\end{split}
\end{equation}
and get 
\begin{equation}\label{eq:SumFourMusProof2}
	\begin{split}
		\sum_{\text{all four}\; \hat{\mu}^{\text{'}}\text{s}} \frac{W\cdot\hat{\lambda}}{W\cdot \mathcal{D}}(-\Delta(f_1f_2)_{\hat{\mu}})\frac{1}{\mathcal{D}\cdot\hat{\mu}}=&\\
		\left(\frac{W\cdot\hat{\lambda}_1}{W\cdot \mathcal{D}}\frac{1}{\mathcal{D}\cdot\hat{\mu}_1}-\frac{W\cdot\hat{\lambda}_2}{W\cdot \mathcal{D}}\frac{1}{\mathcal{D}\cdot\hat{\mu}_2}\right)&\\
		(-\Delta(f_1f_2)_{\hat{\mu}_1}-\Delta(f_1f_2)_{-\hat{\mu}_1}).
	\end{split}
\end{equation}	
A vector identity, valid for any $W$, $\mathcal{D}$ and two pairs
$(\hat{\lambda}_1,\hat{\mu}_1)$ and $(\hat{\lambda}_2,\hat{\mu}_2)$, 
\begin{equation}
	({W\cdot\hat{\lambda}_1})(\mathcal{D}\cdot\hat{\mu}_2)-({W\cdot\hat{\lambda}_2})(\mathcal{D}\cdot\hat{\mu}_1)=(W\cdot \mathcal{D})(\hat{\lambda}_1\cdot\hat{\mu}_2)
\end{equation}
brings the right-hand side of (\ref{eq:SumFourMusProof2}) into the form
\begin{equation}\label{eq:SumFourMusProof3}
	\begin{split}
		\left(\frac{\hat{\lambda}_1\cdot \hat{\mu}_2}{(\mathcal{D}\cdot\hat{\mu}_1)(\mathcal{D}\cdot\hat{\mu}_2)}\right)(-\Delta(f_1f_2)_{\hat{\mu}_1}-\Delta(f_1f_2)_{-\hat{\mu}_1}).
	\end{split}
\end{equation}	
The final step is to prove that
\begin{equation}
	\begin{split}
		(-\Delta(f_1f_2)_{\hat{\mu}_1}-\Delta(f_1f_2)_{-\hat{\mu}_1})(\hat{\lambda}_1\cdot \hat{\mu}_2)=\\
		(\Delta f_1)_{\hat{\mu}_1}(\Delta f_2)_{\hat{\mu}_2}\abs{\hat{\lambda}_1\cdot \hat{\mu}_2}\; .
	\end{split}
\end{equation}

This last equality can be established by going through every of the four cases of pairs $(\hat{\lambda},\hat{\mu})$. Another way is to notice that $-\Delta(f_1f_2)_{\hat{\mu}_1}-\Delta(f_1f_2)_{-\hat{\mu}_1}$ is equal to $(\Delta f)_{\hat{\mu}_1}(\Delta g)_{\hat{\mu}_2}$ up to a sign, i.e. $-\Delta(f_1f_2)_{\hat{\mu}_1}-\Delta(f_1f_2)_{-\hat{\mu}_1}=\pm (\Delta f_1)_{\hat{\mu}_1}(\Delta f_2)_{\hat{\mu}_2}$. This is so, because the left-hand side is invariant under the change in sign $\hat{\mu}_1\rightarrow -\hat{\mu}_1$, whereas the right-hand side changes its sign. The absolute value eliminates this change in sign and so the equality is valid.

\vspace{1em}
\appendix{\bf{Appendix I: The optimization procedure for selecting the dielectric constants}}\label{sec:Optimization}
\begin{table}[h]
\caption{\label{tab:Rayleigh}Mode ordering and Rayleigh frequency for $\theta=17.5^{\circ}$ and $\varphi=0^{\circ}$}
\begin{ruledtabular}
\begin{tabular}{llll}
	$(M_x,M_y)$ & m & Pol.,m & $f_{R}$[GHz]\\
	\colrule\\[-1ex]
	(0,0)&1 & S1=S(0,0),\;\;\;\;P1=P(0,0) & 0.0 \\
	(-1,0)&2 & S2=S(-1,0),\;\;\;P2=P(-1,0) & 36.6827 \\
	(0,-1)&3 & S3=S(0,-1),\;\;\;P3=P(0,-1) & 50.0289 \\
	(0,1)&4 & S4=S(0,1),\;\;\;\:\,P4=P(0,1) & 50.0289 \\
	(-1,-1)&5 & S5=S(-1,-1),\; P5=P(-1,-1) & 56.7146 \\
	(-1,1)&6 &  S6=S(-1,1),\;\;P6=P(-1,1) & 56.7146 \\
	(1,0)&7 &  S7=S(1,0),\;\;\;P7=P(1,0) & 68.23087 \\
	(-2,0)&8 & S8=S(-2,0),\; P8=P(-2,0) & 73.36548 \\
	(-2,-1)&9 &  S9=S(-2,-1),\;P9=P(-2,-1) & 84.68336 \\
	(-2,1)&10 & S10=S(-2,1),\;P10=P(-2,1) & 84.68336\\
\end{tabular}
\end{ruledtabular}
\footnotetext{The odd and even Polarization-Bloch-Floquet No., in column 3, corresponds to the S and P polarization, respectively. It is easier to use the index that runs from 1 to 20 in columns 3, when working with a $20\times 20$ matrix. However, to associate a meaning to the Polarizatio-Bloch-Floquet No., the other two styles  shown in column 3 may be better. }
\end{table}

Construct, out of (\ref{eq:DeffTransfMatr}),  a $8 \times 5$ sub-matrix  composed of transfer matrix elements that couples only the S-polarization of the four modes selected, $(0,0),(-1,0),(0,-1),(0,1)$.  The equations to solve are obtained from formula (\ref{eq:DefinitionTransferElements}). The OUT  set (DIR,POL,BF) is composed of ($\pm$,S,m), with $m=1,\dots,4$ from Table \ref{tab:Rayleigh}. Given the absence of any source at $z=+\infty$, the coefficients $C_\text{POL,BF}^{\text{DIR}}$ are all zero for ($-$,S,m), $m=1,\dots,4$. At IN, the set (dir,pol,bf) contains the incoming ($+$,S,1) and the reflected ($-$,S,m), with $m=1,\dots,4$. There are 8 unknown coefficients, 4 at OUT and another 4 at IN.

The $8\times 5$ matrix transfers the input column vector $(\bm{\mathcal{C}}^{+}_{S1},\bm{\mathcal{C}}^{-}_{S1},\bm{\mathcal{C}}^{-}_{S2},\bm{\mathcal{C}}^{-}_{S3},\bm{\mathcal{C}}^{-}_{S4})$ into the output $(C^{+}_{S1},C^{+}_{S2},C^{+}_{S3},C^{+}_{S4},0,0,0,0)$. For ease of reading, the IN and OUT coefficients are marked using distinct fonts. Within this approximation, we can solve $8$ equations for $8$ unknowns coefficients, by fixing the input $\bm{\mathcal{C}}^{+}_{S1}=1$. The coefficients thus found are the transmission and reflection coefficients for the $4$ modes under consideration. 
To construct an objective function for the optimization process we use a series of properties. The first property is that in the vicinity of the resonant frequency the absolute value of the determinant of the $4\times 4$ matrix $(T^{-,-}_{Sm,Sn})$ attains a minimum. The second one is that the absolute value of the transmission and reflection coefficients of the evanescent modes need to attain high values. The third property is based on the absorptance factor defined as $1-\abs{C^{+}_{S1}}^2-\abs{C^{+}_{S2}}^2-\abs{\bm{\mathcal{C}}^{-}_{S1}}^2-\abs{\bm{\mathcal{C}}^{-}_{S2}}^2$, similarly to the definition from \cite{polo2013electromagnetic}. The absorptance factor goes through a maximum around the resonant frequency.  
The objective function we use is

\begin{gather}\label{eq:OptimizationFunction}
\abs{Det(T^{-,-}_{Sm,Sn})}+\\\nonumber
\abs{\abs{C^{+}_{S1}}^2+\abs{C^{+}_{S2}}^2+\abs{\bm{\mathcal{C}}^{-}_{S1}}^2+\abs{\bm{\mathcal{C}}^{-}_{S2}}^2-0.5}\\\nonumber+
\abs{\abs{C^{+}_{S3}}-10}+\abs{\abs{C^{+}_{S4}}-10}\; .
\end{gather}

Only the transmission coefficients for the evanescent waves are present in (\ref{eq:OptimizationFunction}) because we expect that the reflection coefficients be very close, in absolute value, to the transmission values. We choose $10$ as a target number for this values. For the absorptance factor the target values is chosen at $0.5$. 

For a fixed frequency in the range $37$ GHz to $50$ GHz, we  found a local minimum for  this objective function in the 6-dimensional space of the dielectric constants, each constant constrained to be between 1 and 12. 
After scanning 100 frequencies, the local minimum we selected has $f=41.29$ GHz, $Det(T^{-,-}_{Sm,Sn}) =2.5\times 10^{-5}-3.7\times 10^{-5} i$, $C^{+}_{S1}= 0.59 - 0.0022 i$, $\bm{\mathcal{C}}^{-}_{S1}=  -0.21 - 0.59 i$, $C^{+}_{S2}= 0.041 + 0.31 i$, $\bm{\mathcal{C}}^{-}_{S2}=0.04 + 0.34 i$, $C^{+}_{S3}=-10.03 - 0.15 i$, $\bm{\mathcal{C}}^{-}_{S3}=-10.61 - 0.18 i$,
$C^{+}_{S4}=-3.92 + 9.20 i$ and $\bm{\mathcal{C}}^{-}_{S4}= -4.13 + 9.74 i$. 
We see that the obtained determinant is a small number and that the reflection and the transmissions coefficients are close, in absolute value, to the value 10 we imposed. The absorptance factor came out as $0.04$, far from the value of $0.5$. This is because in the vicinity of the resonant frequency the  transmission and reflection coefficients of the propagative modes attain a minimum at frequencies slightly different than the resonance. 
The dielectric permittivity set for the local minimum is 
\begin{gather}
\epsilon _{\text{1B}}= 4.98655,\epsilon _{\text{2B}}= 2.95855,\epsilon _{\text{1L}}= 7.59549,\\ \nonumber
\epsilon _{\text{2L}} = 5.663,\epsilon _{\text{1R}}=
5.92965,\epsilon _{\text{2R}}= 1.63434
\end{gather}
\appendix{\bf{Appendix J: Complex frequency plane}}

The real resonant frequency is located at the minimum of the determinant on the real $\Omega$-axis.
The zero of the cofactor is located in the upper complex plane, $\text{Im}(f_{zero})>0$ 
\begin{figure}[h]
\includegraphics[scale=1.00]{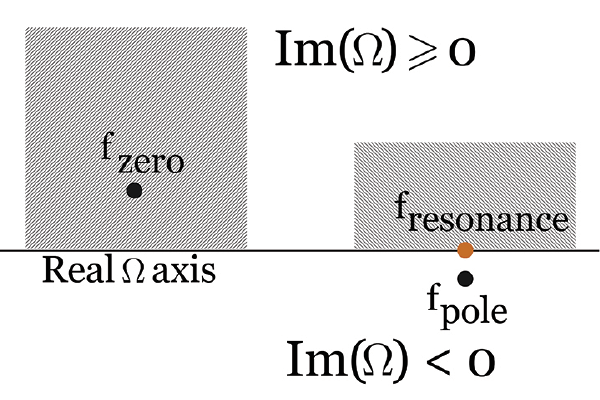}
\caption{\label{fig:Complex_plane_domains_Flattened} The domain used for the cofactor surrounds the zero $f_{zero}$. The domain for determinant does not contain $f_{pole}$, but it contains the resonant frequency. Both domains are above the real $\Omega$ axis}
\end{figure}. A search for the minimum of the absolute value of the $\text{Cofactor}_{S(0,0),S(0,1)}(T^{-,-})$ gives a complex frequency $ \Omega_{zero}= 0.213999052599+0.00014322547392 i$, for which the Cofactor attains the value of $-4.17186\times10^{-7} + 5.51181\times10^{-7} i$. Translated in frequency, using $c=0.25 \text{mm}$, we get $f_{zero}=40.8425 + 0.0273351 i$ in GHz. Once this zero is localized, the parameters $a$ and $b$ are estimated through a linear regression, \cite{Mathematica}.

The linear approximation of the cofactor in the vicinity of $\Omega_{zero}$,
\begin{equation}
\text{Cofactor}_{S(0,0),S(0,1)}(T^{-,-})\cong(a+i b) (\Omega-\Omega_{zero}),
\end{equation}
contains two unknown parameters, $a$ and $b$. The numerical data consists of pairs  $(\Omega,\text{Cofactor}_{S(0,0),S(0,1)}(T^{-,-}))$ obtained from sampling the upper complex-frequency plane  using the analytical formulas for the transfer matrix.

The linear approximation for the determinant starts with localizing  $\text{Re}(\Omega_{resonance})=0.21632449245970709$ on the real axis at the point where the absolute value of the determinant becomes minimum, attaining the value $1.15869$. In terms of the frequency, the resonance is at $\text{Re}(f_{resonance})=41.28635279688047$ GHz. The linear approximation of the determinant in the vicinity of $\Omega_{pole}=0.21632449245970709+ i y_{pole}$,
\begin{equation}
Det(T^{-,-})\cong(a^{'}+i b^{'}) (\Omega-\Omega_{pole}),
\end{equation}
contains three unknown parameters, $a^{'}$, $b^{'}$ and $y_{pole}$. The numerical data consists of pairs  $(\Omega,Det(T^{-,-}))$ obtained from sampling the upper complex-frequency plane around $Re(\Omega_{resonance})$ using the analytical formulas for the transfer matrix. The fit this time is nonlinear because of the product $(a^{'}+i b^{'}) y_{pole}$.
\begin{figure}[h]
\includegraphics[scale=1.0]{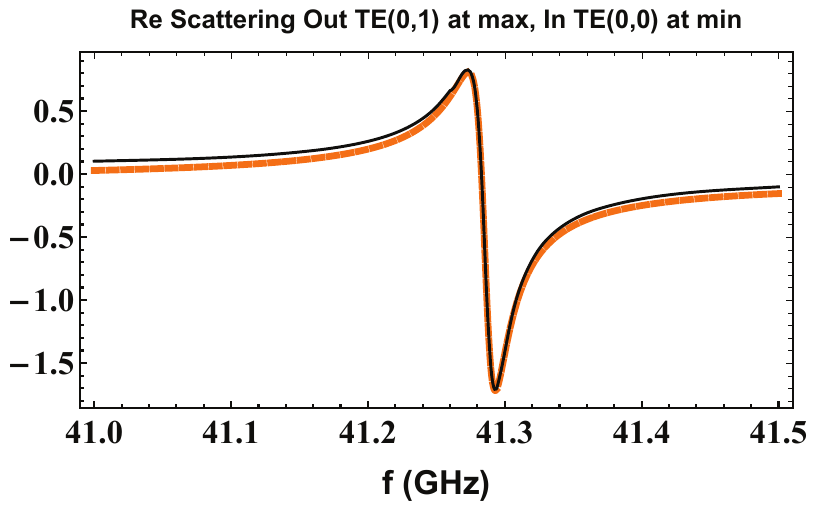}
\caption{\label{fig:ReScatteringOutTE01atmaxInTE00atmin} The real part of the scattering matrix element. The dashed curve (orange) is the zero-pole approximation, whereas the continuous curve (black) is derived from the 20 by 20 transfer matrix using the first 10 Rayleigh-ordered modes.}
\end{figure}
\begin{figure}[h]
\includegraphics[scale=1.0]{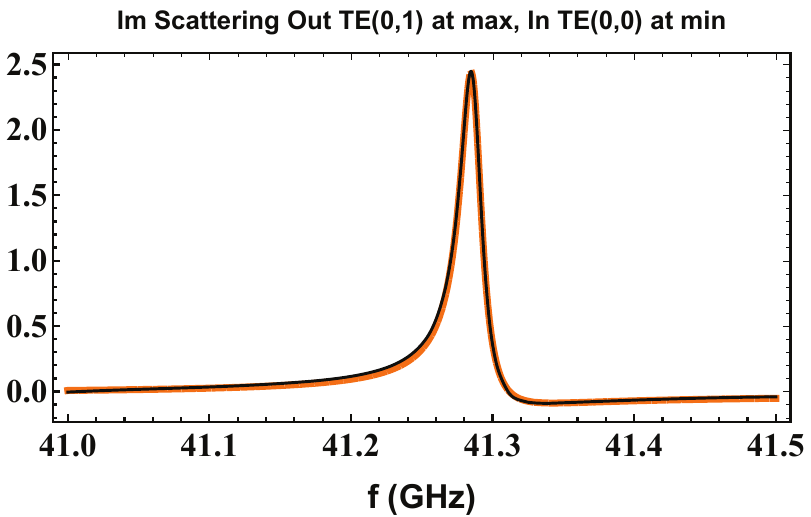}
\caption{\label{fig:ImScatteringOutTE01atmaxInTE00atmin}  The imaginary part of the scattering matrix element. The dashed curve (orange) is the zero-pole approximation, whereas the continuous curve (black) is derived from the 20 by 20 transfer matrix using the first 10 Rayleigh-ordered modes.}
\end{figure}
		
\bibliography{MicroScalingBibliography}

\end{document}